\documentclass[a4paper,fleqn]{cas-dc}

\usepackage[numbers]{natbib}
\usepackage{subcaption}
\usepackage{bm}
\usepackage{algorithmic}
\usepackage{svg}
\usepackage{hyperref}
\usepackage{hyphenat}

\newcommand{\dtau}{\Delta\tau}
\newcommand{\dt}{\Delta t}

\def\tsc#1{\csdef{#1}{\textsc{\lowercase{#1}}\xspace}}
\tsc{WGM}
\tsc{QE}
\tsc{EP}
\tsc{PMS}
\tsc{BEC}
\tsc{DE}

\begin{document}
\let\WriteBookmarks\relax
\def\floatpagepagefraction{1}
\def\textpagefraction{.001}
\shorttitle{HYMOR: An open-source package for global modal, non-modal, and receptivity
analysis in high-enthalpy hypersonic vehicles}
\shortauthors{A. Ant\'on-\'Alvarez et~al.}

\title [mode = title]{HYMOR: An open-source package for modal, non-modal,
  and receptivity analysis in high-enthalpy hypersonic vehicles}
\tnotemark[1,2]




\author[1]{A. Ant\'on-\'Alvarez}[orcid=0000-0002-6434-4211]
\cormark[1]
\ead{aantonal@caltech.edu}

\affiliation[1]{organization={Graduate Aerospace Laboratories, California Institute of Technology},
                city={Pasadena},
                postcode={91125}, 
                state={CA},
                country={USA}}

\author[1,2]{A. Lozano-Dur\'an}[orcid=0000-0001-9306-0261]

\affiliation[2]{organization={Department of Aeronautics and Astronautics, Massachusetts Institute of Technology},
                city={Cambridge},
                postcode={02139}, 
                state={MA},
                country={USA}}

\cortext[cor1]{Corresponding author}



\begin{abstract}
We present HYMOR (HYpersonic MOdal/non-modal, and Receptivity), an
open-source computational framework for the linear stability analysis
of high-enthalpy hypersonic flows. The toolkit includes MATLAB and
Julia implementations and is released under the MIT license. HYMOR
provides global modal, non-modal, and freestream receptivity analyses
capable of capturing interactions among spatially separated physical
mechanisms that are inaccessible to traditional local methods. A
shock-fitting formulation is employed to treat the bow shock as a
sharp discontinuity, ensuring that the interaction of infinitesimal
disturbances with the shock recovers the response predicted by linear
interaction analysis.  The code also solves the nonlinear equations
for base-flow computation and automatically linearizes the resulting
discrete operators for the stability analyses. Several thermochemical
models are available for the treatment of real-gas effects in
high-enthalpy regimes. The numerical implementation is verified
against a collection of benchmark cases that demonstrate the accuracy
and capabilities of the toolkit across its modal, non-modal, and
receptivity analysis modes.\\

\noindent\textbf{Program summary}\\[4pt]
\noindent
\begin{tabular}{@{}p{3.2cm}p{\dimexpr\textwidth-9.7cm}@{}}
\textit{Program Title:}                        & HYMOR --- HYpersonic MOdal/non-modal, and Receptivity \\[2pt]
\textit{CPC Library link to program files:}     & \url{https://doi.org/XX.XXXXX/XXXXXXXXXXX} \\[2pt] 
\textit{Developer's repository link:}           & \url{https://github.com/AdrianAA00/HYMOR} \\[2pt]
\textit{Licensing provisions:}                  & MIT \\[2pt]
\textit{Programming language:}                  & MATLAB, Julia \\[2pt]
\textit{Nature of problem:}                     & Predicting laminar-to-turbulent transition in hypersonic vehicles is critical for thermal protection system design. Existing open-source computational tools for stability analysis in this regime remain limited in scope, lacking an integrated framework for global modal, non-modal, and freestream receptivity analysis with thermochemical modeling and accurate shock–disturbance interaction treatment. \\[2pt]
\textit{Solution method:}                       & HYMOR employs a global stability formulation with shock-fitting to treat the bow shock as a sharp discontinuity, recovering exact linear interaction analysis behavior. The toolkit integrates modal, non-modal, and receptivity analyses with automatic linearization of discrete operators and multiple thermochemical models for high-enthalpy flows. Implementations are provided in MATLAB and Julia. \\[2pt]
\end{tabular}

\end{abstract}




\maketitle

\section{Introduction}
\label{sec::Introduction}

The laminar-to-turbulent transition of hypersonic flows remains one of
the central open problems in high-speed
aerodynamics~\citep{schneider2004hypersonic}. Because the turbulent
state of the boundary layer dramatically increases both skin friction
and, more critically, aerodynamic heating, the ability to predict when
and where transition occurs is essential for sizing thermal protection
systems and establishing vehicle design margins. Understanding the
physical mechanisms by which small disturbances are amplified within
the boundary layer is therefore a prerequisite for transition
prediction, and linear stability analysis has long served as the
principal theoretical tool for this
purpose~\citep{malik1980comparison}.

Early studies of transition mechanisms in compressible flows employed
linear stability theory (LST), which analyzes the growth of
infinitesimal perturbations superimposed on a locally parallel mean
flow through normal-mode solutions of the compressible Navier--Stokes
equations~\citep{lees1946investigation, mack1984boundary}. Subsequent
developments relaxed the parallel-flow assumption to represent more
realistic, spatially evolving boundary-layer configurations. A key
advance in this direction was the introduction of the parabolized
stability equations (PSE) method~\citep{bertolotti1991analysis,
  herbert1997parabolized}, which accounts for the slow streamwise
variation of the base flow and the disturbance envelope while
retaining computational efficiency. To address the role of nonlinear
modal interactions in transition, the PSE method was later generalized
to nonlinear PSE (NPSE)~\citep{bertolotti1992linear,
  chang1993linear}. More recently, global stability analyses have
emerged as powerful tools that resolve the full spatial dependence of
the disturbance field and can capture phenomena, such as interactions
between spatially separated physical mechanisms, that are inaccessible
to the local and weakly non-parallel formulations of LST and
PSE~\citep{theofilis2011global}.

Beyond modal instabilities, non-modal growth of disturbances
constitutes another important alternative route to
transition~\citep{reshotko2001transient}. Transient growth analysis
has proven successful in explaining discrepancies between theoretical
modal predictions and experimental observations in hypersonic
flows~\citep[e.g.][]{Bitter2014, Paredes2016, Dwivedi2020}.
Additionally, experimental evidence has indicated that transition in
high-speed regimes is strongly influenced by freestream
disturbances~\citep{laufer1954factors, kendall1975wind}, making
freestream receptivity a central link between the external disturbance
environment and the onset of transition over hypersonic vehicles. In
this context, the shockwave plays a critical role: freestream
disturbances must traverse the shock before reaching the boundary
layer, and the shock fundamentally modifies the nature of the
perturbations through wave-mode conversion. Beyond the
transmission of incident disturbances, the shock front itself may also
become intrinsically unstable for certain equations of state, giving
rise to D'yakov--Kontorovich instabilities and spontaneous acoustic
emission~\citep{calvorivera2023piston}. For small amplitude
perturbations, this process is studied via linear interaction analysis
(LIA)~\citep{cuadra2025review}. Correctly representing this interaction is essential for any receptivity analysis.  However,
shock-capturing methods, which are the most widespread numerical
treatment of shocks in hypersonic computations~\citep{zhang2016eno,
  vonneumann1950method}, introduce an artificial shock thickness that
has been shown to corrupt the LIA response of infinitesimal
disturbances~\citep{cuadra2025review}. Shock-fitting approaches, in
which the shock is treated as a sharp boundary with exact jump
conditions, eliminate this artifact and recover the correct LIA
behavior~\citep{bonfiglioli2014convergence, cuadra2025review}, making
it the natural choice for a linear stability analysis.

Several software packages have been developed by the community to
perform stability analyses in high-speed flows, including
LASTRAC~\citep{Chang2004_LASTRAC}, STABL~\citep{Johnson1998_STABL},
VESTA~\citep{Pinna2013_VESTA}, EPIC~\citep{Oliviero2015_EPIC}, and
NOLOT~\citep{Hein2005_NOLOT}. These tools have made important
contributions to the analysis of hypersonic-flow stability. However,
they primarily rely on local LST, PSE, or NPSE formulations, and their
applicability to high-enthalpy flows is not uniform, as not all of
them incorporate real-gas, chemical, or thermochemical-non-equilibrium
effects. In addition, not all of these codes are openly available to
the research community, and, to the best of our knowledge, none
provides within a single open-source package an integrated framework
for global stability analysis, transient growth, and freestream
receptivity in high-enthalpy flows.

This paper presents an open-source computational package, referred to
as HYMOR, developed to investigate transition prediction and advance
the physical understanding of transition mechanisms in high-enthalpy
hypersonic flows. The toolkit provides a global formulation for modal,
non-modal, and freestream-receptivity stability analysis that captures
interactions among spatially separated physical mechanisms, a
capability that lies beyond the reach of traditional methods. A
shock-fitting approach is incorporated to ensure accurate treatment of
disturbance--shock interactions across all analysis modes. In its
present form, this shock-fitting capability targets two-dimensional
and axisymmetric flows dominated by a single bow shock. Beyond the
linear tools, the code also provides a solver for the nonlinear
equations governing base flows, together with automatic linearization
of the discrete operators and built-in implementations of the linear
analysis capabilities. The toolkit is applicable to both low- and
high-speed regimes and incorporates several thermochemical models for
high-enthalpy flows in which real-gas effects become significant. All
components of the toolkit are verified against benchmark cases and
compared with selected reference solutions.

The remainder of this paper is structured as
follows. Section~\ref{sec::Methods} presents the continuous
formulation of the physical models, along with their numerical
discretization and implementation. Section~\ref{sec::Verification
  cases} describes several verification cases designed to validate the
different capabilities of the solver against established reference
solutions. Section~\ref{sec::Program documentation} provides program
documentation, including software requirements for both the Julia and
MATLAB versions of the toolkit, as well as a brief introduction to the
tutorial cases distributed with the source
code. Section~\ref{sec::Applications} demonstrates the application of
the toolkit to a realistic hypersonic vehicle geometry. Finally,
section~\ref{sec::Performance} examines the computational performance
and scaling behavior of the different capabilities offered by the
toolkit.

\section{Methods}
\label{sec::Methods}
\begin{figure*}[]
\centering
\includegraphics[width=1\linewidth]{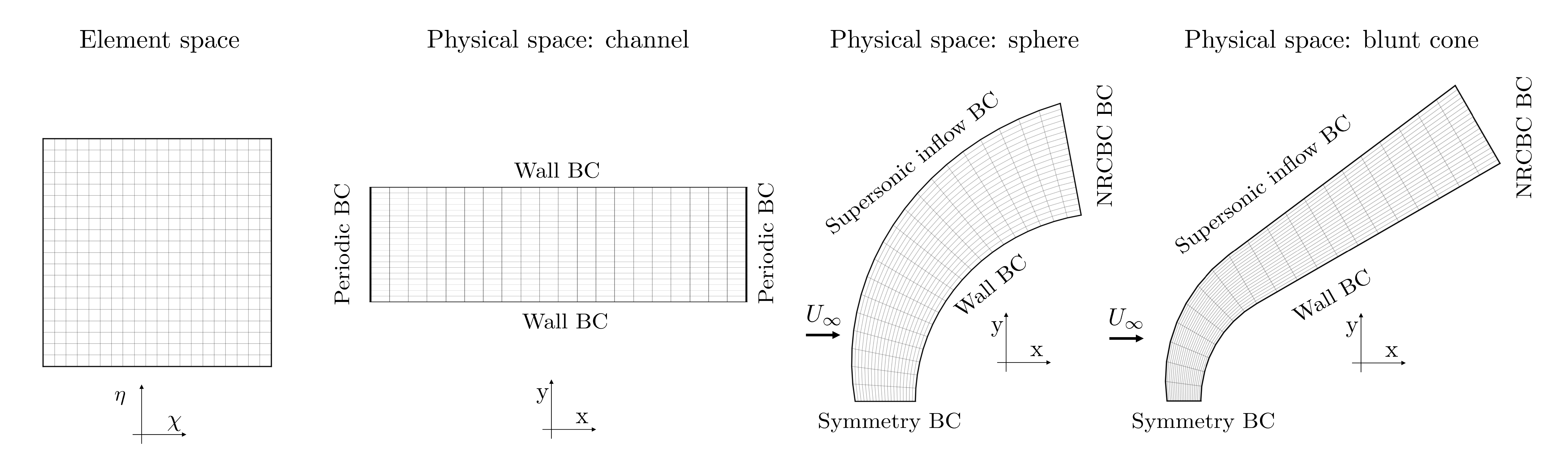}
\caption{Illustration of different possible cases that can be studied
  with HYMOR. The element mesh (coordinates $\chi$--$\eta$) is mapped
  through a curvilinear transformation to generate different meshes in
  physical space (coordinates $x$--$y$). Examples of different
  boundary conditions are also illustrated. }
\label{fig:config_set_up}
\end{figure*}

%
This section presents the theoretical formulation and numerical
implementation underlying HYMOR. An overview of the configurations
available in the framework is illustrated in
figure~\ref{fig:config_set_up}.

\subsection{Modeling equations}
\label{sec::Modeling equations}

\subsubsection{Compressible Navier-Stokes equations}
\label{sec::Compressible Navier-Stokes}

The gas dynamics are modeled using the compressible Navier--Stokes
equations closed with an ideal-mixture equation of state written in
terms of effective thermodynamic properties.  The solver can also
operate in inviscid mode.  We assume thermal equilibrium and use a
single-temperature formulation in which real-gas effects are
incorporated through effective thermodynamic and transport properties.
The governing equations, solved in non-dimensional form, are:
\begin{subequations}\label{eq:ns_compressible_nondim}
\begin{gather}
\frac{\partial \rho}{\partial t} + \frac{\partial (\rho u_{i})}{\partial x_{i}} = 0, \\[3pt]
\frac{\partial (\rho u_{i})}{\partial t} + \frac{\partial (\rho u_{i} u_{j})}{\partial x_{j}}
= -\frac{\partial \left( (\gamma^* - 1)\rho e \right)}{\partial x_{i}}
+ \frac{\partial \tau_{ij}}{\partial x_{j}}, \\[3pt]
\begin{split}
\frac{\partial \left(\tfrac{1}{2}\rho u_i u_i + \rho e\right)}{\partial t}
+ \frac{\partial \left[\left(\tfrac{1}{2}\rho u_i u_i + \gamma^* \rho e\right)u_{j}\right]}{\partial x_{j}}
\\
= \frac{\partial (\tau_{ij} u_{i})}{\partial x_{j}}
- \frac{\partial q_i}{\partial x_i},
\end{split} \\[3pt]
\tau_{ij} = \frac{\mu^*}{\mathrm{Re}_\mathrm{ref}}\left(\frac{\partial u_{j}}{\partial x_{i}} + \frac{\partial u_{i}}{\partial x_{j}}\right)
- \frac{\mu^*}{\mathrm{Re}_\mathrm{ref}}\frac{2}{3}\frac{\partial u_k}{\partial x_k}\,\delta_{ij}, \\[3pt]
q_i = -\frac{\gamma_\mathrm{ref} k^*}{\mathrm{Re}_\mathrm{ref} \,\mathrm{Pr}_\mathrm{ref}} \frac{\partial \left( \tfrac{e}{c_\mathrm{v}^*} \right)}{\partial x_{i}}.
\end{gather}
\end{subequations}
where $x_i$ and $t$ are the spatial and temporal coordinates,
respectively, $u_i$ are the velocity components, $\rho$ is the
density, $e$ is the specific internal energy, $\gamma^* = 1 +
\frac{p}{\rho e}$ is the effective ratio of specific heats,
$c_\mathrm{v}^* = \frac{e}{T}$ is the effective specific heat capacity
at constant volume, $\mu^* = \frac{\mu}{\mu_\mathrm{ref}}$ is the
non-dimensional dynamic viscosity, $k^* = \frac{k}{k_\mathrm{ref}}$ is
the non-dimensional thermal conductivity, $p$ is the pressure, and $T$
is the temperature. The preceding variables are non-dimensionalized
using a reference length $L_\mathrm{ref}$, velocity magnitude
$U_\mathrm{ref}$, density $\rho_\mathrm{ref}$, and specific heat at
constant volume $c_\mathrm{v,ref}$. In the following, all variables
are non-dimensionalized according to this convention unless explicitly
indicated. The resulting non-dimensional numbers are:
\begin{equation*}
\mathrm{Re}_\mathrm{ref} = \frac{\rho_\mathrm{ref} U_\mathrm{ref} L_\mathrm{ref}}{\mu_\mathrm{ref}}, \qquad
\gamma_\mathrm{ref} = \frac{c_\mathrm{p,ref}}{c_\mathrm{v,ref}}, \qquad
\mathrm{Pr}_\mathrm{ref} = \frac{\mu_\mathrm{ref} c_\mathrm{p,ref}}{k_\mathrm{ref}}.
\end{equation*}
where $c_\mathrm{p,ref}$ is the reference specific heat at constant
pressure, $\mu_\mathrm{ref}$ is the reference dynamic viscosity, and
$k_\mathrm{ref}$ is the reference thermal conductivity. The values of
$\gamma^*$, $c_\mathrm{v}^*$, $\mu^*$ and $k^*$ are modeled differently
depending on the thermochemical and transport model employed as
explained in the following section.

\subsubsection{Thermochemical and transport models}
\label{sec::Thermochemical and transport models}

\begin{figure}[]
\centering
\begin{subfigure}{\columnwidth}
\centering
\includegraphics[width=0.9\columnwidth]{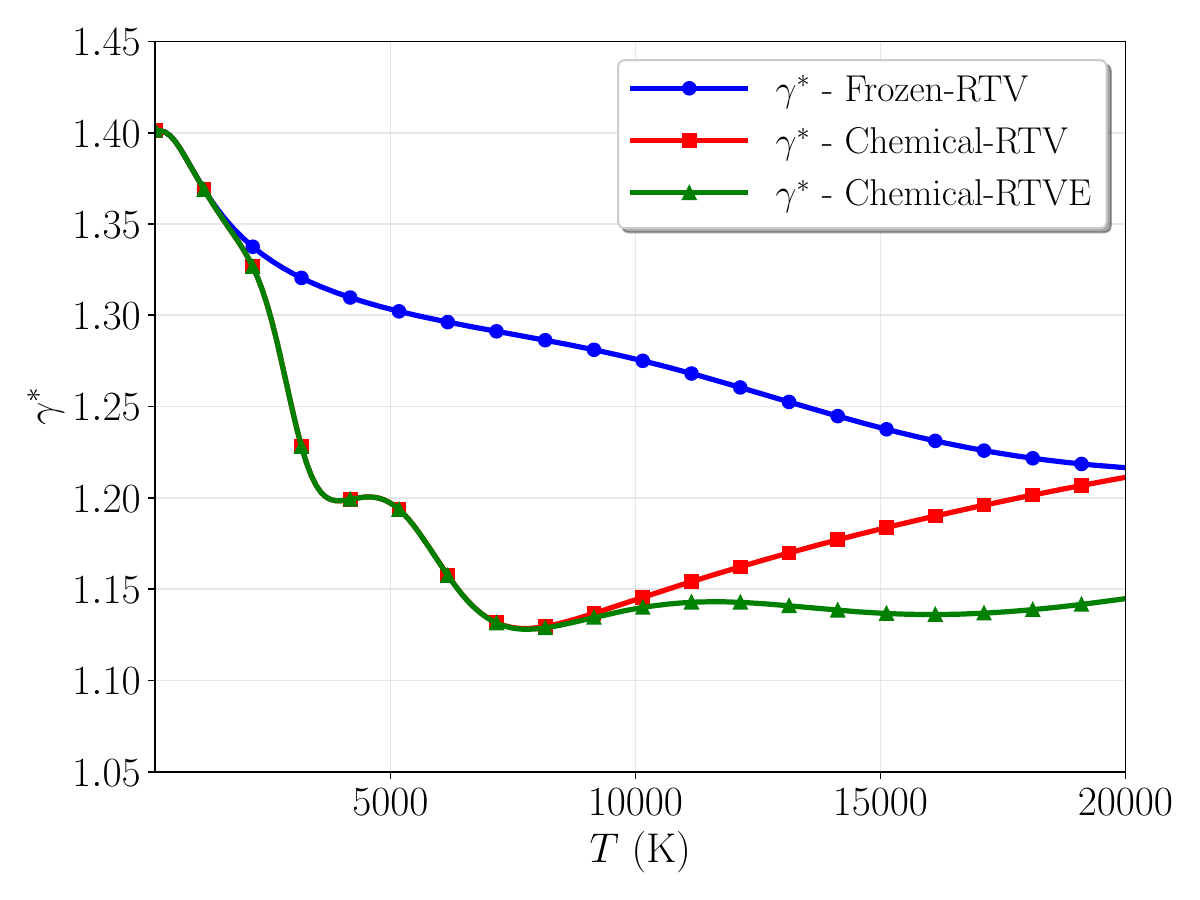}
\caption{}
\label{fig:A01a}
\end{subfigure}
\begin{subfigure}{\columnwidth}
\centering
\includegraphics[width=0.9\columnwidth]{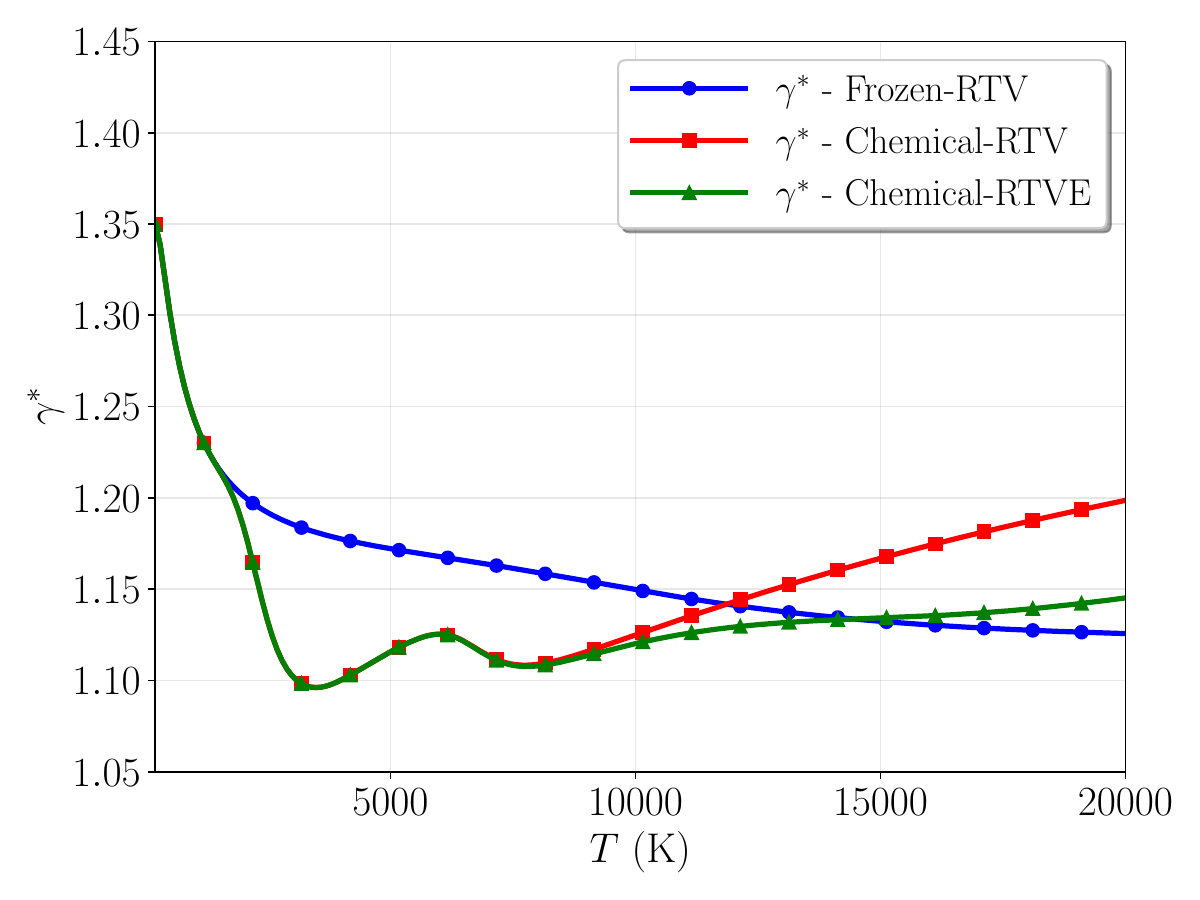}
\caption{}
\label{fig:A01b}
\end{subfigure}
\caption{Effective specific-heat ratio $\gamma^{*}$ as a function of
  temperature $T$ for different thermochemical models at
  $\rho=0.02~\mathrm{kg\,m^{-3}}$. The models are: Frozen-RTV (frozen
  chemistry with translational\hyp{}rotational\hyp{}vibrational
  equilibrium), Chemical-RTV (chemical and
  translational\hyp{}rotational\hyp{}vibrational equilibrium), and
  Chemical-RTVE (chemical and
  translational\hyp{}rotational\hyp{}vibrational\hyp{}electronic
  equilibrium). Gas molar fractions are (\textit{a}) Earth atmosphere:
  $X_{N_2}=0.7812$, $X_{O_2}=0.2095$, $X_{Ar}=0.0093$; and
  (\textit{b}) Mars atmosphere: $X_{CO_2}=0.9556$, $X_{N_2}=0.0270$,
  $X_{Ar}=0.0160$, $X_{O_2}=0.0014$.}
\label{fig:A01}
\end{figure}

The selection of appropriate thermochemical models is critical for the
accurate prediction of hypersonic flow conditions. High-temperature
effects, including vibrational excitation, chemical dissociation, and
ionization, can significantly alter fundamental flow properties such
as the bow-shock stand-off distance, which directly influences
stability characteristics. In the formulation presented in
Eqs.~(\ref{eq:ns_compressible_nondim}), all thermochemical effects in
the inviscid flow are condensed into the effective ratio of specific
heats, $\gamma^*$. Its value can be calculated using any of the
following five thermochemical models:
\begin{itemize}
    \item Calorically perfect gas (CPG).
    \item Frozen chemistry with
      translational\hyp{}rotational\hyp{}vibrational equilibrium
      (Frozen\hyp{}RTV).
    \item Chemical and translational\hyp{}rotational\hyp{}vibrational
      equilibrium (Chemical\hyp{}RTV).
    \item Chemical and
      translational\hyp{}rotational\hyp{}vibrational\hyp{}electronic
      equilibrium (Chemical\hyp{}RTVE).
   \item Reduced chemical non-equilibrium surrogate with
 translational\hyp{}rotational\hyp{}vibrational\hyp{}electronic
     equilibrium (NonEq-RTVE).
\end{itemize}
%
The thermochemical model must be chosen to match the flight regime of
interest, as each model entails a distinct set of physical assumptions
that are only valid over specific ranges of temperature and pressure.
Figure~\ref{fig:A01} compares $\gamma^*$ for three of the models
considered, for both Earth and Mars atmospheric compositions.  More
details on the implementation of each of the models can be found in
the appendix~\ref{sec::Physical_models}.

Regarding transport properties, the code provides two modeling
options. The first employs Sutherland's laws for the dynamic
viscosity, $\mu^*$, and thermal conductivity,
$k^*$~\citep{sutherland1893lii}. The second relies on
collision-integral-based models implemented through
Cantera~\citep{cantera}. In this case, the species viscosities and
thermal conductivities are computed from collision
integrals~\citep{wright2005recommended}, and the corresponding mixture
properties are obtained using Wilke's mixing rule for
viscosity~\citep{wilke1950viscosity} and the Mathur--Saxena rule for
thermal conductivity~\citep{mason1958approximate}.

\subsubsection{Boundary conditions}
\label{sec::Boundary conditions}

The following boundary conditions are available in the solver:
\begin{itemize}
\item Shock-compatibility boundary conditions. These conditions
  effectively create an immersed interface that imposes the
  Rankine--Hugoniot jump conditions as boundary conditions, ensuring
  consistency between the freestream and post-shock flow. Projecting
  the jump conditions onto the shock-normal direction yields:
\begin{align}
    \rho_1 (U_1 + u_s) &= \rho_2 (U_2 + u_s), \nonumber \\
    p_1 + \rho_1 (U_1 + u_s)^2 &= p_2 + \rho_2 (U_2 + u_s)^2, \nonumber \\
    e_1 + \frac{p_1}{\rho_1} + \frac{1}{2} (U_1 + u_s)^2 &= e_2 + \frac{p_2}{\rho_2} + \frac{1}{2} (U_2 + u_s)^2
    \label{eq:RankineHugoniot}
\end{align}
   where subscripts 1 and 2 denote conditions immediately upstream and
   downstream of the shock, respectively. Here, $U$ denotes the
   velocity component normal to the shock surface in the lab frame,
   and $u_s$ is the shock-normal speed, defined as positive when the
   shock moves opposite to the freestream direction.  These conditions
   impose the Rankine--Hugoniot jump relations at the fitted shock,
   treating the shock as an internal boundary between the prescribed
   upstream state and the computed post-shock flow. The pressure is
   closed through the effective thermodynamic relation
   $p=(\gamma^*-1)\rho e$, where $\gamma^*$ is evaluated according to
   the selected thermochemical model. For non-calorically-perfect
   models, the resulting nonlinear system is solved iteratively as
   described in appendix~\ref{sec::shock_jump_both}.
 \item Wall boundary conditions. For the velocity field, the code
   supports no-slip, slip, and Navier-slip boundary conditions. For
   the temperature field, both adiabatic and isothermal wall
   conditions are available.
   \item Symmetry boundary conditions can be used on a plane of
     geometric symmetry $(y=0)$.
   \item Non-reflecting characteristic outflow boundary
     conditions~\citep{poinsot1992boundary}.
   \item Supersonic and subsonic inflow and outflow boundary
     conditions. For supersonic inflow, all flow variables are
     prescribed from the freestream state. For subsonic inflow, all
     variables except pressure are imposed. At supersonic outflow
     boundaries, no variables are imposed. For subsonic outflow, a
     back-pressure condition is used in which the static pressure is
     prescribed.
\end{itemize}

\subsection{Numerical Implementation}
\label{sec::Numerical Implementation}

\subsubsection{Spatial and temporal discretization}
\label{sec::Spatial and temporal discretization}

The code performs direct numerical simulation (DNS) of the 2D or
axisymmetric compressible Navier--Stokes
equations~\eqref{eq:ns_compressible_nondim}. The equations are
discretized on a structured grid. To conform the computational domain
to a given geometry, curvilinear coordinate mappings are employed,
transforming a uniform Cartesian grid $(\chi, \eta)$ into the physical
space $(x, y)$, see figure \ref{fig:config_set_up}. Mappings are
currently implemented for the following configurations, which are
commonly encountered in high-enthalpy flow applications:
\begin{itemize}
\item Blunt wedge with a cylindrical tip (2D) / blunt cone with a
  spherical tip (3D-axisymmetric).
\item Cylinder (2D) / sphere (3D-axisymmetric).
\item Wedge (2D) / cone (3D-axisymmetric).
\end{itemize}
Additional geometries can be accommodated by defining custom
curvilinear mappings following the structure of the existing
implementations.

The spatial discretization employs a second-order finite-volume
scheme. Integrating the governing equations over each control volume
and applying the divergence theorem yields the semi-discrete form
\begin{equation}\label{eq:fv_semidiscrete}
\begin{split}
V_{i,j}\,\frac{\mathrm{d}\,\mathbf{U}_{i,j}}{\mathrm{d} t}
&+ \left(\hat{\mathbf{F}}_{i+\frac{1}{2},\,j}
       - \hat{\mathbf{F}}_{i-\frac{1}{2},\,j}\right) \\
&+ \left(\hat{\mathbf{G}}_{i,\,j+\frac{1}{2}}
       - \hat{\mathbf{G}}_{i,\,j-\frac{1}{2}}\right)
 = \mathbf{S}_{i,j}.
\end{split}
\end{equation}
where $V_{i,j}$ is the cell volume and $\mathbf{U}_{i,j}$ is the
vector of cell-averaged conserved variables:
\begin{equation}\label{eq:state_vector_2d}
\mathbf{U}_{i,j} =
\begin{pmatrix}
\rho \\[3pt] \rho u \\[3pt] \rho v \\[3pt] \rho E
\end{pmatrix}_{i,j}, \qquad
\rho E = \tfrac{1}{2}\,\rho\!\left(u^{2}+v^{2}\right) + \rho\,e,
\end{equation}
where $u$ and $v$ denote the velocity components in the $x$- and
$y$-directions, respectively. In the axisymmetric formulation, $x$
corresponds to the axial coordinate and $v$ to the radial velocity
component. The source term $\mathbf{S}_{i,j}$ accounts for the
geometric contributions arising from the cylindrical coordinate
system; it vanishes identically in the 2D Cartesian formulation. Each
numerical flux comprises inviscid and viscous contributions,
\begin{equation}\label{eq:flux_split}
\hat{\mathbf{F}}_{i+\frac{1}{2},\,j}
= \hat{\mathbf{F}}^{\,\mathrm{inv}}_{i+\frac{1}{2},\,j}
- \hat{\mathbf{F}}^{\,\mathrm{vis}}_{i+\frac{1}{2},\,j},
\end{equation}
and likewise for $\hat{\mathbf{G}}$. Flow variables at each cell face
are obtained by linear interpolation from adjacent cell centers, and
the face integral to compute the flux is approximated with midpoint
quadrature. The gradients required for the viscous fluxes and the heat
flux are computed using central differences on the curvilinear mesh.

Time integration is performed using either an explicit fourth-order
Runge--Kutta (RK4) scheme or an implicit backward-Euler method. The
RK4 scheme is used when accurate time evolution is required, whereas
the backward-Euler method is primarily employed to compute steady-state
base flows, toward which the implicit solvers converge directly. The
time step may be prescribed or determined adaptively from a
Courant--Friedrichs--Lewy (CFL) condition. For the implicit
formulation, two nonlinear solution strategies are available: a Picard
fixed-point iteration and a Newton method with pseudo-transient
continuation. Details of the local CFL estimate, the self-adaptive CFL
ramp, and the two nonlinear solvers are provided in
appendix~\ref{sec:time-marching}.

\subsubsection{Shock-fitting approach}
\label{sec::Shock-fitting_method}

The solver includes the option of running simulations with a
shock-fitting algorithm~\citep{hussaini1983} to avoid artifacts
associated with shock capturing, including artificial shock thickness
in linear shock--disturbance interactions and, in nonlinear base-flow
computations, possible carbuncle-type
instabilities~\citep{pandolfi2001numerical}. The shock-fitting
procedure imposes the boundary conditions of the finite-volume solver
using the Rankine--Hugoniot jump conditions described in
section~\ref{sec::Boundary conditions}. The bow shock is represented
through a cubic-spline parametrization,
$(x^n,y^n)_{\text{shock-spline}} = F^n(s)$, where $s$ denotes the
arc-length coordinate along the shock and $n$ indicates the current
time step.  The splines evolve independently of the computational
grid, allowing the shock to move continuously rather than being
constrained to discrete grid points. This independence provides
subgrid shock positioning, avoids carbuncle-type shock-capturing
artifacts, and helps preserve the nominal order of the spatial
discretization near the fitted shock.

The coupling between the splines and the finite-volume cells occurs
through designated shock cells, $(x_{ij}^n,
y_{ij}^n)_{\text{shock-cell}}$, defined as the cell centers located
nearest to the shock-parametrizing spline. These shock cells act as
ghost cells to impose shock boundary conditions in the finite-volume
discretization, with the Rankine--Hugoniot jump conditions
(section~\ref{sec::Boundary conditions}) solved subject to the Riemann
invariant as a constraint. The methodology for solving the
Rankine--Hugoniot relations while incorporating chemical effects is
detailed in
appendix~\ref{sec::shock_jump_both}. Figure~\ref{fig:shock_fitting}
illustrates an example of the freely moving spline configuration and
its corresponding shock cells. It is worth delineating the regime of
applicability of the present shock-fitting treatment. Because the
shock is parametrized as a one-dimensional arc-length spline, the
methodology applies to two-dimensional and axisymmetric flows
dominated by a single bow shock.  Extending it to fully
three-dimensional bow shocks would require a two-dimensional surface
parametrization, deformation of the computational mesh, and the
treatment of shock-topology changes.  Moreover, only the primary bow
shock is fitted: secondary or embedded shocks, such as those arising
from recompression, flow separation, or corner interactions in
cone--flare, double-cone, and inlet configurations, are not considered
within this framework.
\begin{figure}[]
\centering
\includegraphics[width=0.5\linewidth]{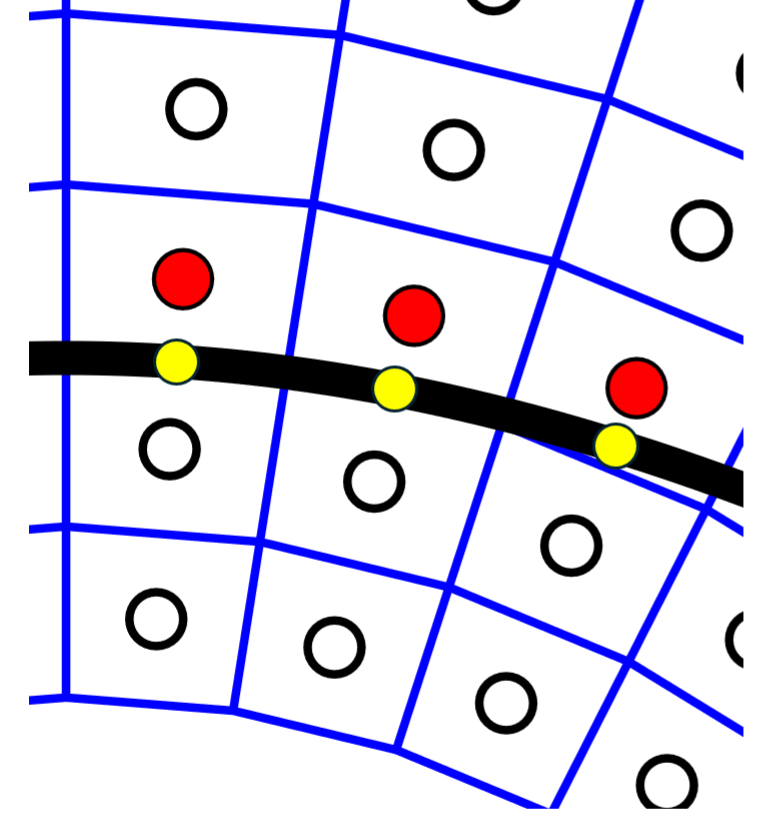}
\caption{Approach and notation used to exchange information between
  the shock and the finite-volume grid at time step $n$. Blue lines:
  cell faces; white circles: finite-volume cell centers
  $(x^n_{ij},y^n_{ij})$; red circles: shock-cell centers
  $(x^n_{ij},y^n_{ij})_{\text{shock-cell}}$; yellow circles:
  shock points $(x^n_{k},y^n_{k})_{\text{shock-point}}$; black line:
  shock spline $(x^n(s),y^n(s))_{\text{shock-spline}} = F^n(s)$.}
\label{fig:shock_fitting}
\end{figure}

A summary of the steps followed by the shock-fitting algorithm is
given below: 
\begin{enumerate}
    \item Initial flow conditions at time step $n$ in cell
      $(x_{ij},y_{ij})$: $\rho^n_{ij}$, $\rho u^n_{ij}$, $\rho
      v^n_{ij}$, and $\rho E^n_{ij}$. The initial shock position is
      parametrized with
      splines: \[(x^n(s),y^n(s))_{\text{shock-spline}} = F^n(s).\]

    \item Compute the positions of the shock cells, defined as the
      finite-volume cell centers closest to the shock spline. Each
      shock cell $(x_{ij}^n, y_{ij}^n)_{\text{shock-cell}}$ has an
      associated discrete point on the shock spline $(x_{k}^n,
      y_{k}^n)_{\text{shock-point}}$, defined as the closest point on
      the spline to the given shock cell.

    \item Interpolate the variables from the finite-volume grid
      $(x_{ij}^n, y_{ij}^n)$ to the associated shock points $(x_{k}^n,
      y_{k}^n)_{\text{shock-point}}$. Note that both states 1 and 2
      must be interpolated at the shock points.

    \item Determine the state 2 from the new Rankine--Hugoniot
      solution at the shock points $(x_{k}^n,
      y_{k}^n)_{\text{shock-point}}$ using the algorithm described in
      appendix~\ref{sec::shock_solution}. The shock speed is also part
      of the solution and determines how the shock spline moves for
      the next time step.

    \item Interpolate the Rankine--Hugoniot solutions computed on the
      shock points $(x_{k}^n, y_{k}^n)_{\text{shock-point}}$ back to
      the shock cells $(x_{ij}^n, y_{ij}^n)_{\text{shock-cell}}$.

    \item Compute fluxes at each face and advance the solution in time
      to obtain $\rho^{n+1}_{ij}$, $\rho u^{n+1}_{ij}$, $\rho
      v^{n+1}_{ij}$, and $\rho E^{n+1}_{ij}$ for the next time step.

    \item Update the shock position $(x_{k}^{n+1},
      y_{k}^{n+1})_{\text{shock-point}}$ using the calculated shock
      velocity from step 4. Fit the new spline to the updated shock
      points: \[(x^{n+1}(s), y^{n+1}(s))_{\text{shock-spline}} =
      F^{n+1}(s).\]
\end{enumerate}

Section~\ref{sec::shock_fitting_verification} presents verification
cases for calorically perfect-gas flows in both 2D and 3D axisymmetric
configurations. Verification cases involving real-gas effects are
presented in section~\ref{sec::verification_real_gas}. In addition,
verification of the Rankine--Hugoniot conditions in the presence of
thermochemical effects is provided in
section~\ref{sec::Rankine_verification}.

\subsection{Linear stability analysis}
\label{sec::lin_stab_analysis_implementation}

Three approaches are implemented in the software to characterize
linear instability mechanisms: modal analysis, non-modal
(transient-growth) analysis, and freestream receptivity
analysis. Here, we describe the implementation of each method. The
following notation convention is adopted: non-bold variables (e.g.,
$q$) refer to the continuous formulation, whereas bold variables
(e.g., $\bm{q}$) denote discretized quantities. For example, $q =
[\rho,\rho u,\rho v,\rho E]^T$ denotes the state vector, whereas
$\bm{q}$ denotes the array collecting all finite-volume cell
variables. The total flow field is decomposed as $q = q_0 + q'$, where
the prime (${}'$) denotes infinitesimal disturbances and the subscript
$0$ denotes the steady base flow.

We denote the continuous Navier--Stokes equations as
\[
\frac{\partial q}{\partial t} = N(q),
\]
and introduce infinitesimal perturbations of the conservative
variables,
\[
q' = \big[\rho',\, (\rho u)',\, (\rho v)',\, (\rho E)'\big]^T. 
\]
The discrete perturbation state is
\[
\bm{q}' = \big[\bm{\rho}',\, \bm{(\rho u)}',\, \bm{(\rho v)}',\, \bm{(\rho E)}'\big]^T.
\]
In the shock-fitting formulation, $\bm{q}'$ denotes only the
disturbances downstream of the shock. The perturbation of the shock
location is also included within the linearization and is represented
by the vector $\bm{\eta}'$, which collects the shock-front
displacements in the local normal direction. Therefore, in these
cases, the extended state disturbance reads $\bm{q}'_e =
\big[\bm{q}',\, \bm{\eta}'\big]^T$. In cases without shock fitting, we
define $\bm{q}'_e = \bm{q}'$.

With the previous notation, the linearized semi-discrete system reads
\begin{equation}
\frac{\mathrm{d} \bm{q}'_e}{\mathrm{d} t}
=
\bm{L}(\bm{q}_{e,0})\,\bm{q}'_e,
\qquad
\bm{L}(\bm{q}_{e,0})
\equiv
\left.\frac{\delta \bm{N}}{\delta \bm{q}_e}\right|_{\bm{q}_{e,0}},
\label{eq:linearized_system}
\end{equation}
where $\bm{N}$ denotes the spatial discretization of the
Navier--Stokes equations and $\bm{L}$ is the numerical linearization
of $\bm{N}$ about the base flow $\bm{q}_{e,0}$ computed using finite
differences, denoted by $\delta / \delta \bm{q}_e$. The linearized
operator is constructed using the same spatial discretization employed
to compute the base flow.  To compute the base flow $\bm{q}_{e,0}$, the
code offers the numerical schemes described in
section~\ref{sec::Numerical Implementation} to solve the compressible
Navier--Stokes equations with any of the available thermochemical
models (see section~\ref{sec::Thermochemical and transport models}).

\subsubsection{Global modal analysis}
\label{sec::modal_analysis_implementation}

For the global modal analysis, we solve the eigenvalue problem,
$\lambda \bm{q}'_e = \bm{L} \bm{q}'_e$.  

In shock-fitting modal calculations, the upstream freestream is
prescribed and is not included in the eigenproblem. Because the
upstream base flow is uniform and supersonic relative to the body,
downstream perturbations cannot feed back into the upstream domain
through the characteristics~\citep{schmid2012stability}.  The
eigenproblem therefore includes only the post-shock flow and the
fitted-shock displacement. Any exponential instability that might
occur can therefore arise only downstream of the shock. Several cases
are presented to validate the linear stability analysis in
section~\ref{sec::stability_analysis_verification}.

The size of the matrix $\bm{L}$ increases rapidly with grid size. For
a grid with $N \times N$ cells and four conservative variables,
$\text{size}(\bm{L}) = O(16\,N^4)$.  This means that, even for the
relatively small grids used for validation in figures~\ref{fig:07}
and~\ref{fig:08}, $\bm{L}$ would occupy approximately $160\,\text{GB}$
of RAM in double precision.

A popular approach to solve this eigenvalue problem is the Arnoldi
iteration~\citep{theofilis2011global}. However, the Arnoldi iteration
converges to the eigenvalues with the largest absolute value, whereas
only the eigenvalues with the largest real part are of interest. To
address this, matrix transformations are applied to the linearized
operator.  A common choice is the shift-and-invert (S-I)
transformation \citep{lehoucq1998arpack}, $\bm{\hat{L}} = (\bm{L} -
\sigma \bm{I})^{-1}$, where $\sigma$ is the user-prescribed
shift. This approach is implemented in packages such as ARPACK,
LAPACK, and SLEPc. Once the shift is applied, the shifted matrix
$\bm{L}^* = (\bm{L} - \sigma \bm{I})$ is typically factorized via $LU$
or $QR$ decompositions to solve the inverse problem. The main drawback
is that, although $\bm{L}$ is very sparse, the factorizations are
typically dense due to fill-in, so that memory requirements scale
again as $O(16\,N^4)$.

\begin{figure}[]
\centering
\includegraphics[width=1\linewidth]{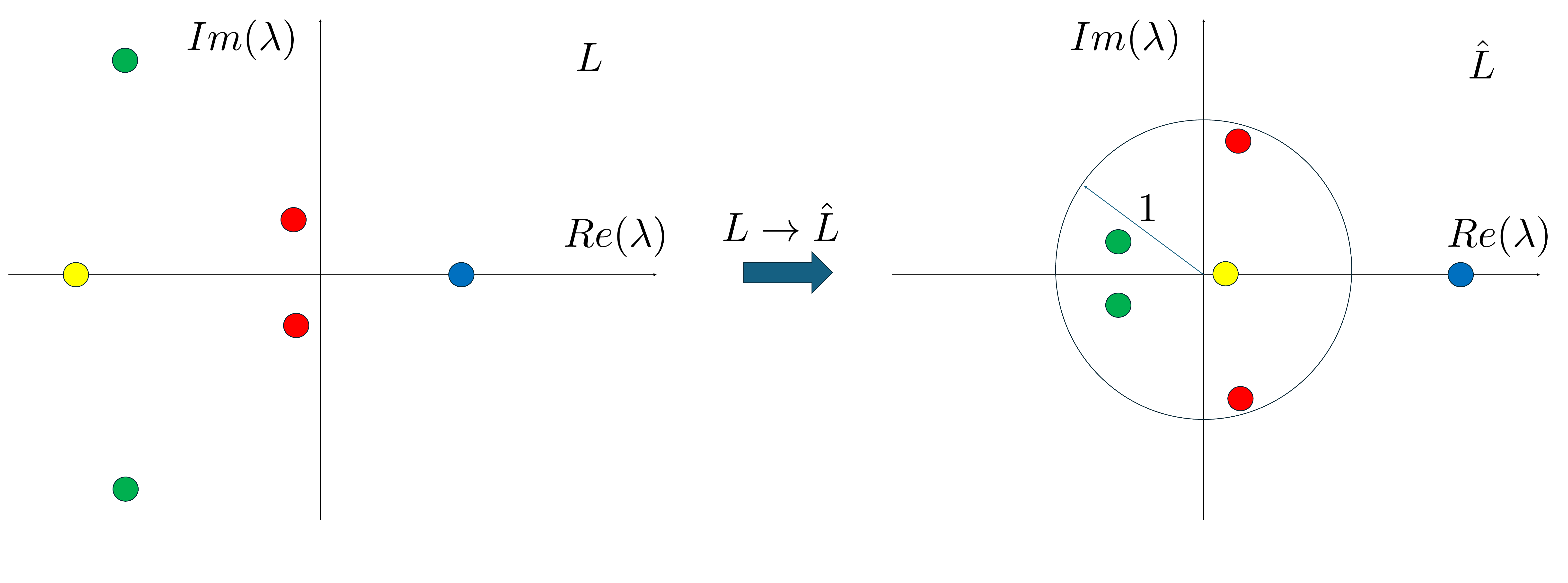}
\caption{Illustration of the spectrum transformation applied to a
  matrix $\bm{L}$ to make the Arnoldi iteration converge to the
  right-half-plane eigenvalues. Transformation: $\bm{\hat{L}} =
  \sum_{k=0}^{M} \frac{(\bm{L}\,\tau)^k}{k!} \approx
  \exp(\bm{L}\,\tau)$.}
\label{fig:eigenvalue_compute}
\end{figure}

To avoid this limitation, a sparsity-preserving spectral
transformation is applied instead. The modified operator maps each eigenvalue $\lambda$ of
$\bm{L}$ to $\exp(\lambda\,\tau)$, see
figure~\ref{fig:eigenvalue_compute}. After this mapping, the
eigenvalues with the largest magnitude, to which the Arnoldi iteration
converges most rapidly, correspond to eigenvalues in the right
half-plane of the original operator. As a result, the iteration
preferentially retrieves the most unstable modes of $\bm{L}$. The
transformation is approximated as
\begin{equation}
    \bm{\hat{L}} \equiv \sum_{k=0}^{M} \frac{(\bm{L}\,\tau)^k}{k!} \approx \exp(\bm{L}\,\tau),
\end{equation}
where $M$ and $\tau$ are parameters to be chosen.

The key to this transformation is to preserve accurately the
eigenvalue structure so that the mapping $\lambda \rightarrow
\exp(\lambda\,\tau)$ can be inverted by selecting the correct
logarithm branch. The approximation error is controlled via the
parameters $M$ and $\tau$:
\begin{equation*}
    |\bm{\hat{L}} - \exp(\bm{L}\,\tau)| \approx \left|\frac{(\bm{L}\,\tau)^{M+1}}{(M+1)!}\right|.
\end{equation*}
For maximum accuracy, $M$ should be increased and $\tau$
decreased. However, increasing $M$ raises the computational cost,
while reducing $\tau$ causes the transformed eigenvalues to cluster
together, slowing the convergence of the Arnoldi iteration. A good
estimate of $\tau$ is given by the CFL-limited time step $\tau\approx
\Delta t_{\text{CFL}}$ such that $\text{CFL} = \tau
\max(|\lambda_i|)$. Introducing this estimate yields
\begin{equation*}
    |\bm{\hat{L}} - \exp(\bm{L}\,\tau)| \approx \frac{(\text{CFL})^{M+1}}{(M+1)!},
\end{equation*}
so that the error in a converged eigenvalue satisfies
\begin{equation}
    |\exp(\lambda^* \tau) - \exp(\lambda \tau)| \approx \frac{(\text{CFL})^{M+1}}{(M+1)!},
\end{equation}
where $\lambda$ is the true eigenvalue and $\lambda^*$ the eigenvalue
retrieved by the Arnoldi iteration. In practice, we observed that
setting $\tau$ with $\text{CFL} = \tfrac{1}{2}$ and $M = 5$ yields
well-converged eigenvalues at a reasonable computational cost.

The transformation is implemented within the Implicitly Restarted
Arnoldi Method using ARPACK~\citep{lehoucq1998arpack}, with the
exponential approximation in the inner loop offloaded to the GPU (if
available). Within the Arnoldi
iterations, $\bm{\hat{L}}$ is never formed explicitly, as this would
require multiplying $N^2 \times N^2$ matrices at $O(N^6)$
cost. Instead, the operation $\bm{v}^* = \sum_{k=0}^{M}
\frac{(\bm{L}\,\tau)^k}{k!}\,\bm{v}$ is reorganized as a sequence of
sparse matrix--vector multiplications, each costing $O(N^2)$. The
total cost for the Arnoldi iteration with the exponential
transformation scales as $O(N^3)$, because $\tau$ must decrease with
finer mesh size, increasing the number of Arnoldi iterations by
$O(N)$.
\begin{table}
\caption{Comparison of memory and computation requirements for
  shift-and-invert (S-I) and exponential (Exp) transformations with
  the Implicitly Restarted Arnoldi Method (ARPACK
  implementation~\citep{lehoucq1998arpack}). The eigenvalue problem
  consists of finding the 40 most unstable eigenvalues of a flow on an
  $N \times N$ grid. Computations were performed with a Ryzen 7945HX
  (16 cores, 32\,GB) and an NVIDIA RTX~4090 Mobile GPU (16\,GB). NA
  indicates cases that could not be completed due to memory
  limitations. Memory estimates for S-I assume 100\% fill-in during
  decomposition.}
\label{tab:eigen_compute}
{\setlength{\aboverulesep}{0pt}
\setlength{\belowrulesep}{0pt}
\begin{tabular*}{\tblwidth}{@{} L|LL|LL@{} }
 & \multicolumn{2}{L|}{Wall-clock time (seconds)} & \multicolumn{2}{L}{Memory (GB)} \\
\midrule
$N$ & S-I & Exp & S-I & Exp \\
\midrule
50  & 2.4   & 0.8   & 0.8     & $8 \times 10^{-5}$ \\
100 & 20.4  & 6.4   & 12.8    & $3.2 \times 10^{-4}$ \\
200 & 260.8 & 34.7  & 204.8   & $1.28 \times 10^{-3}$ \\
400 & NA    & 283.4 & 3,276 & $5.12 \times 10^{-3}$ \\
800 & NA    & 2,103.9 & 52,428 & $2.05 \times 10^{-2}$ \\
\bottomrule
\end{tabular*}}
\end{table}
Table~\ref{tab:eigen_compute} shows that the current implementation
enables larger and faster computations than the shift-and-invert
approach, since in the latter the memory requirements associated with
fill-in quickly become unaffordable as the grid size increases. Note
that the memory estimates for shift-and-invert assume 100\% fill-in,
corresponding to the worst-case scenario; for this specific problem,
the fill-in was around 15\%, which explains why the case with $N =
200$ was able to run even though the 100\% fill-in estimate exceeded
the available memory. This GPU-accelerated and memory-efficient
implementation is key to carrying out modal stability analyses
efficiently at larger grid sizes.

\subsubsection{Global transient growth}
\label{sec::transient_post_shock_implementation}

To perform the transient growth analysis, the energy gain for a given
disturbed flow is measured using perfect-gas Chu energy
norm~\citep{chu1965energy} evaluated with local effective
thermodynamic properties (see appendix~\ref{sec:eq_models}):
\begin{equation}
\begin{split}
   E = \int \bigg[ & \underbrace{\frac{\rho_0 a_0^2}{2(\gamma_0^*p_0)^2}p'^2}_{\text{pressure}} 
   + \underbrace{\frac{\rho_0}{2}u_i'u_i'}_{\text{kinetic}}
    + \underbrace{\frac{(\gamma_0^*-1)p_0}{2\gamma_0^*}
\left(\frac{s'}{R_{g,0}}\right)^2}_{\text{entropic}} \bigg] \mathrm{d}V_D,
\end{split}
   \label{eq::Chu_norm}
\end{equation}
where $V_D$ denotes the flow domain, $s'$ is the entropy disturbance,
$a_0$ is the speed of sound of the base flow, and $R_{g,0}$ is the gas
constant of the base flow. In the cases with shock-fitting, $V_D$
refers only to the region downstream of the shock wave. This energy
norm allows us to differentiate contributions due to acoustic,
kinetic, and entropic energy modes.

The original perturbation state vector,
\begin{equation*}
   q' = \left[\rho',\, (\rho u)',\, (\rho v)',\, (\rho E)'\right],
\end{equation*}
is converted to Chu's variables,
\begin{equation*}
   \tilde{q}' = \left[p',\, u',\, v',\, \frac{s'}{R_{g,0}}\right],
\end{equation*}
through a linear
transformation matrix $Q_V$. The Chu energy is then defined as
\begin{equation}
E \;=\; \int_{V_D} q'^T Q_V^T\, W_V\, Q_V\, q' \, dV_D,
\end{equation}
where $W_V$ denotes the metric term that accounts for the weights in
the Chu energy norm. The discrete
energy gain, computed using the post-shock finite-volume cells to
characterize the transient growth of downstream perturbations, is
\begin{equation*}
G_D(t) \;=\; \frac{E(t)}{E(0)}
\;=\;
\frac{\bm{q}'_e(t)^T \bm{P}^T \, \bm{Q}^T \bm{M}\, \bm{Q}\, \bm{P} \, \bm{q}'_e(t)}
{\bm{q}'_e(0)^T \bm{P}^T \bm{Q}^T \bm{M}\, \bm{Q}\, \bm{P} \, \bm{q}'_e(0)} ,
\end{equation*}
where $D$ in $G_D(t)$ denotes `downstream' gain and $\bm{P}$ projects
the extended state $\bm{q}'_e$ onto the flow variables used in Chu's
energy norm. See appendix~\ref{sec::transient_growth_appendix} for the
detailed definition of every discrete operator.

The gain for the linearized system is obtained by formulating the
optimization as a generalized Rayleigh quotient. The perturbation
dynamics satisfy
\begin{equation}
  \frac{\mathrm{d} \bm{q}'_e}{\mathrm{d} t} = \bm{L}\bm{q}'_e,
  \qquad
  \bm{q}'_e(t) = \exp(\bm{L}t)\,\bm{q}'_e(0),
\end{equation}
and the discrete energy gain is therefore
\begin{equation}
  G_D(t)
  = \frac{\bm{q}'_e(0)^T \bm{C}(t)\,\bm{q}'_e(0)}
  {\bm{q}'_e(0)^T \bm{D}\,\bm{q}'_e(0)},
\end{equation}
where
\begin{align*}
  \bm{C}(t) =  \exp(\bm{L}t)^T \bm{P}^T \bm{Q}^T \bm{M}\, \bm{Q}\, \bm{P} \, \exp(\bm{L}t),
  \\
  \bm{D} = \bm{P}^T \bm{Q}^T \bm{M}\, \bm{Q}\, \bm{P}.
\end{align*}
The resulting generalized eigenvalue problem is solved using a Lanczos
iteration (see appendix~\ref{sec::transient_growth_appendix}). Solving
this problem yields the optimal initial disturbance $\bm{q}'_e(0)$ that
maximizes the energy growth at a prescribed time $t$. The user can
specify the discrete time points at which these gains are optimized.
To determine the maximum over all times, denoted by $G_D^\text{opt}$,
a sufficient number of discrete times must be explored. The
optimization is designed to obtain the solution for several times
simultaneously, with almost negligible additional cost compared to
solving for a single time. For further refinement, bisection or Newton
methods can be used to accelerate convergence to the absolute maximum
gain over time.

\subsubsection{Freestream receptivity}
\label{sec::freestream_receptivity_implementation}

The freestream receptivity analysis implemented in the code is
targeted at cases in which strong shock waves form, defining two
physical regions: the freestream domain and the post-shock domain.  In
this scenario, the amplification of disturbances can be conceptually
divided into two regimes: shock transmission and post-shock
amplification. We consider a freestream disturbance $q'_\infty$ with
an initially undisturbed post-shock field, i.e., $q'_e = 0$ at $t=0$.
The goal of this analysis is to investigate how freestream
disturbances excite the post-shock flow until the forced asymptotic
response is reached. For a single forcing frequency this response is
periodic; for multiple non-commensurate frequencies it is
multi-frequency or quasiperiodic. This yields an input--output system
in which the input is the prescribed harmonic freestream forcing
$q'_\infty$, and the output is the resulting post-shock response $q'$.

The freestream input is represented only through its trace on the
upstream side of the fitted shock. The imposed upstream disturbance is
expanded as a finite sum of harmonic components,
\begin{equation}
  q_\infty'(s,t)
  =
  \Re\left\{
  \sum_{l=0}^{N_\omega}
  \widehat{q}'_{\infty,l}(s)
  \exp(-i\omega_l t)
  \right\},
  \label{eq:freestream_trace_expansion}
\end{equation}
where $s$ denotes the arc length along the shock surface and the
temporal frequencies $\omega_l$ are specified by the user. The
optimization variables are the complex amplitude functions
$\widehat{q}'_{\infty,l}(s)$ in
\eqref{eq:freestream_trace_expansion}. After discretization of the
shock surface, each complex amplitude $\widehat{\bm q}_{\infty,l}'$
contains the conservative-variable perturbations at the shock points
for the $l$th frequency.

The discretized forcing can be expressed as
\begin{equation}
  \widehat{\bm q}_\infty'(t)
  =
  \exp(-i\bm{\Omega}t)
  \widehat{\bm q}_\infty'(0),
  \qquad
  \bm{q}_\infty'(t)
  =
  \bm{\Sigma}\exp(-i\bm{\Omega}t)
  \widehat{\bm q}_\infty'(0).
  \label{eq:freestream_reconstruction_operator}
\end{equation}
where $\bm{\Omega}$ applies the prescribed temporal
frequency to each harmonic forcing, and $\bm{\Sigma}$ sums the frequency
components to recover the instantaneous upstream perturbation at the shock in physical space.

Substitution into \eqref{eq:linearized_system} yields the forced
post-shock system
\begin{equation}
  \frac{\mathrm{d}\bm{q}'_e}{\mathrm{d} t}
  =
  \bm{L}\bm{q}'_e
  +
  \bm{F}(t)\widehat{\bm q}_\infty'(0),
  \qquad
  \bm{F}(t)
  =
  \bm{B}\bm{\Sigma}\exp(-i\bm{\Omega}t),
  \label{eq:forced_system}
\end{equation}
where the initial post-shock disturbances are set to zero:
$\bm{q}'_e(0)=\bm{0}$. The operator $\bm{B}$ represents the coupling
between freestream and post-shock disturbances, mediated by their
interaction across the bow shock. It is also linearized using finite
differences, as in equation~\ref{eq:linearized_system}; see
appendices~\ref{sec::linearization_details} and
\ref{sec::freestream_receptivity_appendix} for more details. A
verification of the linearized shock--disturbance interaction model is
discussed in section~\ref{sec::Verification_Disturbances_Shock}.
  
The evolution of the discrete freestream disturbances
$\bm{\widehat{q}}'_\infty$
(equation~\ref{eq:freestream_reconstruction_operator}) and the forced
downstream perturbations $\bm{q}_e'$ (equation~\ref{eq:forced_system})
can be expressed as an extended system
\begin{equation}
    \dfrac{\mathrm{d}}{\mathrm{d} t} \begin{bmatrix}
        \bm{\widehat{q}}'_\infty \\
        \bm{q}_e' \\
    \end{bmatrix} =
    \begin{bmatrix}
        -i\bm{\Omega} & \bm{0} \\
        \bm{B}\bm{\Sigma} & \bm{L} \\
    \end{bmatrix} \begin{bmatrix}
        \bm{\widehat{q}}_\infty' \\
        \bm{q}_e' \\
    \end{bmatrix}
    \equiv \dfrac{\mathrm{d} \bm{\overline{q}}' }{\mathrm{d} t} = \bm{\overline{L}} \bm{\overline{q}}' \, ,
\end{equation}
where $\bm{\overline{q}}'$ denotes the state that contains freestream
and downstream disturbances and $\bm{\overline{L}}$ is the linearized
operator of the full system.  

The gain is defined as
\begin{equation}
   G_T(t) = \frac{E(t)}{E_\infty^{\mathrm{ref}}},
      \label{eq::gain_ref}
\end{equation}
where the subscript $T$ denotes the \emph{total} gain, i.e., the gain
accounting for disturbance amplification both across the shock wave
and downstream of the bow shock, measured relative to the freestream
input energy $E_\infty^{\mathrm{ref}}$.  More precisely,
$E_\infty^{\mathrm{ref}}$ is defined as the disturbance energy that
has crossed the shock after a reference time $t^{\text{ref}}$. We
define $t^{\text{ref}}$ as the time required for the base-flow mass
flux through the shock, $\dot{m}_\infty$, to fill the total mass
contained in the domain downstream of the shock, $m_D$, namely
$t^{\text{ref}} = m_D/\dot{m}_\infty$.  The gain can be written as a
generalized Rayleigh quotient:
\begin{equation}
   G_T(t)   =
  \frac{
  \widehat{\bm q}_\infty'(0)^\dagger
  \bm{C}_\infty(t)
  \widehat{\bm q}_\infty'(0)}
  {
  \widehat{\bm q}_\infty'(0)^\dagger
  \bm{D}_\infty
  \widehat{\bm q}_\infty'(0)},
\end{equation}
where $^\dagger$ denotes the conjugate transpose. As in
section~\ref{sec::transient_post_shock_implementation}, the
generalized Rayleigh quotient is computed via the Lanczos iteration;
further details on the construction of the discrete operators are
given in appendix~\ref{sec::freestream_receptivity_appendix}. Solving
this problem yields the optimal initial freestream disturbance
$\bm{\widehat{q}}'_\infty(0)$ that maximizes the energy amplification
downstream of the shock at a prescribed time horizon~$t$.  The time
horizon can be varied to examine how the optimal disturbance evolves
over different intervals. Provided that the post-shock linear operator
is modally stable, the sustained harmonic freestream forcing drives
the system toward a time-periodic response.  The optimal disturbance
associated with this asymptotic regime can be obtained by selecting a
sufficiently large~$t$.

\section{Verification cases}
\label{sec::Verification cases}

\subsection{Verification of shock fitting code}
\label{sec::shock_fitting_verification}

\subsubsection{2D geometries}

\begin{figure}[]
\centering
\begin{subfigure}[t]{1\columnwidth}
\includegraphics[width=\columnwidth]{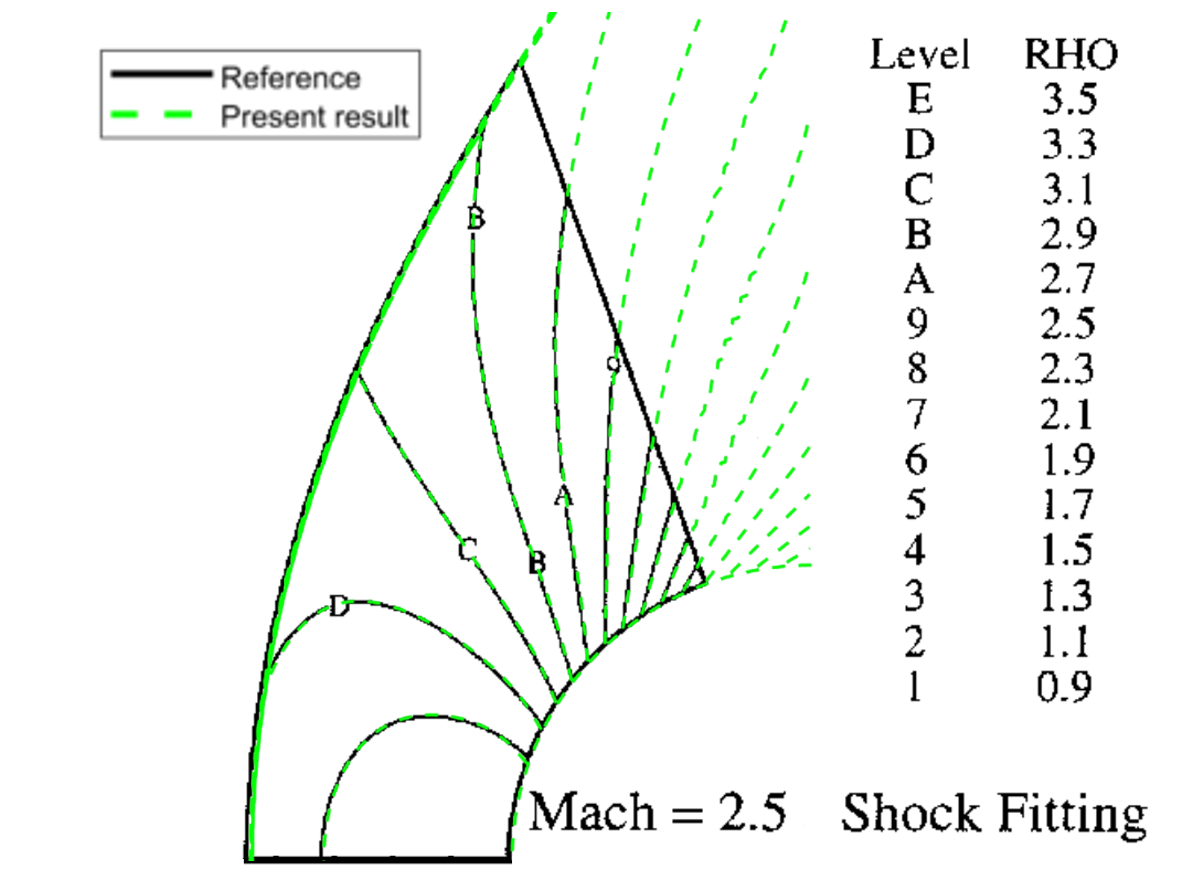}
\caption{}
\label{fig:07a}
\end{subfigure}
\begin{subfigure}[t]{1\columnwidth}
\includegraphics[width=\columnwidth]{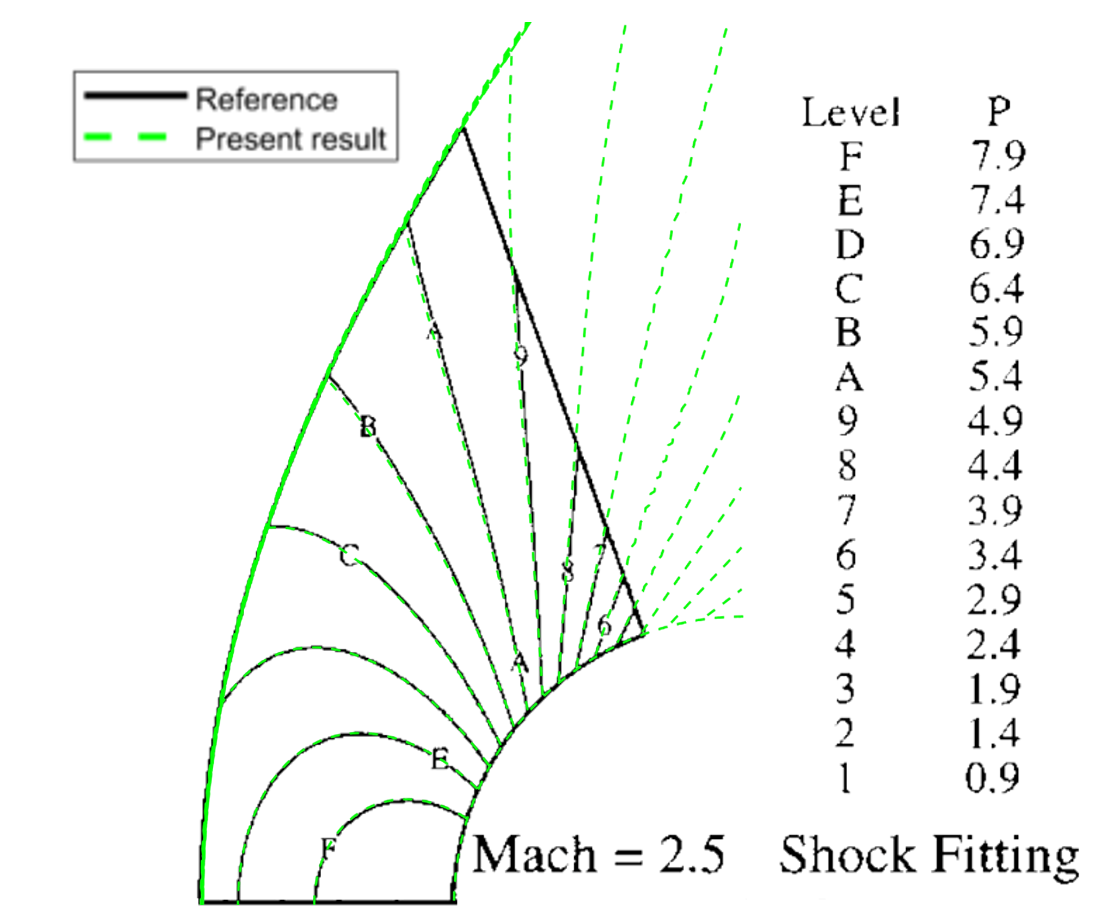}
\caption{}
\label{fig:07b}
\end{subfigure}
\caption{Verification test on a Mach 2.5 bow shock over a
  cylinder. CPG, $\gamma = 1.4$. Comparison of HYMOR with results from
  \citet{carpenter1999accuracy}. (\textit{a}) Density
  isocontours. (\textit{b}) Pressure isocontours.}
\label{fig:07}
\end{figure}

The results from HYMOR are compared with the solution of the Euler
equations reported by \citet{carpenter1999accuracy} for a Mach 2.5
flow over a cylinder of radius $R$. Figure~\ref{fig:07} shows that the
density and pressure contours obtained with the present solver are in
good agreement with the reference solution.

\begin{figure}[]
\centering
\begin{subfigure}[t]{1\columnwidth}
\includegraphics[width=\columnwidth]{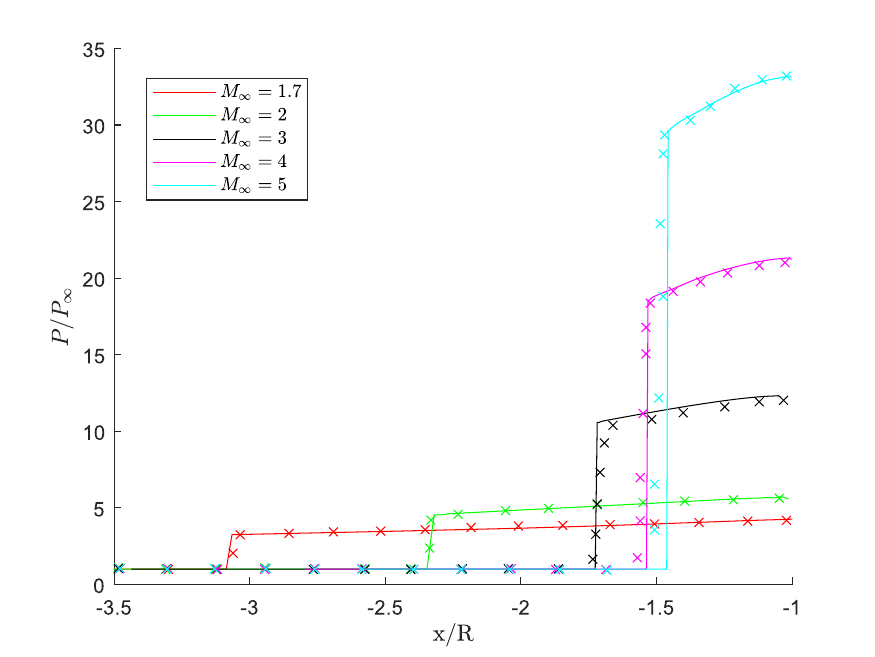}
\caption{}
\label{fig:08a}
\end{subfigure}
\begin{subfigure}[t]{1\columnwidth}
\includegraphics[width=\columnwidth]{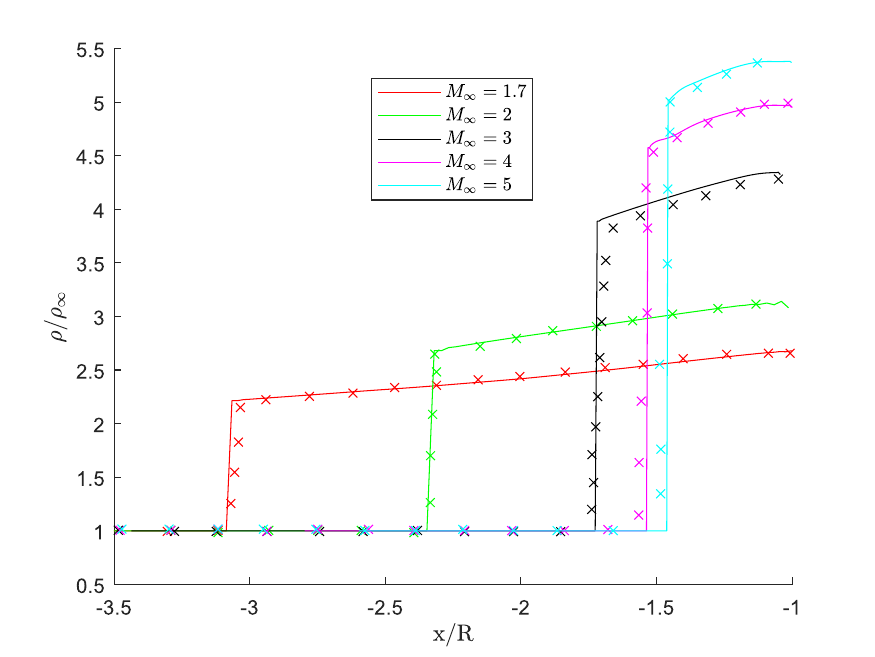}
\caption{}
\label{fig:08b}
\end{subfigure}
\caption{Verification test along the stagnation line of a
  cylinder. CPG, $\gamma = 1.4$. Comparison of HYMOR (solid lines) with results from
  \citet{sinclair2017theoretical} (crosses). (\textit{a})
  Pressure. (\textit{b}) Density.}
\label{fig:08}
\end{figure}

A subsequent test case compares our solution with the shock-capturing
inviscid solution from \citet{sinclair2017theoretical}, with the
results shown in figure~\ref{fig:08}. This comparison considers flow
over a cylinder at Mach numbers 1.7, 2, 3, 4, and 5, following the
methodology of \citet{sinclair2017theoretical}. Although
shock-capturing approaches inherently suffer from accuracy limitations
due to the artificial thickness over which shocks are
resolved~\citep{carpenter1999accuracy}, the comparison shows
satisfactory agreement across all tested Mach numbers when evaluated
against the reference solution.

\subsubsection{3D axisymmetric geometries}

\begin{figure}[]
  \begin{center}
  \def~{\hphantom{0}}
  \begin{subfigure}[c]{0.80\columnwidth}
    \centering
    \includegraphics[width=0.8\linewidth]{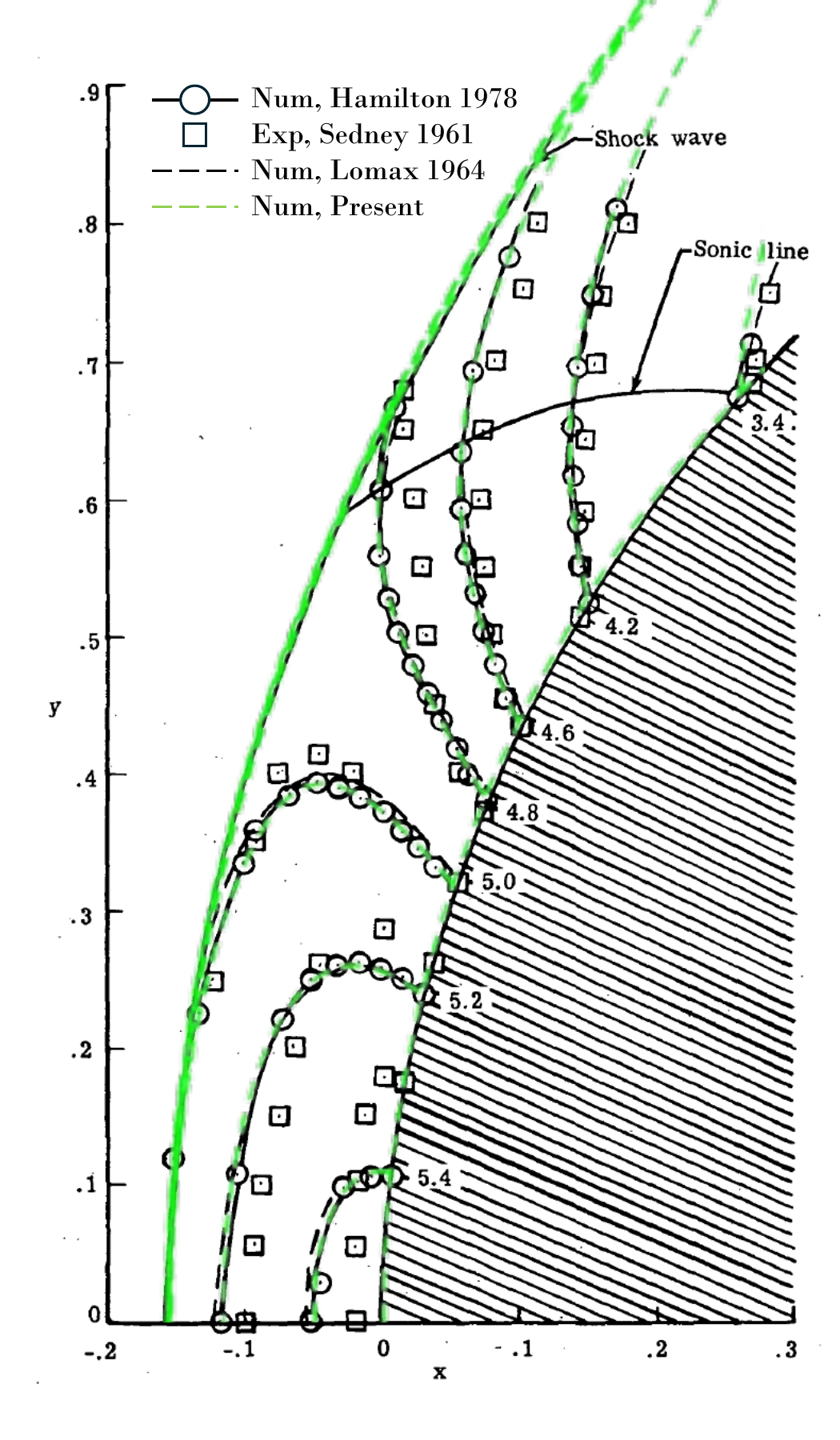}
    \caption{}
    \label{fig:3D_density}
  \end{subfigure}
  \hfill
  \begin{subfigure}[c]{0.45\columnwidth}
    \centering
    \begin{tabular}{ccc}
        \toprule
       $\psi$ (deg) & $p/p_\infty$ Reference & $p/p_\infty$ Present \\[3pt] \midrule
       ~0~~~~ & 0.9246 & 0.9243\\
       ~5.625 & 0.9141 & 0.9146\\
       11.250 & 0.8831 & 0.8834\\
       16.875 & 0.8330 & 0.8335\\
       22.500 & 0.7676 & 0.7677\\
       28.125 & 0.6894 & 0.6898\\
       33.750 & 0.6037 & 0.6043\\
       39.375 & 0.5148 & 0.5155\\
       45.000 & 0.4278 & 0.4282\\
       \bottomrule
    \end{tabular}
    \caption{}
    \label{tab:3D_pressure_wall}
  \end{subfigure}
  \caption{Verification test on spherical geometry. $\mathrm{M}_\infty
    = 8.06$ and $\gamma = 1.4$. Euler equations with an ideal
    gas. (\textit{a}) Density contours and shock position
    \citep{hamilton1978solution}. (\textit{b}) Pressure at the wall,
    at spherical angle $\psi$ [deg] (stagnation line starts at $\psi =
    0$ [deg]), reference data from \citet{hamilton1978solution}.  }
  \label{fig:3Daxym}
  \end{center}
\end{figure}

To verify the axisymmetric solution capability, we consider as a
reference the Mach 8.06 flow over a sphere reported by
\citet{hamilton1978solution}. The density contours and shock position
are compared against both numerical and experimental results in
figure~\ref{fig:3D_density}. Good agreement is observed, particularly
with the numerical results from the reference solution. The wall
pressure is reported in table~\ref{tab:3D_pressure_wall}, together
with the numerical results from \citet{hamilton1978solution}. Once
again, the results of the present solver agree well with the
reference.

\subsection{Testing real gas models}
\label{sec::verification_real_gas}
\begin{figure}[]
\centering
\includegraphics[width=0.6\linewidth]{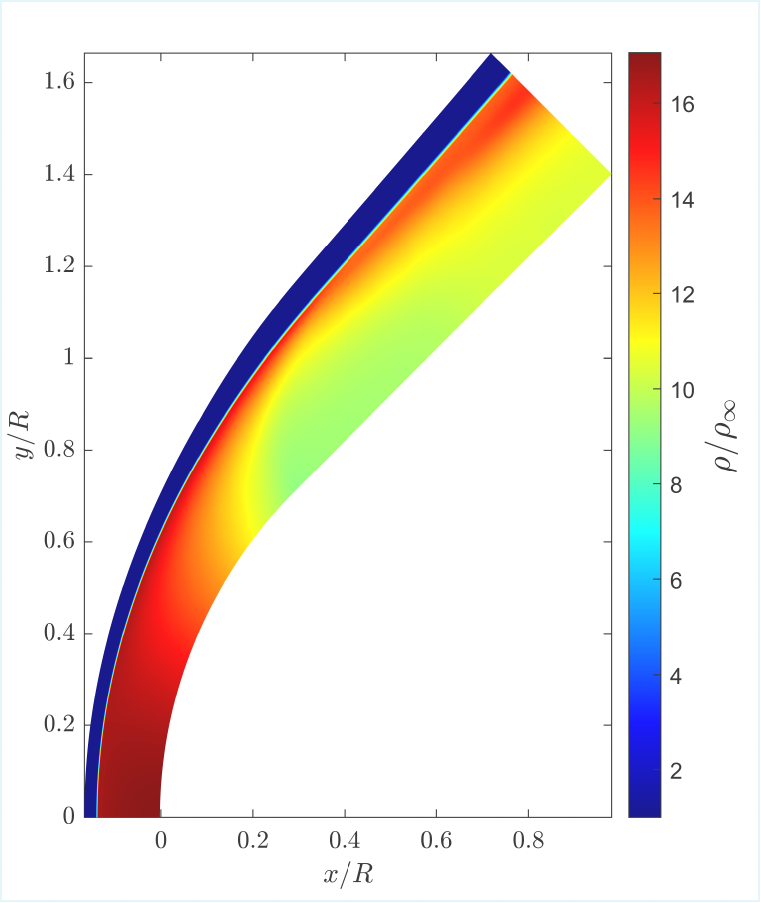}
\caption{Verification of real-gas models using the 2D Euler equations
  over a $45^\circ$ wedge. Density is non-dimensionalized with the
  freestream density. Results are obtained with HYMOR using
  Chemical-RTVE.}
\label{fig:Eilmer_geo}
\end{figure}

We test the implementation of the Chemical-RTVE and NonEq-RTVE models
in HYMOR against finite-rate chemistry with thermal relaxation
computations performed with Eilmer~\citep{gibbons2023eilmer}.  The
test case consists of inviscid flow over a $45^\circ$ wedge with a
cylindrical tip of radius $R$, under the freestream conditions listed
in table~\ref{tab:Eilmer_setup}. This case is selected because it
entails thermal and chemical non-equilibrium conditions immediately
behind the shock wave.  As an example, the solution obtained under
chemical and thermal equilibrium conditions is shown in
figure~\ref{fig:Eilmer_geo}.
\begin{table*}
  \begin{center}
\def~{\hphantom{0}}
  \caption{Freestream test conditions used to compare HYMOR Chemical-RTVE and
    NonEq-RTVE models with finite-rate chemistry from Eilmer.}
  \label{tab:Eilmer_setup}
  \begin{tabular}{cccccccc}
    \toprule
      Model & Wedge angle & Freestream composition & $R$ & $T_\infty$ & $p_\infty$ & $U_\infty$ & $\mathrm{M}_\infty$ \\[3pt] 
      \midrule
      Inviscid & $45^\circ$, 2D & $X_{CO_2} = 1$ & 0.045 m & 1,403 K & 10,000 Pa & 4,425 m/s & 7.95 \\
      \bottomrule
  \end{tabular}

  \end{center}
\end{table*}

\begin{figure*}[]
\centering
\includegraphics[width=0.9\linewidth]{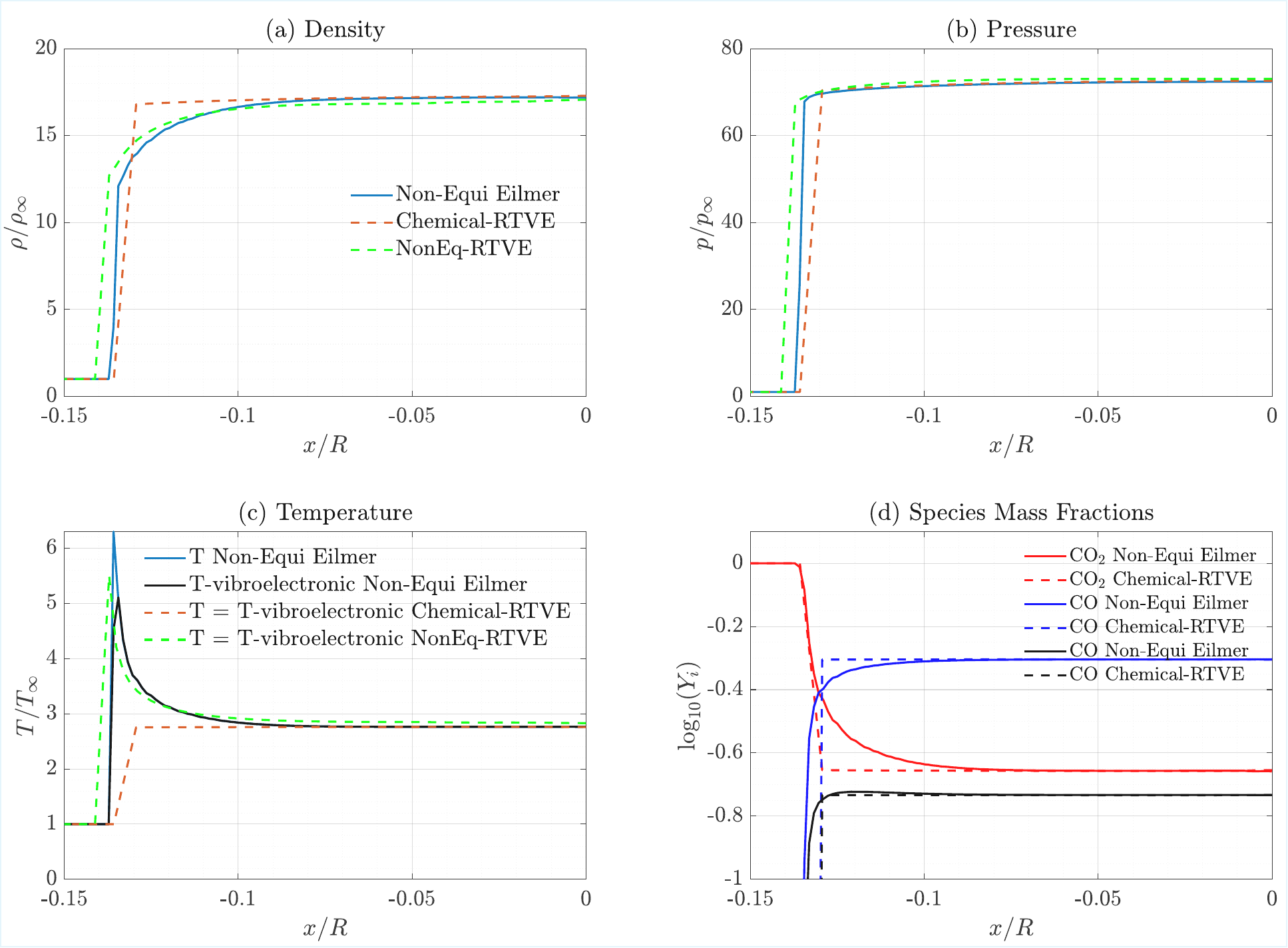}
\caption{Comparison of equilibrium and non-equilibrium models. HYMOR
  solutions with Chemical-RTVE and NonEq-RTVE are shown with dashed
  lines, while results from Eilmer are shown with solid lines.
  Properties are computed along the stagnation line. The coordinate
  $x$ is zero at the stagnation point. In (\textit{c}), $T$ denotes
  the translational--rotational temperature, and
  $T_{\mathrm{vib\text{-}electronic}}$ denotes the vibroelectronic
  temperature associated with vibrational and electronic degrees of
  freedom.}
\label{fig:Eilmer_comparison}
\end{figure*}

The comparison between Eilmer and HYMOR is presented in
figure~\ref{fig:Eilmer_comparison}. The most prominent feature
distinguishing the Eilmer results from the equilibrium model
(HYMOR-Chemical-RTVE) is the large translational--rotational
temperature spike immediately behind the shock. This is consistent with
thermochemical non-equilibrium: energy is not transferred instantaneously
to the vibrational and chemical modes, so it is initially concentrated
in translation--rotation. By including chemical relaxation
(HYMOR-NonEq-RTVE), HYMOR reproduces this non-equilibrium behavior near
the shock front.

A further observation from the Eilmer results is that the
vibrational--electronic mode relaxes faster than the chemical
composition for this case, motivating the assumption of thermal
equilibrium adopted in HYMOR-NonEq-RTVE while retaining chemical
non-equilibrium. Note that species concentrations are not shown for
HYMOR-NonEq-RTVE because this model does not advect species
explicitly; instead, only the effective thermodynamic variables
$\gamma^*$ and $c_\mathrm{v}^*$ are evolved. The shock standoff
distance in HYMOR-NonEq-RTVE deviates slightly from the full Eilmer
solution because the reduced model employs only two progress variables
and accounts solely for chemical relaxation.  Despite this,
HYMOR-NonEq-RTVE captures the principal post-shock trends, and the two
solutions converge farther downstream as the flow progressively
relaxes toward equilibrium.

\subsection{Verification of Rankine--Hugoniot jump conditions with thermochemical effects}
\label{sec::Rankine_verification}

\begin{figure*}[]
\centering
\includegraphics[width=0.9\linewidth]{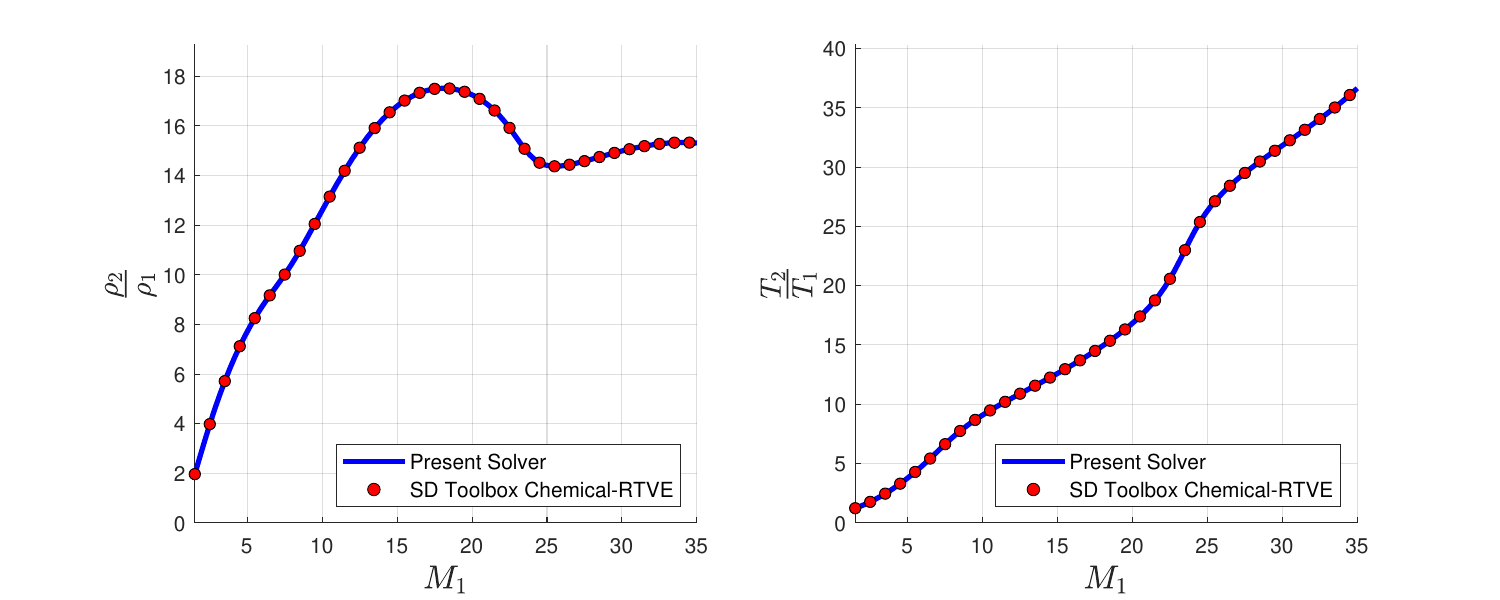}
\caption{Comparison of the Rankine--Hugoniot solver implementation in
  HYMOR with Shock and Detonation Toolbox results
  \citep{edl_sdtoolbox_2023} for Chemical-RTVE equilibrium. Case with
  $u_s = 0$. Mars atmosphere: $X_{CO_2}:0.9556$, $X_{N_2}:0.0270$,
  $X_{Ar}:0.0160$, $X_{O_2}:0.0014$.  $p_1 = 0.1$ atm, $T_1 = 293.15$
  K. Left: density ratio. Right: temperature ratio.}
\label{fig:comparisonSDtoolbox}
\end{figure*}

We compare the results from HYMOR with Chemical-RTVE against the
solutions obtained using the Shock and Detonation Toolbox (SD
Toolbox)~\citep{edl_sdtoolbox_2023} for a one-dimensional normal
shockwave. Because the SD Toolbox evaluates the same thermochemical
model through Cantera, this comparison constitutes a verification of
implementation correctness rather than a validation of the chemical
model itself: the reference and HYMOR share identical thermochemical
formulations, so no discrepancies associated with the modeling are
expected and any deviation would be attributable solely to the
implementation. Independent toolkits such as the Combustion Toolbox of
\citet{cuadra2025combustion} could instead be used to assess the
agreement between alternative chemical models, a distinct exercise
that probes the modeling rather than its implementation and lies
outside the present verification scope.
Figure~\ref{fig:comparisonSDtoolbox} confirms that good agreement with
the reference solution is indeed achieved.

\subsection{Stability analysis verification}
\label{sec::stability_analysis_verification}

\begin{figure*}[]
\centering
\begin{subfigure}[t]{1\columnwidth}
\includegraphics[width=\columnwidth]{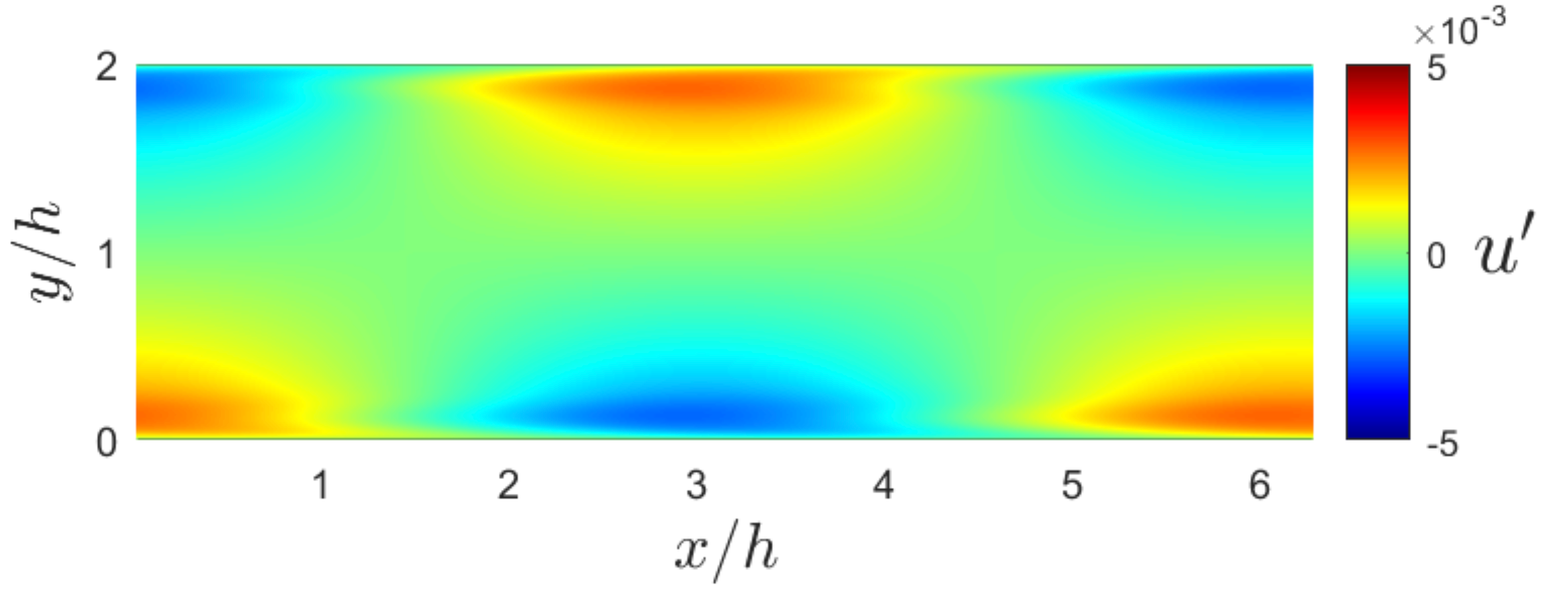}
\caption{}
\label{fig:09a}
\end{subfigure}
\begin{subfigure}[t]{1\columnwidth}
\includegraphics[width=\columnwidth]{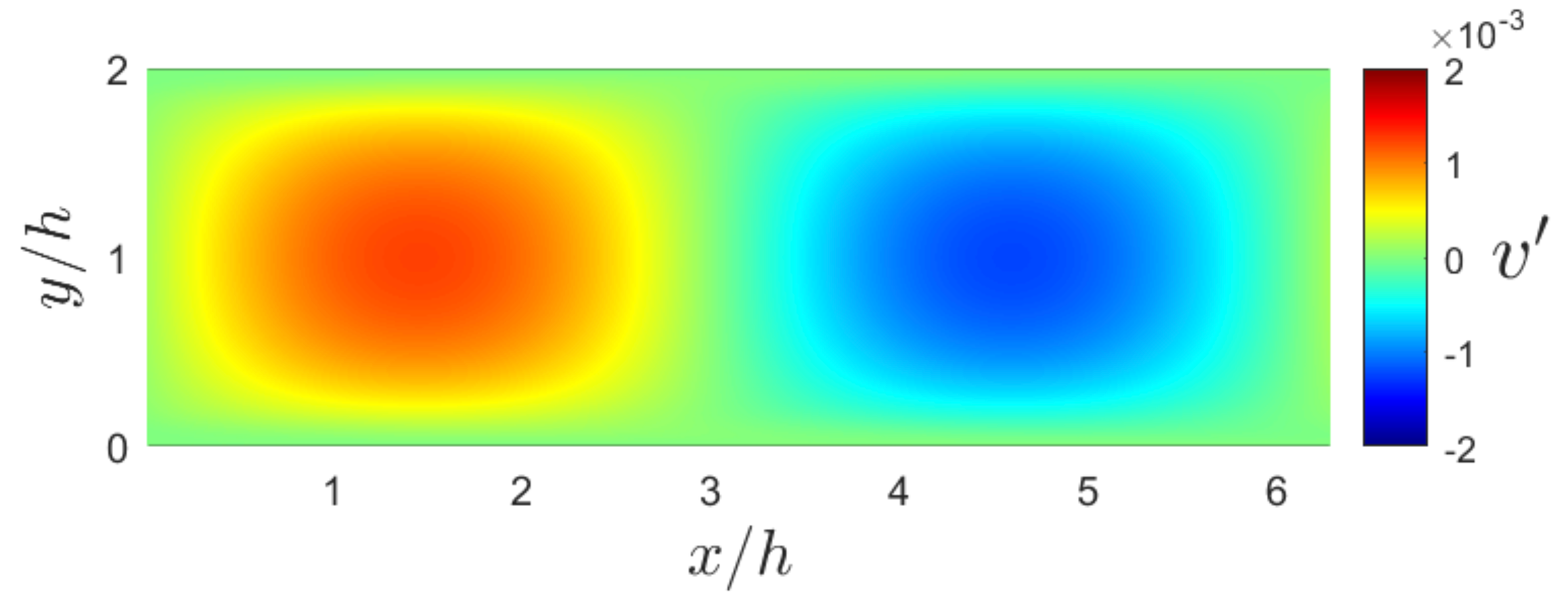}
\caption{}
\label{fig:09b}
\end{subfigure}
\begin{subfigure}[t]{0.85\columnwidth}
\includegraphics[width=\columnwidth]{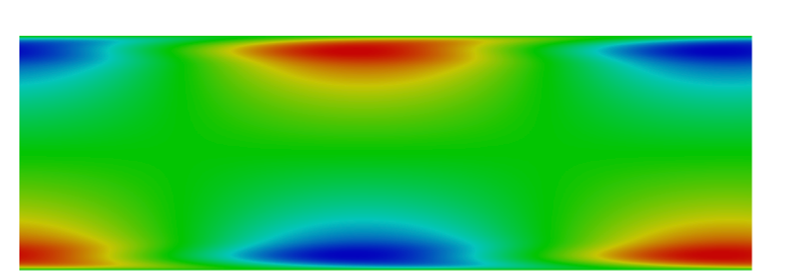}
\caption{}
\label{fig:09c}
\end{subfigure}\hspace{0.1\columnwidth}
\begin{subfigure}[t]{0.85\columnwidth}
\includegraphics[width=\columnwidth]{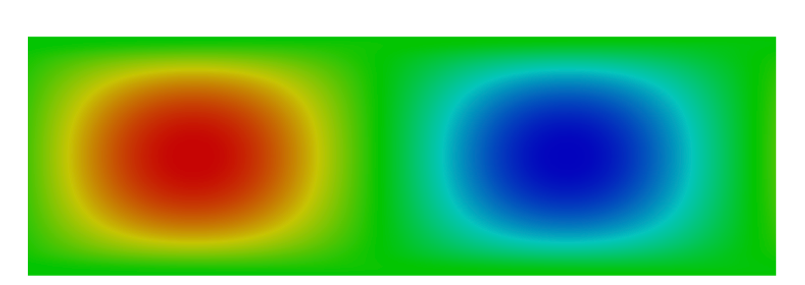}
\caption{}
\label{fig:09d}
\end{subfigure}
\caption{Verification test for most unstable eigenmode for the
  incompressible channel flow at $\mathrm{Re} = 7,500$. (\textit{a})
  Horizontal velocity mode. Present results. (\textit{b}) Vertical
  velocity mode. Present results. (\textit{c}) Horizontal velocity
  mode. Reference \citet{nektar}. (\textit{d}) Vertical velocity
  mode. Reference \citet{nektar}.}
\label{fig:09}
\end{figure*}

We consider first an incompressible channel flow. The computational
results are benchmarked against the reference solutions of
\citet{canuto2007spectral} and \citet{nektar}. Following the
specifications of those references, the Reynolds number is set to
$7,500$. To recover incompressible flow conditions, the Mach number is
set to $\mathrm{M}=0.001$. Table~\ref{tab:channeltab} shows the convergence
behavior of the most unstable eigenvalue. The results exhibit
excellent agreement with the reference values. The corresponding
eigenmode comparison is shown in figure~\ref{fig:09}, which likewise
demonstrates strong agreement with the reference solution.
\begin{table*}
  \begin{center}
\def~{\hphantom{0}}
\caption{ Convergence of the most unstable eigenvalue for
  incompressible channel flow at $Re = 7,500$. The channel half-height
  is $h$.}
  \label{tab:channeltab}
  \begin{tabular*}{\tblwidth}{@{} LLL@{} }
    \toprule
      Mesh domain $2h \times 2 \pi h$ & Present eigenvalue $U/h$ & Difference with reference \citet{canuto2007spectral} (\%) \\[3pt] \midrule
       $200 \times 400$ & $0.002219 + 0.249764\mathrm{i}$ & 0.051\\
       $400 \times 800$ & $0.002231 + 0.249892\mathrm{i}$ & 0.000080\\
       \bottomrule
  \end{tabular*}
  \end{center}
\end{table*}

\begin{figure*}[]
\centering
\begin{subfigure}[t]{0.91\columnwidth}
\includegraphics[width=\columnwidth]{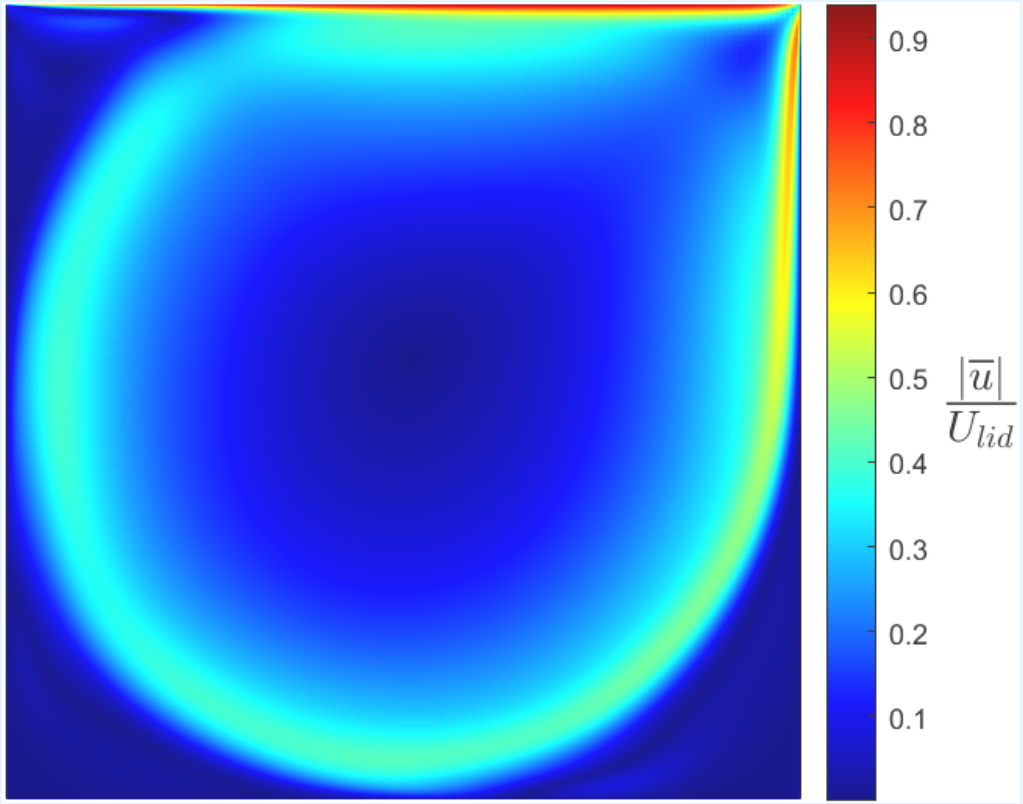}
\caption{}
\label{fig:10a}
\end{subfigure}
\begin{subfigure}[t]{1\columnwidth}
\includegraphics[width=\columnwidth]{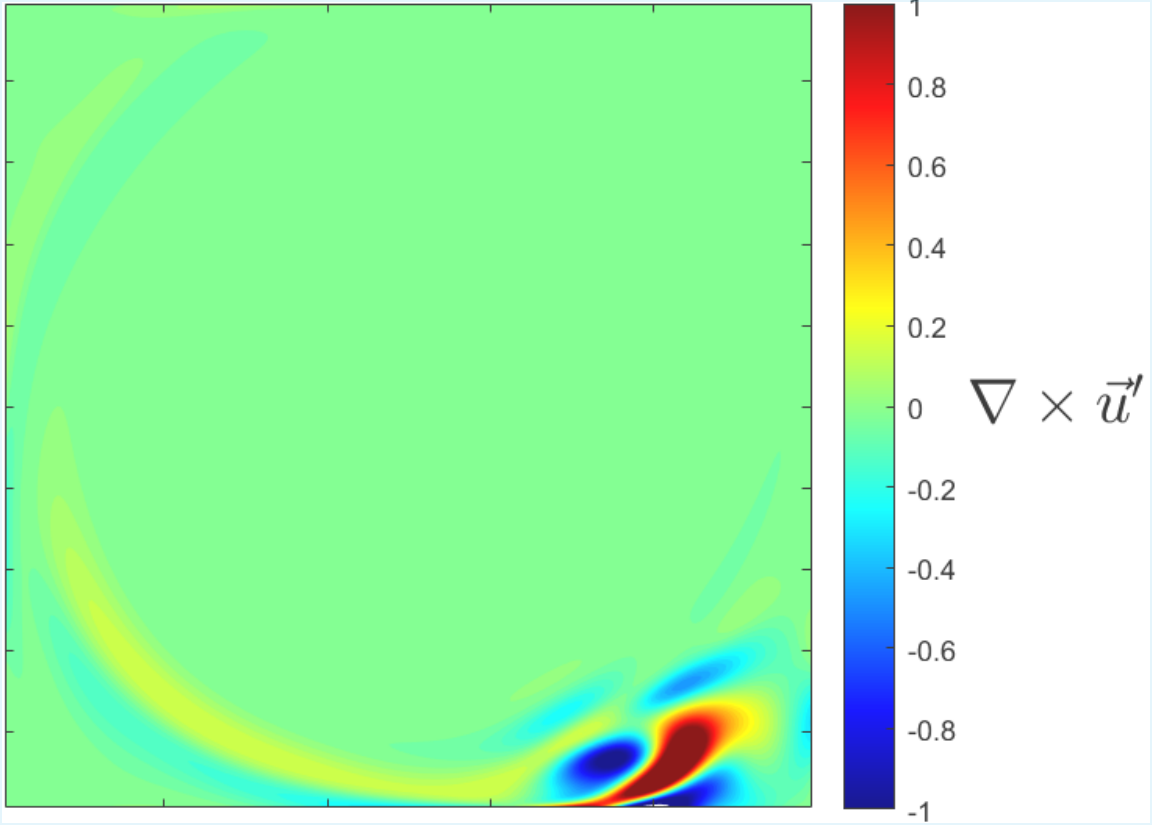}
\caption{}
\label{fig:10b}
\end{subfigure}
\caption{Verification test for stability analysis of incompressible
  lid-driven cavity flow. (\textit{a}) Steady state velocity magnitude
  at $\mathrm{Re} = 8,000$. Grid $400 \ \times \ 400$. (\textit{b}) Most
  unstable vorticity eigenmode $\mathrm{Re} = 8,030$.}
\label{fig:10}
\end{figure*}

The next verification case considers a lid-driven cavity flow
configuration. The analysis is conducted for incompressible flow
conditions, achieved by setting the Mach number to $\mathrm{M}=0.001$.
Previous studies indicate that the first Hopf bifurcation appears at
approximately $\mathrm{Re} = 8,000$. Reported critical values vary across the
literature: \citet{fortin1997localization} reported $\mathrm{Re} = 8,000$,
\citet{auteri2002numerical} found $\mathrm{Re} = 8,018$, and
\citet{pan2000projection} identified $\mathrm{Re} = 8,500$. Through global
stability analysis of the base flow, shown in figure~\ref{fig:10a},
the present study determines the critical Reynolds number to be $\mathrm{Re}_c
= 8,030$, as reported in table~\ref{tab:liddriventab}. This result
shows good agreement with previously published values. The analysis of
the most unstable eigenmode reveals the instability mechanism
responsible for the generation of limit-cycle oscillations. The
instability is localized in the bottom-right corner region, consistent
with the observations of \citet{bruneau20062d} regarding the spatial
localization of the unstable mode.
\begin{table}
  \begin{center}
\def~{\hphantom{0}}
  \caption{Verification test for stability of incompressible
    lid-driven cavity flow. Convergence of the critical Reynolds
    number at which the first Hopf bifurcation appears.}
  \label{tab:liddriventab}
  \begin{tabular}{cc}
    \toprule
       Mesh size & $\mathrm{Re}_c$ \\[3pt]
    \midrule
       $100 \times 100$ & 7,450\\
       $200 \times 200$ & 8,010\\
       $400 \times 400$ & 8,030\\
    \bottomrule
  \end{tabular}
  \end{center}
\end{table}

\begin{figure*}[]
\centering
\begin{subfigure}[t]{0.80\columnwidth}
\includegraphics[width=\columnwidth]{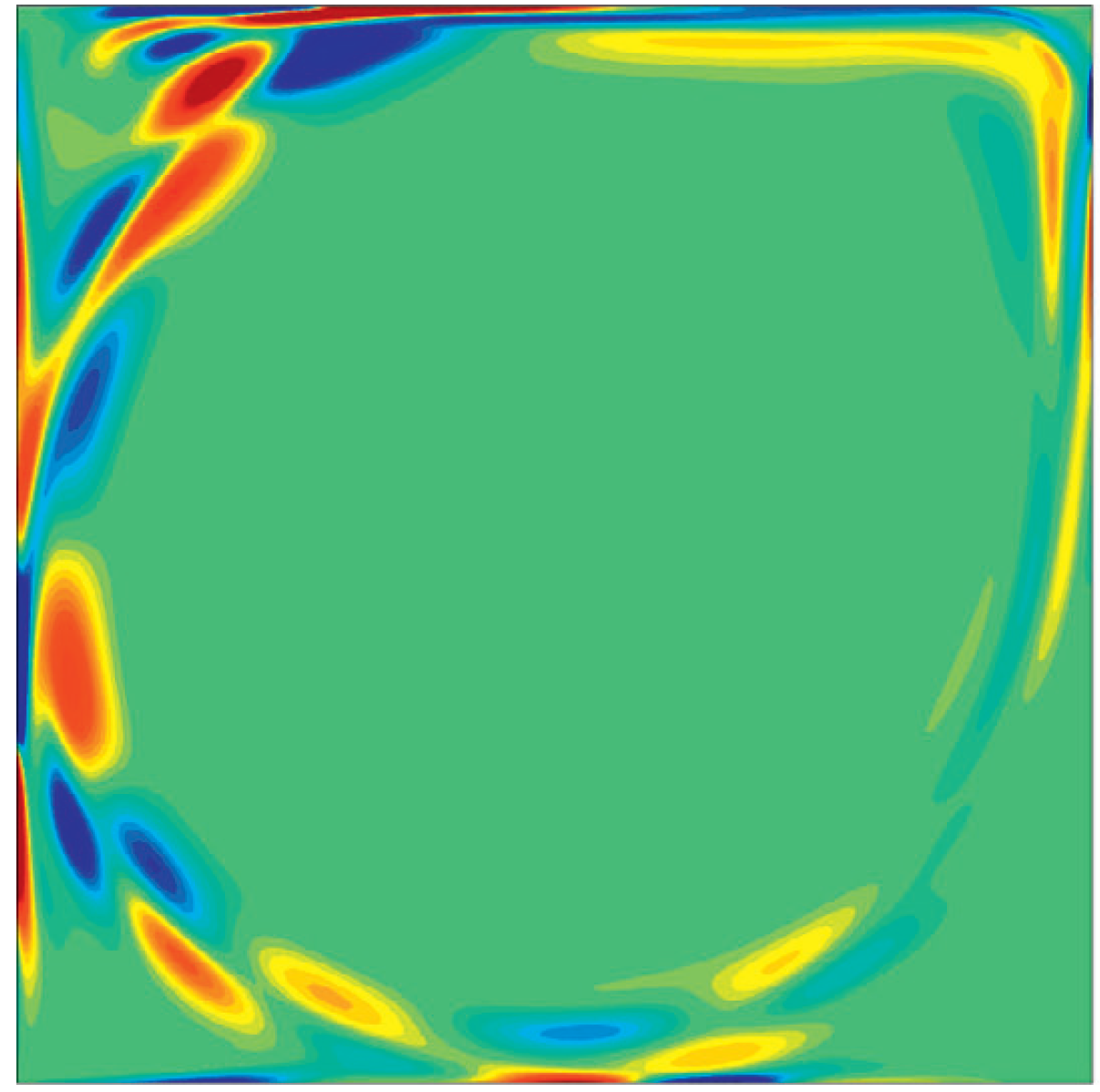}
\caption{}
\label{fig:11a}
\end{subfigure}
\begin{subfigure}[t]{0.99\columnwidth}
\includegraphics[width=\columnwidth]{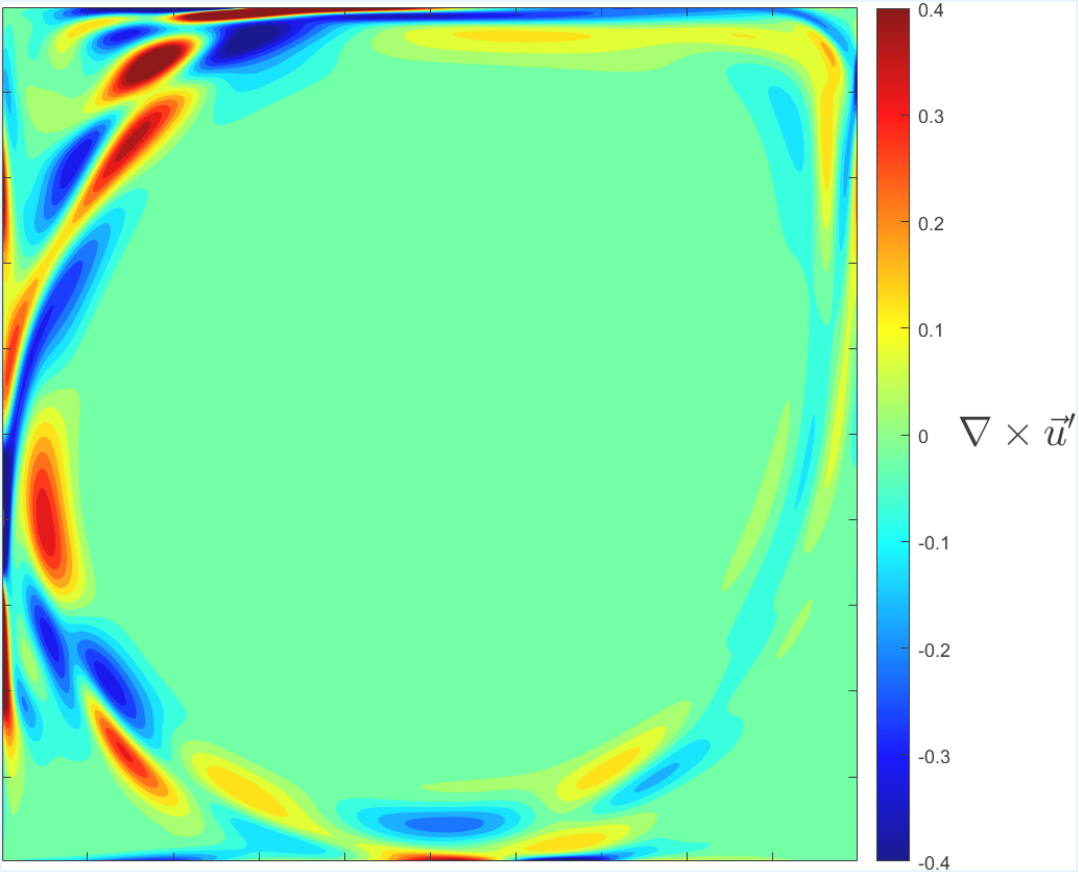}
\caption{}
\label{fig:11b}
\end{subfigure}
\caption{Verification test for stability analysis in a compressible
  lid-driven cavity flow test case at $\mathrm{M} = 0.95$ and
  $\mathrm{Re} = 11,200$. Comparison of the vorticity eigenmode
  associated with the leading instability. (\textit{a}) Reference
  \citet{ohmichi2017compressibility}. (\textit{b}) Present result.}
\label{fig:11}
\end{figure*}

Finally, we test a compressible lid-driven cavity flow under the test
conditions of \citet{ohmichi2017compressibility} at $\mathrm{M} = 0.95$ and $\mathrm{Re}
= 11,200$. Under these conditions, the flow becomes unstable, and the
corresponding unstable mode is compared with the reference results in
figure~\ref{fig:11}. This test case is particularly challenging
because, at such a high Reynolds number, the flow undergoes multiple
bifurcations~\citep{bruneau20062d}. Despite this complexity, the
present results reproduce all the intricate features of the reference
solution.

\subsection{Verification of linearized shock-disturbance interactions}
\label{sec::Verification_Disturbances_Shock}

In this section, we verify the correct implementation of the linear
shock--disturbance interaction between freestream and downstream
disturbances at the shock. We consider a base flow (denoted with subscript $0$) consisting of a
stationary normal shock separating two constant-velocity states:
upstream (state 1) and downstream (state 2). For this specific test
case, the flow is assumed to be calorically perfect, with a Mach
number of 28 and a specific heat ratio of 1.18. Under these
conditions, the base flow can be determined from the standard
nonlinear Rankine--Hugoniot relations.

Small perturbations are superimposed on this base flow. In
particular, the disturbance in the freestream ($x < 0$) is specified
as a pure entropy wave. In the freestream region, entropy waves
convect at the local flow velocity $u_1$ and, in the linear limit, do
not generate pressure or velocity fluctuations. Therefore, the
incident wave takes the form:
\begin{equation}
    \mathbf{q}'_1 = \begin{pmatrix} \rho'_1 \\
    u'_1 \\
    p'_1  
    \end{pmatrix} = 
    \begin{pmatrix}
    \varepsilon \, \rho_{1,0} \, e^{i k_x (x - u_{1,0} \, t)} \\
    0 \\
    0
    \end{pmatrix},
\end{equation}
where the disturbance is characterized by the shock-normal wavenumber
$k_x$ and has small amplitude $\varepsilon$ relative to the base-flow
density $\rho_{1,0}$. 

The solution to the interaction problem follows the linear interaction
analysis (LIA) derived by \citet{mckenzie1968interaction}. A
fundamental result of LIA is that the shockwave acts as a coupling
mechanism between flow modes. Consequently, a pure incident
disturbance, such as an entropy or acoustic wave, generally does not
remain pure after transmission; instead, it excites a superposition of
all admissible downstream modes. In the present case, this leads to
the excitation of both entropy and acoustic modes:
\begin{equation}
    \mathbf{q}'_2 = A_{ac} \mathbf{v}_{ac} e^{i k_{ac} (x - (u_{2,0} + a_{2,0}) t)} + A_{ent} \mathbf{v}_{ent} e^{i k_{ent} (x - u_{2,0} t)},
\end{equation}
where the eigenvectors (normalized by pressure and density
respectively) are:
\begin{equation}
    \mathbf{v}_{ac} = \begin{pmatrix} 1/a_{2,0}^2 \\ 1/(\rho_{2,0} a_{2,0}) \\ 1 \end{pmatrix}, \quad
    \mathbf{v}_{ent} = \begin{pmatrix} 1 \\ 0 \\ 0 \end{pmatrix}
\end{equation}
where $a$ is the sound speed, and the wavenumbers are $k_{ac} =
\frac{u_{1,0} k_x}{u_{2,0} + a_{2,0}}$ and $k_{ent} = \frac{u_{1,0}
  k_x}{u_{2,0}}$. Substituting the wave ansatz into the jump
conditions yields the transmission matrix system derived by
\citet{mckenzie1968interaction}. Solving this system yields the
analytic LIA solution for the verification case.

\begin{figure}[]
\centering
\includegraphics[width=\columnwidth]{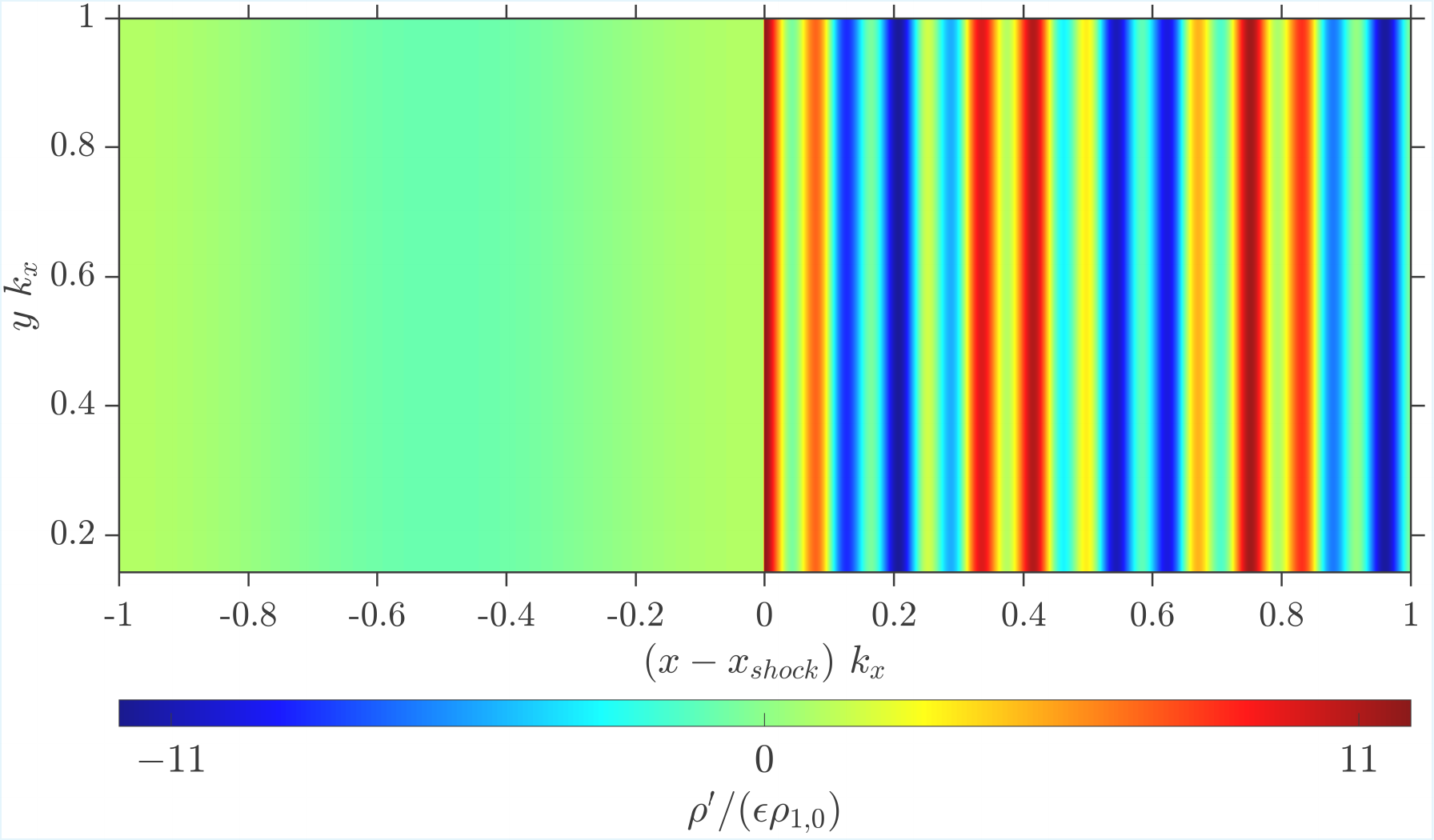}
\caption{Verification of the mode-conversion mechanism across a 1D
  shock wave. Density contours of an entropy disturbance placed in the
  freestream at $\mathrm{M}_1 = 28$, $\gamma = 1.18$, and calorically
  perfect gas.}
\label{fig:LIA_image_rho}
\end{figure}

The numerical solution is computed with HYMOR (see
section~\ref{sec::freestream_receptivity_implementation}). The
resulting density disturbance is shown in
figure~\ref{fig:LIA_image_rho}. Across the shock, its magnitude
increases by more than one order of magnitude, owing to the high Mach
number and low specific-heat ratio of the test case. In addition, the
spatial wavenumber of the density disturbance increases significantly
downstream of the shock. This behavior is a direct consequence of the
flow deceleration across the shock, which leads to larger wavenumbers
(particularly for the entropic mode).

\begin{figure*}[]
\centering
\includegraphics[width=0.9\linewidth]{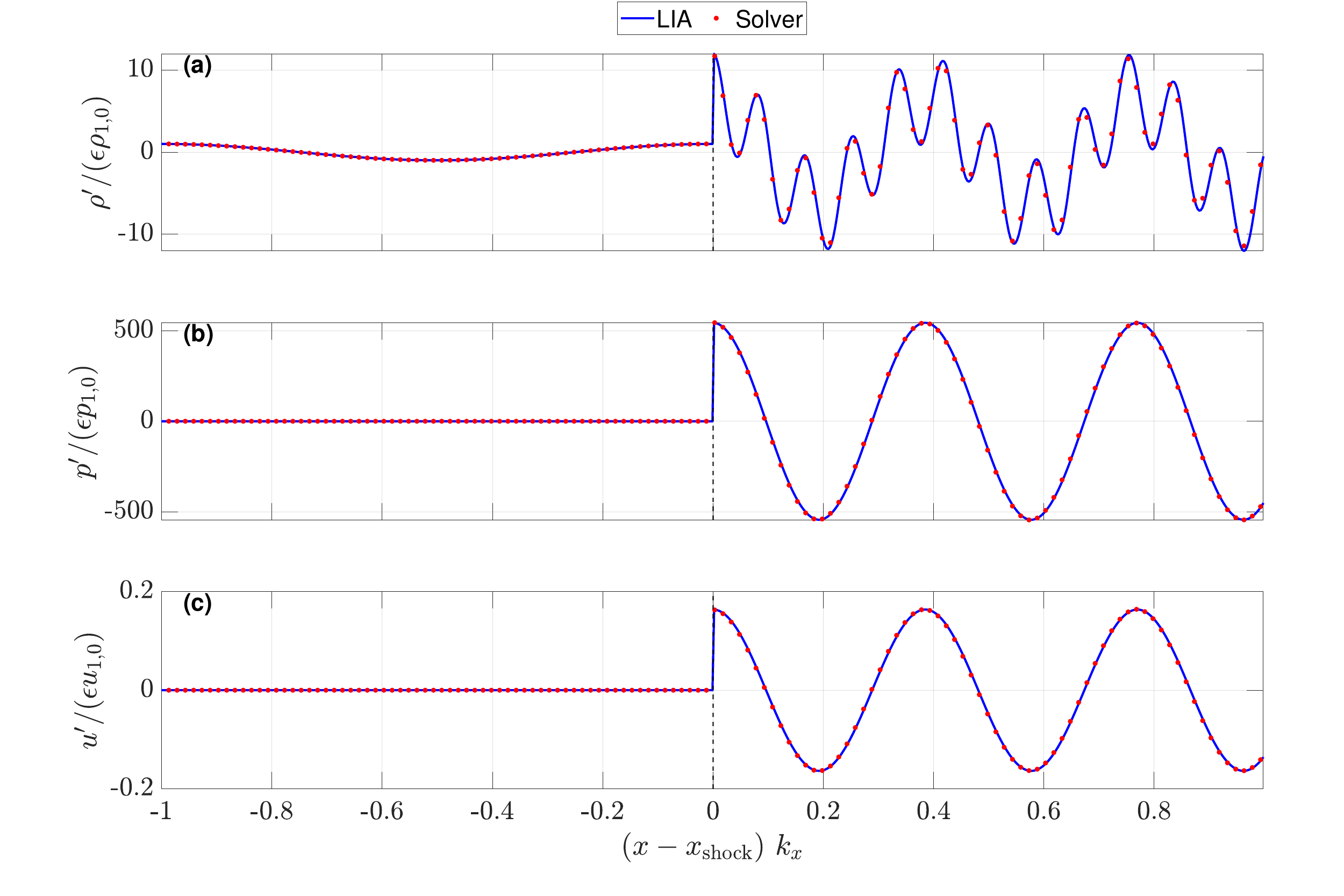}
\caption{Verification of the mode-conversion mechanism across a 1D
  shock wave. Comparison of HYMOR against the LIA analytical solution
  for $\mathrm{M}_1 = 28$, $\gamma = 1.18$, and a calorically
  perfect gas. (\textit{a}) Density disturbance. (\textit{b}) Pressure
  disturbance. (\textit{c}) Velocity disturbance.}
\label{fig:LIA_comparison}
\end{figure*}

In figure~\ref{fig:LIA_comparison}, the solution from HYMOR is
compared against the analytical solution from LIA. Good agreement is
observed among the density, pressure, and velocity results. The
comparison shows that the initial entropy wave, which initially
contains only density fluctuations, generates pressure and velocity
oscillations through the induced acoustic mode. The pressure and
velocity fields are excited only by the acoustic mode and display a
nearly sinusoidal pattern.

\section{Program documentation}
\label{sec::Program documentation}

The entire codebase is implemented in both MATLAB and Julia with
identical functionality, enabling users to choose the language that
best suits their workflow. Table~\ref{tab:languages} summarizes the
languages and their roles, while tables~\ref{tab:top_level} and
\ref{tab:utils} detail the directory layout.
%
\begin{table*}
  \begin{center}
  \def~{\hphantom{0}}
  \caption{Programming languages used in the HYMOR framework.}
  \label{tab:languages}
  \begin{tabular*}{\tblwidth}{@{} LLL@{} }
    \toprule
    Language & Files & Role \\[3pt]
    \midrule
    MATLAB  & 108 & Primary solver, stability analysis, and tutorials \\
    Julia   & 111 & Mirror implementation of the full solver framework \\
    \bottomrule
  \end{tabular*}
  \end{center}
\end{table*}
%
\begin{table*}
  \begin{center}
  \def~{\hphantom{0}}
  \caption{Top-level directory structure of the HYMOR repository.}
  \label{tab:top_level}
  \begin{tabular*}{\tblwidth}{@{} LLL@{} }
    \toprule
    Directory / File & Type & Description \\[3pt]
    \midrule
    \texttt{src/}         & Julia module  & Julia package entry point (\texttt{HypersonicsStability.jl}) \\
    \texttt{chemistry/}   & Data + code   & Thermochemistry models, gas properties, and shock-jump tables \\
    \texttt{utils/}       & Code          & Core solver utilities (mesh, operators, stability, time marching) \\
    \texttt{tutorials/}   & Examples      & Three tutorial cases with input files and reference figures \\
    \texttt{profiling/}   & Benchmarks    & Eigenvalue-solver and time-marching performance tests \\
    \texttt{Project.toml} & Config        & Julia package dependencies \\
    \texttt{README.md}    & Documentation & Project overview and quick-start guide \\
    \bottomrule
  \end{tabular*}
  \end{center}
\end{table*}
%
\begin{table*}
  \begin{center}
  \def~{\hphantom{0}}
  \caption{Contents of the \texttt{utils/} directory.}
  \label{tab:utils}
  \begin{tabular*}{\tblwidth}{@{} LLL@{} }
    \toprule
    Sub-directory & Files & Description \\[3pt]
    \midrule
    \texttt{Initialization/}    & 10 & Input loading, solution initialization, and restart logic \\
    \texttt{Mesh/}              & ~8 & Curvilinear mesh generation and element-space mappings \\
    \texttt{Operators/}         & 22 & Spatial derivatives, flux computations (2-D and 3-D), and BCs \\
    \texttt{Shock\_fitting/}    & 15 & Shock evolution, upstream conditions, and spline fitting \\
    \texttt{Time\_marching/}    & ~7 & Explicit RK4 and implicit Backward-Euler integrators \\
    \texttt{Energy\_budgets/}   & ~6 & Chu energy norm computation and budget decomposition \\
    \texttt{Postprocessing/}    & ~6 & Plotting routines for base flow and perturbation modes \\
    \texttt{Stability\_analysis/} & 38 & Modal and non-modal stability solvers and receptivity \\
    \bottomrule
  \end{tabular*}
  \end{center}
\end{table*}

\subsection{Software requirements}
\label{sec::software_requirements}

\paragraph{MATLAB.}
MATLAB R2023b or later is required. The Optimization Toolbox and
Curve Fitting Toolbox are also required. The Parallel Computing
Toolbox is optional for GPU-accelerated stability computations. No
additional toolboxes are necessary.

\paragraph{Julia.}
Julia 1.9 or later is required. Package dependencies are declared in
\texttt{Project.toml}. All dependencies are installed automatically
via \texttt{Pkg.instantiate()}.

\paragraph{Drivers.}
An NVIDIA GPU with CUDA support is recommended for the
stability-analysis modules, including the eigenvalue solvers and
transient-growth computations, in order to accelerate computations.
These computations can also run on CPU, but they take significantly
longer.

\subsection{Running a tutorial}
\label{sec::running_tutorial}

Each tutorial directory contains a driver script (\texttt{main.m} or
\texttt{main.jl}), a parameter file (\texttt{input\_file.m} or
\texttt{input\_file.jl}), and a \texttt{README.md} with a step-by-step
walkthrough. To run a tutorial:
\begin{enumerate}
  \item Set the \texttt{solver\_dir} variable in the driver script to
    point to the repository root.
  \item Execute \texttt{main.m} in MATLAB or \texttt{main.jl} in
    Julia.
\end{enumerate}

\section{Applications}
\label{sec::Applications}

To illustrate the capabilities of the code, we present a
representative case study. A detailed tutorial that replicates this
example can be found in the toolkit:
\texttt{HYMOR/tutorials/hypersonic\_blunt\_cone}. The geometry
consists of a spherical nose tip of radius $R$ followed by a
$20^\circ$ half-angle cone. The freestream conditions correspond to a
high-enthalpy flow at $\mathrm{M}_\infty = U_\infty / a_\infty =
12.0$, where $a_\infty$ is the freestream speed of sound, and
$\mathrm{Re}_\infty = \rho_\infty U_\infty R / \mu_\infty = 100,000$,
representative of a hypersonic vehicle flying through Earth's
atmosphere at an altitude of approximately 50~km. The detailed
freestream conditions are listed in
table~\ref{tab:freestream_conditions_example}.

The flow field is obtained with HYMOR by solving the compressible
Navier--Stokes equations. For the thermochemical model, the
Chemical-RTVE formulation is adopted together with transport
properties based on collision integrals (see
section~\ref{sec::Thermochemical and transport models}). This model
assumes thermal and chemical equilibrium, which is a reasonable
approximation under the present conditions: the vibrational and
chemical Damk\"{o}hler numbers, based on the post-normal-shock
velocity and the nose radius, are of order $10^3$ and $10^2$,
respectively.

At the shock wave, Rankine--Hugoniot jump conditions are imposed. At
the wall, adiabatic and no-slip boundary conditions are enforced.
Axial symmetry is imposed along the symmetry axis, and the outflow
boundary is treated with a non-reflecting characteristic boundary
condition.

\begin{figure}[]
\centering
\includegraphics[width=0.75\columnwidth]{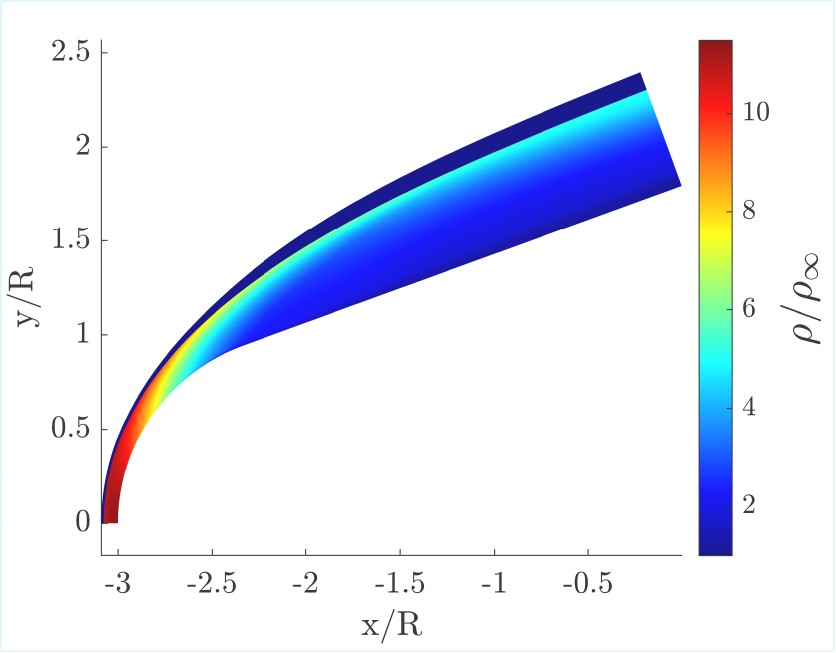}
\caption{Application of HYMOR to a blunt cone. Non-dimensional density
  field of the steady-state solution at $\mathrm{M}_\infty = 12.0$ and
  $\mathrm{Re}_\infty = 100,000$.}
\label{fig:example_rho}
\end{figure}

The nonlinear solver reaches a steady state, which is visualized in
figure~\ref{fig:example_rho}. Near the stagnation point, the flow
experiences a rapid compression, resulting in a very thin shock
stand-off distance. Downstream, the flow undergoes a strong expansion
around the spherical tip.
\begin{table*}
  \begin{center}
  \def~{\hphantom{0}}
  \caption{Freestream conditions used for the present application case.}
  \label{tab:freestream_conditions_example}
  \begin{tabular*}{\tblwidth}{@{} LLLLLL@{} }
    \toprule
      $U_\infty$ [m/s] & $\rho_\infty$ [kg/m$^3$] & $T_\infty$ [K] & Freestream composition & $\mathrm{M}_\infty$ & $\mathrm{Re}_\infty$ \\[3pt]
       4,000 & 0.001 & 270 & $X_{\mathrm{N}_2}\!:\!0.7812$, 
       $X_{\mathrm{O}_2}\!:\!0.2095$,
       $X_{\mathrm{Ar}}\!:\!0.0093$ & 12.0 & 100,000 \\
       \bottomrule
  \end{tabular*}
  \end{center}
\end{table*}

\subsection{Global modal stability analysis}

The modal stability analysis indicates that the flow is modally
stable: the least-damped eigenmode, i.e., the one whose eigenvalue
$\lambda$ lies closest to the imaginary axis, decays at a rate $\sigma
= \mathrm{Re}(\lambda) = -0.139$, with temporal evolution
$\mathbf{q}(t) = \exp(\lambda\, t\, U_\infty / R)\,\mathbf{q}(0)$.
The leading eigenvalues are listed in
table~\ref{tab:unstable_eigenvalues}. These results are consistent
with previous studies of hypersonic boundary layers over blunt
bodies~\citep{paredes2016linear, paredes2019nose,
  paredes2020mechanism}.
\begin{table}
  \begin{center}
  \def~{\hphantom{0}}
  \caption{Global modal stability analysis: Leading eigenvalues,
    ordered by decreasing growth rate $\sigma =
    \mathrm{Re}(\lambda)$.}
  \label{tab:unstable_eigenvalues}
  \begin{tabular*}{\tblwidth}{@{} LLL@{} }
    \toprule
    Mode & $\sigma = \mathrm{Re}(\lambda)$ & $\omega = \mathrm{Im}(\lambda)$ \\[3pt]
    \midrule
    1  & $-0.139$ & $~0.000$ \\
    2  & $-0.242$ & $~0.000$ \\
    3  & $-0.265$ & $~0.000$ \\
    4  & $-0.312$ & $~0.092$ \\
    5  & $-0.312$ & $-0.092$ \\
    6  & $-0.391$ & $~0.222$ \\
    7  & $-0.391$ & $-0.222$ \\
    8  & $-0.457$ & $~0.333$ \\
    \bottomrule
  \end{tabular*}
  \end{center}
\end{table}

\begin{figure}[]
\centering
\includegraphics[width=\columnwidth]{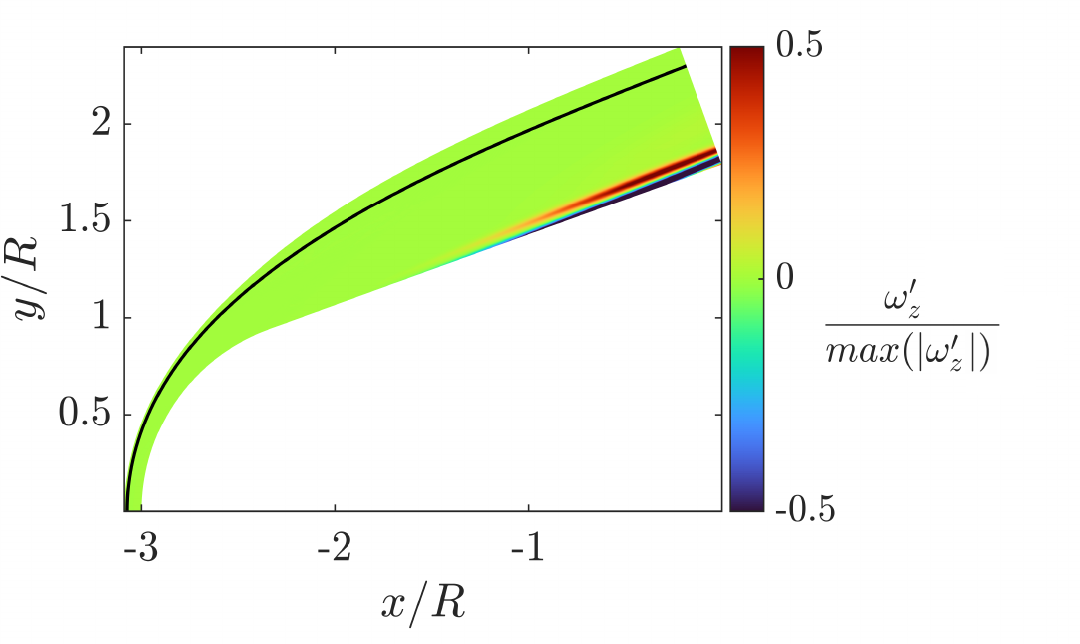}
\caption{Global modal stability analysis: vorticity of the
  least-damped disturbance.}
\label{fig:modal_vort}
\end{figure}

The spatial structure of the least-damped mode is shown in
figure~\ref{fig:modal_vort}.  A thin shear layer of vorticity is
concentrated in the vicinity of the boundary layer, indicating that
boundary-layer transition is the physical mechanism closest to
instability onset, a finding that is well established in the
literature.

\subsection{Global transient growth analysis}

\begin{figure}[]
\centering
\includegraphics[width=0.8\columnwidth]{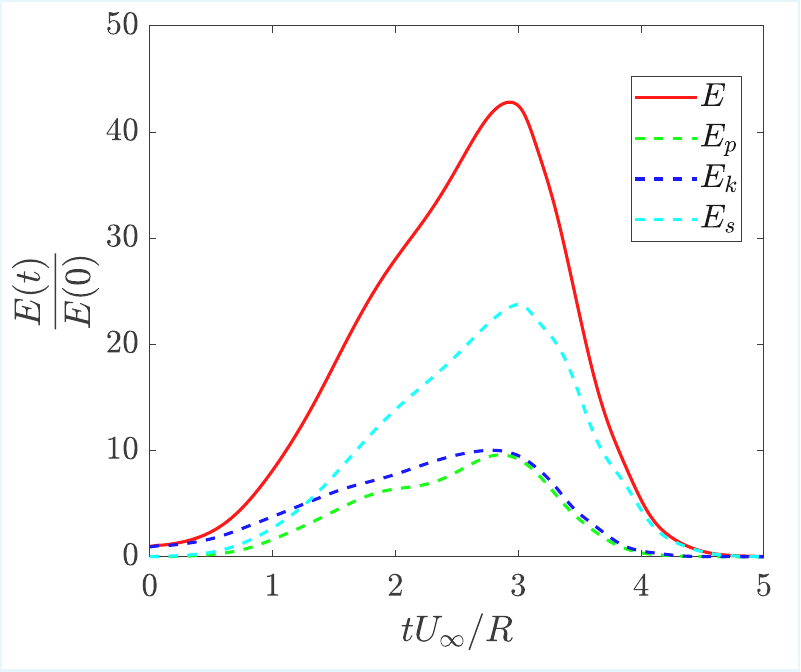}
\caption{Transient growth analysis: energy amplification of the Chu
  energy norm for the disturbance that produces optimal growth at
  $t\,U_\infty / R = 3$. }
\label{fig:transient_energy}
\end{figure}

The objective of the transient-growth analysis is to identify the
optimal initial disturbance that maximizes the energy amplification
ratio $E(t)/E(0)$. For the present example, the optimization horizon
is set to $t\,U_\infty / R = 3$. The temporal evolution of the energy
growth for this optimal disturbance is shown in
figure~\ref{fig:transient_energy}. The total energy amplification
peaks near the target time $t\,U_\infty / R = 3$, reaching a value of
approximately 40. A decomposition into the individual components of
the Chu energy norm reveals that the entropic contribution $E_s$
accounts for approximately $50\%$ of the total growth, while the
kinetic $E_k$ and acoustic $E_p$ components each contribute roughly
$25\%$.

\begin{figure*}[]
\centering
\begin{subfigure}[t]{\columnwidth}
\includegraphics[width=\columnwidth]{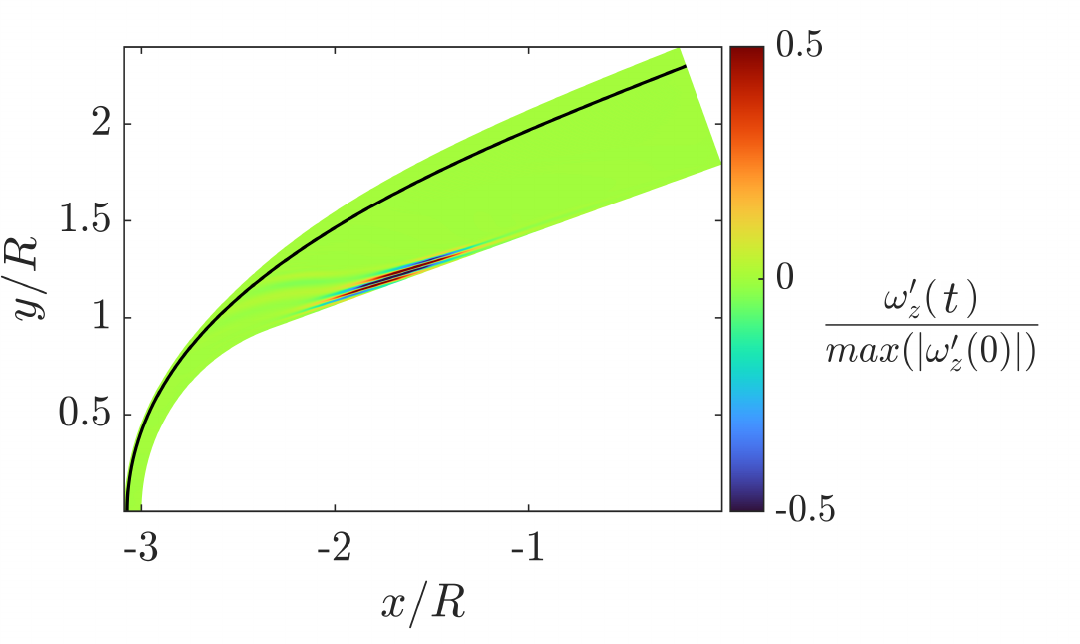}
\caption{}
\label{fig:example_non_modal_vort_0}
\end{subfigure}
\begin{subfigure}[t]{\columnwidth}
\includegraphics[width=\columnwidth]{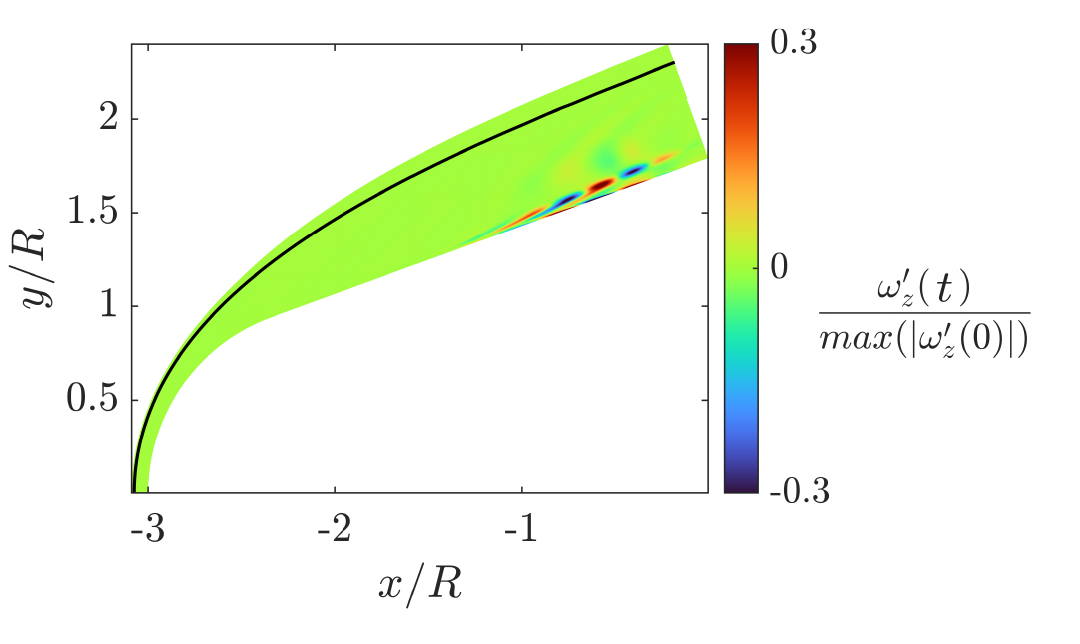}
\caption{}
\label{fig:example_non_modal_vort_2_5}
\end{subfigure}
\caption{Transient growth analysis. Disturbance that produces optimal
  energy growth at $t\,U_\infty / R = 3$. (\textit{a})~Non-dimensional
  vorticity of the initial disturbance at $t =
  0$. (\textit{b})~Non-dimensional vorticity at $t =
  2.5\,R/U_\infty$. }
\label{fig:example_non_modal_vort}
\end{figure*}

The spatial structure of the optimal initial disturbance is depicted
in figure~\ref{fig:example_non_modal_vort_0}. The perturbation is
localized within the boundary layer, downstream of the favorable
pressure-gradient region associated with the spherical tip. The
disturbance consists of vortical shear streaks oriented at
approximately $-45^\circ$. At later time $t = 2.5\,R/U_\infty$
(figure~\ref{fig:example_non_modal_vort_2_5}), these structures have
been advected along the boundary layer and, more importantly, tilted
by the mean shear. This behavior is reminiscent of the
well-characterized Orr mechanism, which is known to produce transient
energy amplification in shear-dominated flows~\citep{jiao2021orr}.

\subsection{Freestream receptivity analysis}

\begin{figure}[]
\centering
\includegraphics[width=0.8\columnwidth]{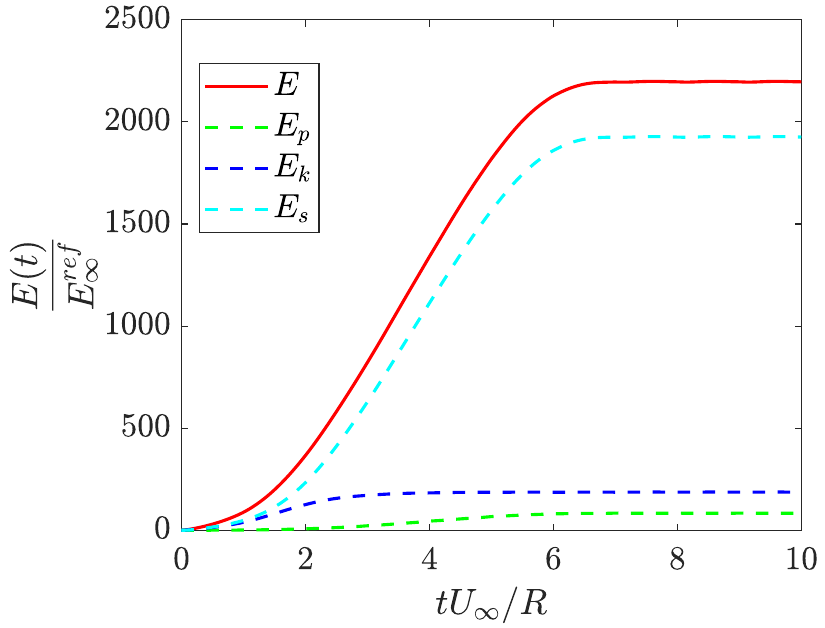}
\caption{Freestream receptivity analysis: energy amplification of the
  Chu energy norm for the freestream disturbance that produces optimal
  growth at $t\,U_\infty / R = 10$.}
\label{fig:freestream_energy}
\end{figure}

Finally, we illustrate the freestream receptivity capability of the
code. In this analysis, the objective is to determine the optimal
freestream disturbance that maximizes the post-shock energy growth at
a target time $t = 10\,R/U_\infty$. The resulting energy amplification
history is shown in figure~\ref{fig:freestream_energy}.  In contrast
to the transient growth case, the energy amplification is
substantially larger, reaching a value close to $2,000$. This larger
gain has two sources: amplification and mode conversion during shock
transmission, followed by non-modal amplification in the post-shock
shear layers. LIA predicts that shock transmission and generation
coefficients depend strongly on Mach number
\citep{mckenzie1968interaction}, so the strong bow shock provides a
large input--output gain for selected incident disturbances. In the
present case, the Chu-energy decomposition identifies the entropic
energy component $E_s$ as the dominant contribution.

\begin{figure*}[]
\centering
\begin{subfigure}[t]{\columnwidth}
\includegraphics[width=\columnwidth]{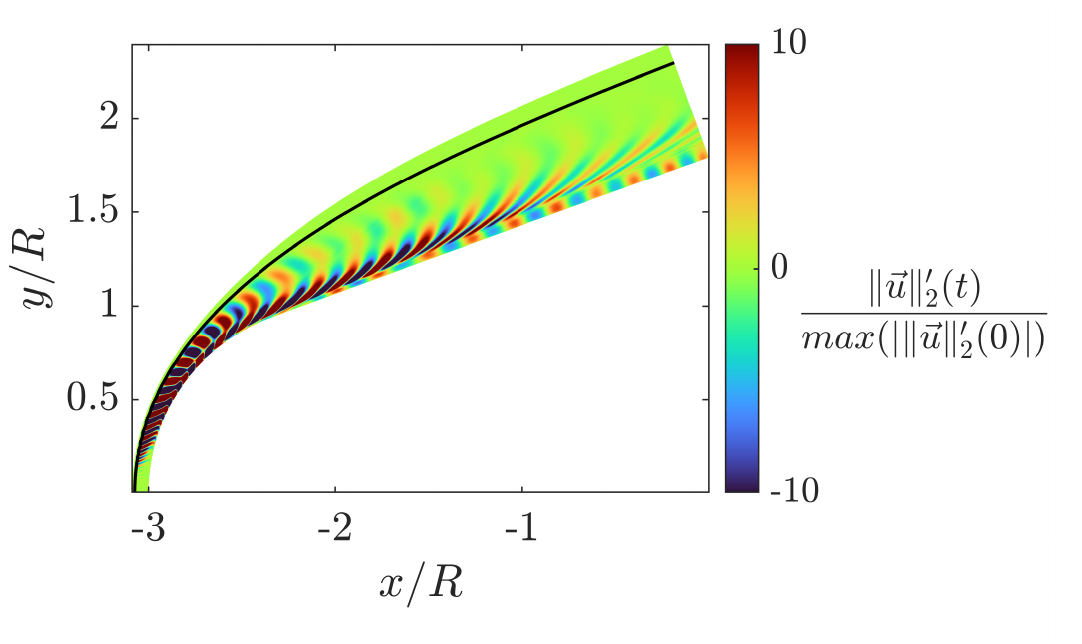}
\caption{}
\label{fig:example_freestream_velocity}
\end{subfigure}
\begin{subfigure}[t]{\columnwidth}
\includegraphics[width=\columnwidth]{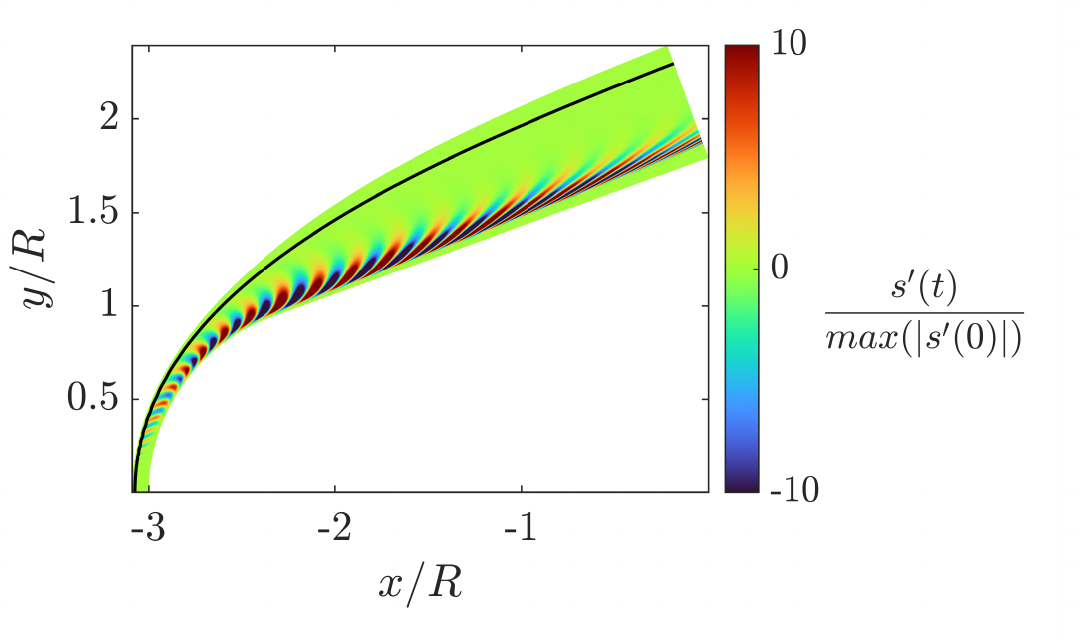}
\caption{}
\label{fig:example_freestream_vort_zoom}
\end{subfigure}
\caption{Freestream receptivity analysis. Post-shock disturbances
  induced by the optimal freestream perturbation at $t =
  10\,R/U_\infty$. (\textit{a})~Non-dimensional velocity
  magnitude. (\textit{b})~Non-dimensional entropy. }
\label{fig:example_freestream_modes}
\end{figure*}

The spatial distribution of the disturbances induced by the optimal
freestream perturbation after $t\,U_\infty/R = 10$ of freestream
forcing is shown in figure~\ref{fig:example_freestream_modes}.  The
velocity-magnitude field
(figure~\ref{fig:example_freestream_velocity}) reveals that the energy
amplification is concentrated near the stagnation region, where
boundary modes close to the wall are excited by acoustic waves
generated within the entropy layer, though their intensity decays
further downstream. The entropy field
(figure~\ref{fig:example_freestream_vort_zoom}), by contrast, shows
that entropy disturbances propagate along the entropy layer induced by
the bow-shock curvature and are tilted by the mean shear.

\section{Performance}
\label{sec::Performance}

The software comprises two main computational components: time
marching of the nonlinear Navier--Stokes equations for the computation
of base flows, and iterative stability solvers. The performance of
each component is assessed in this section.

\subsection{Time-marching solver}

To evaluate the performance of the time-marching algorithm, a test
case with a periodic domain and the Chemical-RTVE thermochemical model
is considered. Performance is quantified using the performance index
(PID), as defined in previous work~\citep{fehn2019matrix, krais2021flexi}:
\begin{equation}
    \mathrm{PID} = \frac{t_{\mathrm{wall}} \cdot
      N_{\mathrm{cores}}}{N_{\mathrm{DOF}} \cdot N_{\mathrm{steps}}
      \cdot N_{\mathrm{RK}}},
\end{equation}
where $t_{\mathrm{wall}}$ is the wall-clock time, $N_{\mathrm{cores}}$
the number of CPU cores,
$N_{\mathrm{DOF}}$ the number of degrees of freedom,
$N_{\mathrm{steps}}$ the number of time steps, and $N_{\mathrm{RK}}$
the number of Runge--Kutta stages per time step. This metric provides a
computational efficiency measure scaled by the problem size.

\begin{figure}[]
\centering
\includegraphics[width=\columnwidth]{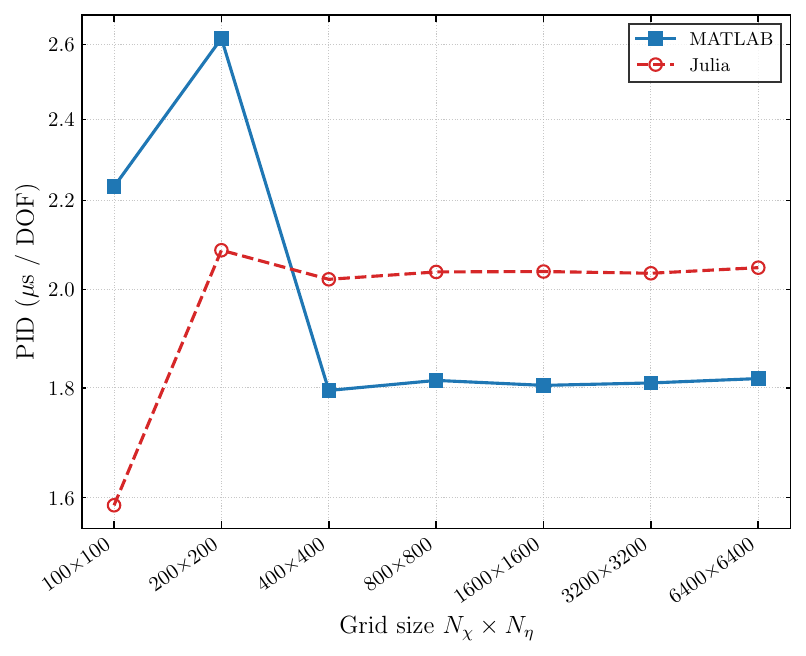}
\caption{PID performance of the nonlinear Navier--Stokes time-marching
  solver in HYMOR for different grid sizes. A fourth-order
  Runge--Kutta scheme is used with the Chemical-RTVE thermochemical
  model. Results for the MATLAB and Julia implementations are
  compared. All measurements were obtained on an AMD Ryzen~9 7945HX
  processor.}
\label{fig:performance_time_marching}
\end{figure}

The results, presented in figure~\ref{fig:performance_time_marching},
show that both the MATLAB and Julia implementations achieve nearly
size-independent scaling of approximately
$2~\mu\mathrm{s}/\mathrm{DOF}$ for sufficiently large grids; that is,
a single core advances one degree of freedom (one finite-volume cell)
per RK stage in approximately $2~\mu\mathrm{s}$. The Julia version was
compiled with just-in-time (JIT) compilation using the \texttt{O3}
optimization flag; nevertheless, it exhibits similar performance to
MATLAB. This is because both implementations rely on vectorized
operations that, at the lowest level, execute within highly optimized
BLAS/LAPACK libraries. For comparison, \citet{krais2021flexi} reported
a PID of approximately $1~\mu\mathrm{s}/\mathrm{DOF}$ for a similar
problem using a native C++ implementation, which is roughly twice as
fast as the present code. It should be noted, however, that the
hardware used in their study was not specified, and therefore the
comparison is provided only as a reference.  Even so, these results
show that, despite the use of high-level languages, the vectorization
strategy delivers competitive performance relative to highly optimized
low-level implementations.

Both Julia and MATLAB implementations support multithreading within a
single node. In practice, performance is limited by memory bandwidth,
and peak throughput is typically reached with only a fraction of the
available threads. Given that the code operates on 2-D grids (either
planar Cartesian or axisymmetric), the current performance provides
reasonable wall-clock times for numerical simulations across a wide
range of configurations.

It should be emphasized, however, that the solver does not currently
support distributed-memory parallelism: there is no MPI layer, and
execution is therefore confined to a single shared-memory node.  For
the two-dimensional configurations targeted in this work, this is not
restrictive.  It does constitute a clear limitation for prospective
extensions of the framework: simulations at substantially higher
Reynolds numbers, or the eventual generalization to fully
three-dimensional base flows, would entail memory footprints and
times-to-solution for which distributed-memory parallelism becomes
essential. Extending the time-marching solver to a hybrid
MPI$+$multithreading model is therefore identified as a priority for
future development.

\subsection{Stability solvers}

The second computational component comprises the linear stability
solvers: modal analysis, transient growth, and freestream
receptivity. Performance is assessed using the example case described
in section~\ref{sec::Applications}. In all three analyses, the
innermost loop of the iterative solver is offloaded to the GPU, in
this case an NVIDIA RTX~4090 Mobile GPU, when available.  This
strategy yields a significant speed-up, as the primary bottleneck in
the stability solvers is memory bandwidth, which is substantially
higher on modern GPUs.

\begin{figure}[]
\centering
\includegraphics[width=\columnwidth]{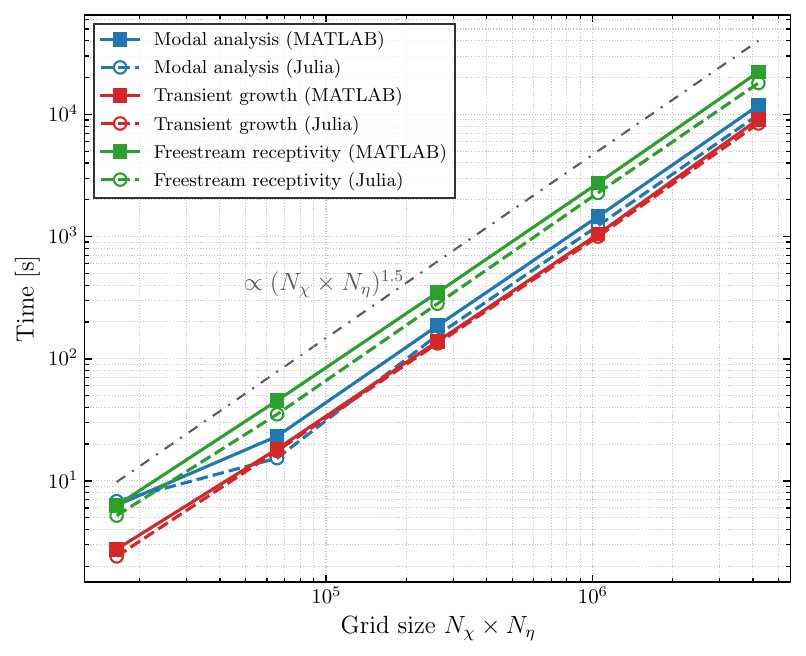}
\caption{Wall-clock time required to compute the ten leading modes for
  each of the three linear stability analyses. All measurements were
  obtained on an AMD Ryzen~9 7945HX processor with an NVIDIA RTX~4090
  mobile GPU as accelerator. The predicted theoretical scaling is
  indicated by the dash-dotted line.}
\label{fig:performance_eigen_solvers}
\end{figure}

The computational cost as a function of grid resolution is compared
for both the Julia and MATLAB implementations in
figure~\ref{fig:performance_eigen_solvers}. The observed scaling
follows the predicted theoretical rate of $(N_\chi \times
N_\eta)^{3/2}$. This exponent arises from the combination of two
contributions: the spatial operations scale linearly with the number
of grid points $N_\chi \times N_\eta$, while the maximum allowable
time step is governed by a stability limit. For an advection-dominated
problem, the CFL-limited time step introduces an additional factor
proportional to $(N_\chi \times N_\eta)^{1/2}$ for roughly isotropic
grids, yielding the overall $3/2$ scaling. The computational results
agree well with this prediction. In this case, the Julia
implementation is slightly faster than MATLAB, owing to finer control
over GPU operations, which enables more efficient offloading to the
accelerator.

\section{Conclusions}
\label{sec::Conclusions}

We have introduced HYMOR, an open-source computational toolkit for
global linear stability analysis of high-enthalpy hypersonic
flows. The toolkit addresses a gap in the existing landscape of
stability analysis software by integrating modal, non-modal, and
freestream receptivity analyses within a single, publicly available
package built on a global formulation.

The global modal analysis capability resolves the full spatial
dependence of the disturbance field, enabling the identification of
instability mechanisms that involve interactions between spatially
separated flow regions, phenomena that cannot be captured by local or
weakly non-parallel methods such as LST and PSE. The transient growth
module provides a tool for quantifying non-modal energy amplification,
offering insight into transition scenarios where classical modal
predictions are insufficient. The freestream receptivity analysis
completes the transition prediction chain by connecting the freestream
disturbances to the internal boundary-layer response.

A central design choice of the toolkit is the adoption of a
shock-fitting formulation, which treats shock waves as sharp
discontinuities and has been verified, for the tested linear
interaction case, to reproduce the LIA response to infinitesimal
disturbances.  This reduces the shock-capturing artifacts associated
with artificial shock thickness, which is particularly important in
high-enthalpy regimes where strong shocks are present. The code
additionally provides nonlinear operators for base-flow computation
with automatic linearization of the discrete operators, and
incorporates several thermochemical models to account for real-gas
effects.

The capabilities of HYMOR have been verified against a collection of
benchmark cases spanning its modal, non-modal, and receptivity
analysis modes. The results are in agreement with reference data,
confirming the accuracy of the numerical implementation and the
correctness of the shock-fitting treatment.

The toolkit is released under the MIT license and provides both Julia
and MATLAB implementations. HYMOR is intended to serve as a foundation
for further development by the hypersonic stability community. The
open-source nature of HYMOR is designed to facilitate these
developments and to promote reproducibility in hypersonic transition
research.

\section*{Acknowledgements}
\label{sec::Acknowledgements}
The authors gratefully acknowledge Professor Hans G. Hornung for providing non-equilibrium computation results to validate the thermochemical models employed in this work. We also thank Professor Joseph E. Shepherd for his guidance in implementing thermochemical models and for generously sharing thermochemical data files.

\section*{Funding}
\label{sec::Funding}
This work was supported by an Early Career Faculty grant from NASA's
Space Technology Research Grants Program (Grant Number 80NSSC23K1498).

\appendix
\section{Details of problem formulation}
\label{sec::Details_of_problem_formulation}

\subsection{Physical models}
\label{sec::Physical_models}

\subsubsection{Chemical equilibrium and frozen models}
\label{sec:eq_models}

Chemical-equilibrium and chemically frozen models have been
implemented in the solver. Chemical-equilibrium models assume that the
species have sufficient time to react compared with the flow time
scale, and also assume that no inter-species diffusion occurs.
Chemically frozen flow, on the other hand, assumes that the chemical
composition remains unchanged because the reaction rates are too slow
relative to the flow time scales. The following thermal and chemical
model variants are available:
\begin{enumerate}
    \item Frozen chemistry with
      translational\hyp{}rotational\hyp{}vibrational equilibrium
      (Frozen-RTV). Chemical composition is considered constant and
      equal to the freestream conditions. Thermodynamic properties for
      the translational\hyp{}rotational\hyp{}vibrational modes are evaluated
      from NASA9 polynomial fits~\citep{mcbride2002nasa}.  \\
    \item Chemical and translational\hyp{}rotational\hyp{}vibrational
      equilibrium (Chemical-RTV).  Chemical equilibrium is computed
      with Cantera \citep{cantera}. The chemistry mechanisms used can
      be found in \citet{shepherd_cti_mech}, file
      \textit{airNASA9noions.yaml}. Translational\hyp{}rotational\hyp{}vibrational
      equilibrium data are obtained from NASA9 least-squares fits
      \citep{mcbride2002nasa}. \\
    \item Chemical and
      translational\hyp{}rotational\hyp{}vibrational\hyp{}electronic
      equilibrium (Chemical-RTVE). Chemical equilibrium is also
      computed with Cantera \citep{cantera}. The chemistry mechanisms
      used can be found in \citet{shepherd_cti_mech}, file
      \textit{airNASA9ions.yaml}. Translational\hyp{}rotational\hyp{}vibrational\hyp{}electronic
      equilibrium data are obtained from NASA9 least-squares fits
      \citep{mcbride2002nasa}.
\end{enumerate}

\begin{figure*}[]
\centering
\begin{subfigure}[t]{0.49\textwidth}
\includegraphics[width=\textwidth]{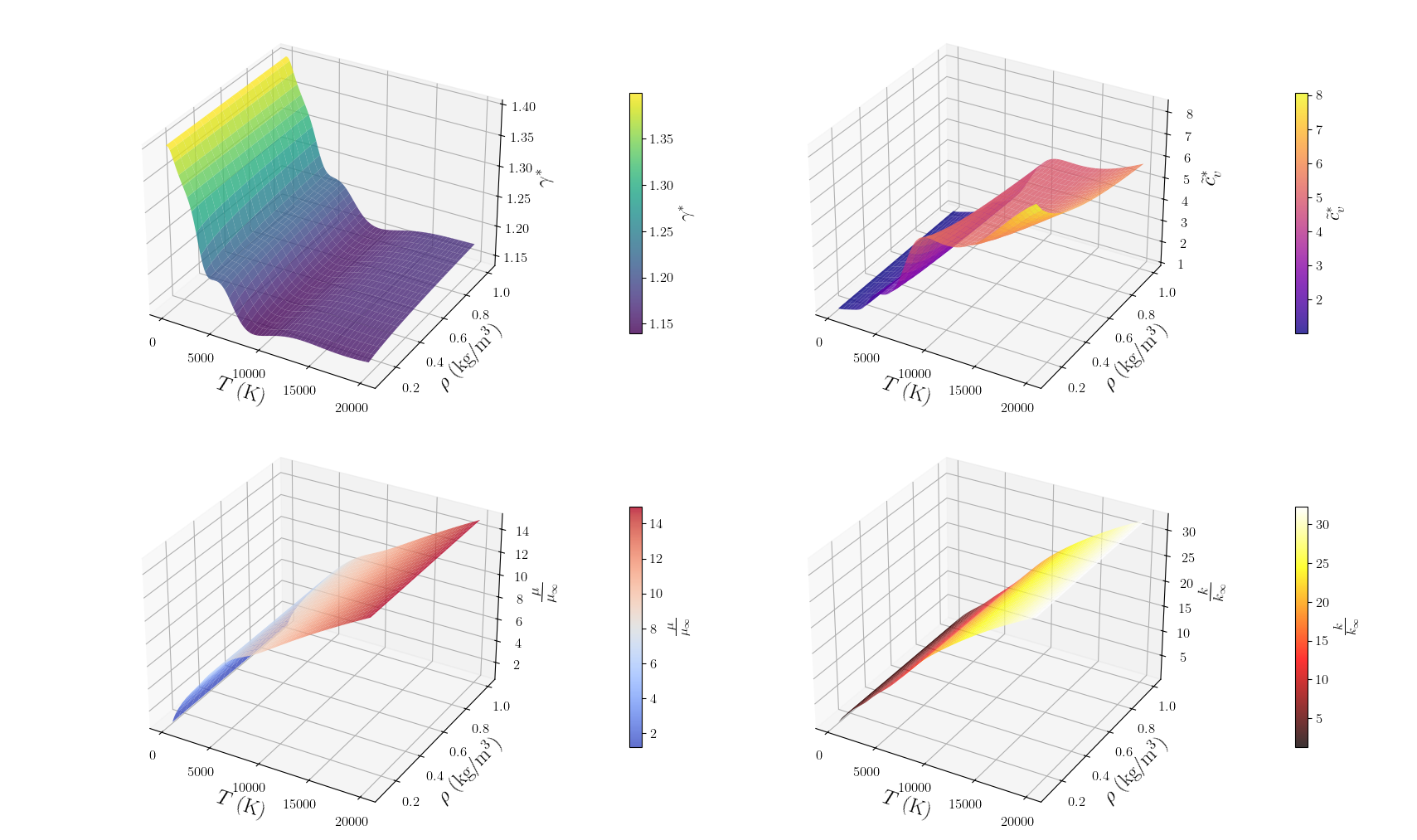}
\caption{}
\label{fig:A02a}
\end{subfigure}
\hfill
\begin{subfigure}[t]{0.49\textwidth}
\includegraphics[width=\textwidth]{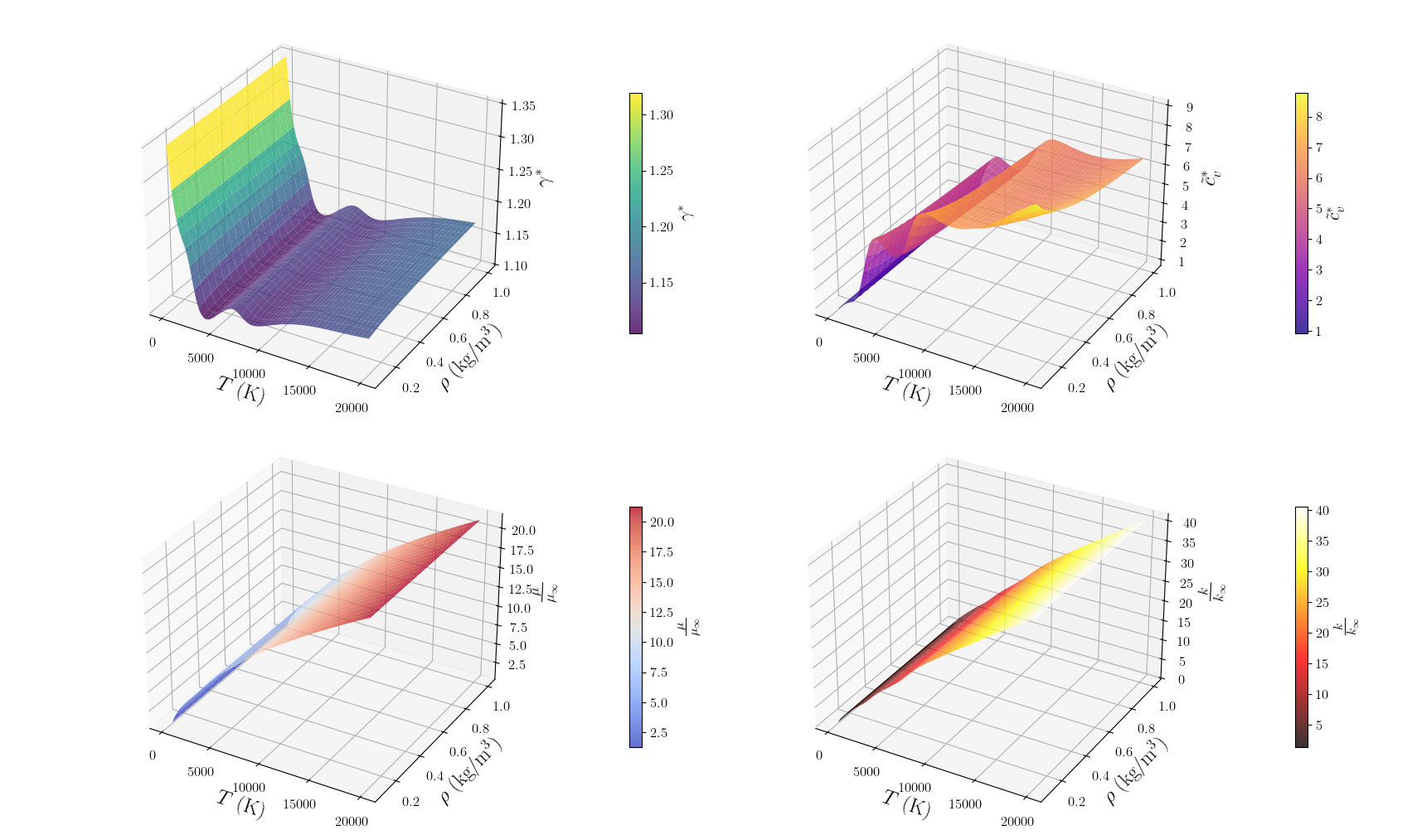}
\caption{}
\label{fig:A02b}
\end{subfigure}
\caption{Physical model:
  chemical-translational\hyp{}rotational\hyp{}vibrational\hyp{}electronic
  equilibrium. Fits to $\gamma^*$, $c_\mathrm{v}^*$, $\mu/\mu_\infty$ and
  $k/k_\infty$.
  Freestream conditions are set to $T=300\,\mathrm{K}$ and
  $\rho=1.225\,\mathrm{kg\,m^{-3}}$ for non-dimensionalization. The
  fits are constructed as functions of $\rho$ and $e$. $T$ is shown in
  the plots for easier interpretation.  (\textit{a}) Earth atmosphere:
  $X_{N_2}:0.7812$, $X_{O_2}:0.2095$, $X_{Ar}:0.0093$. (\textit{b})
  Mars atmosphere: $X_{CO_2}:0.9556$, $X_{N_2}:0.0270$,
  $X_{Ar}:0.0160$, $X_{O_2}:0.0014$.}
\label{fig:A02}
\end{figure*}

The thermochemical properties $\gamma^*(e,\rho)$ and
$c_\mathrm{v}^*(e,\rho)$ are computed differently depending on the selected
physical model. They are parametrized over a range of values of $e$
and $\rho$ expected to arise in the Navier--Stokes solutions of the
problems of interest. A radial-basis-function least-squares fit is
then constructed for each variable; see
figure~\ref{fig:A02}. These fits are subsequently used within the
solver. The reason for using fitted models instead of solving the
equilibrium equations in every finite-volume cell at every time step
is computational cost. The fitting approach accelerates the code by a
factor of approximately 1000; see table~\ref{tab:Cantera_speedup}.
\begin{table}
\caption{Average time required to compute one equilibrium state with
  the Chemical-RTVE model. Mars atmosphere: $X_{CO_2}:0.9556$,
  $X_{N_2}:0.0270$, $X_{Ar}:0.0160$, $X_{O_2}:0.0014$.}
\label{tab:Cantera_speedup}
{\setlength{\aboverulesep}{0pt}
\setlength{\belowrulesep}{0pt}
\begin{tabular*}{\tblwidth}{@{} L|LL@{} }
 & Cantera & Present Fits \\
\midrule
Time per solution ($\mu$s) & 2665 & 0.086 \\
\bottomrule
\end{tabular*}}
\end{table}

\subsubsection{Chemical non-equilibrium model}
\label{sec:no-eq_models}

\begin{figure}[]
\centering
\begin{subfigure}[t]{0.9\columnwidth}
\includegraphics[width=\columnwidth]{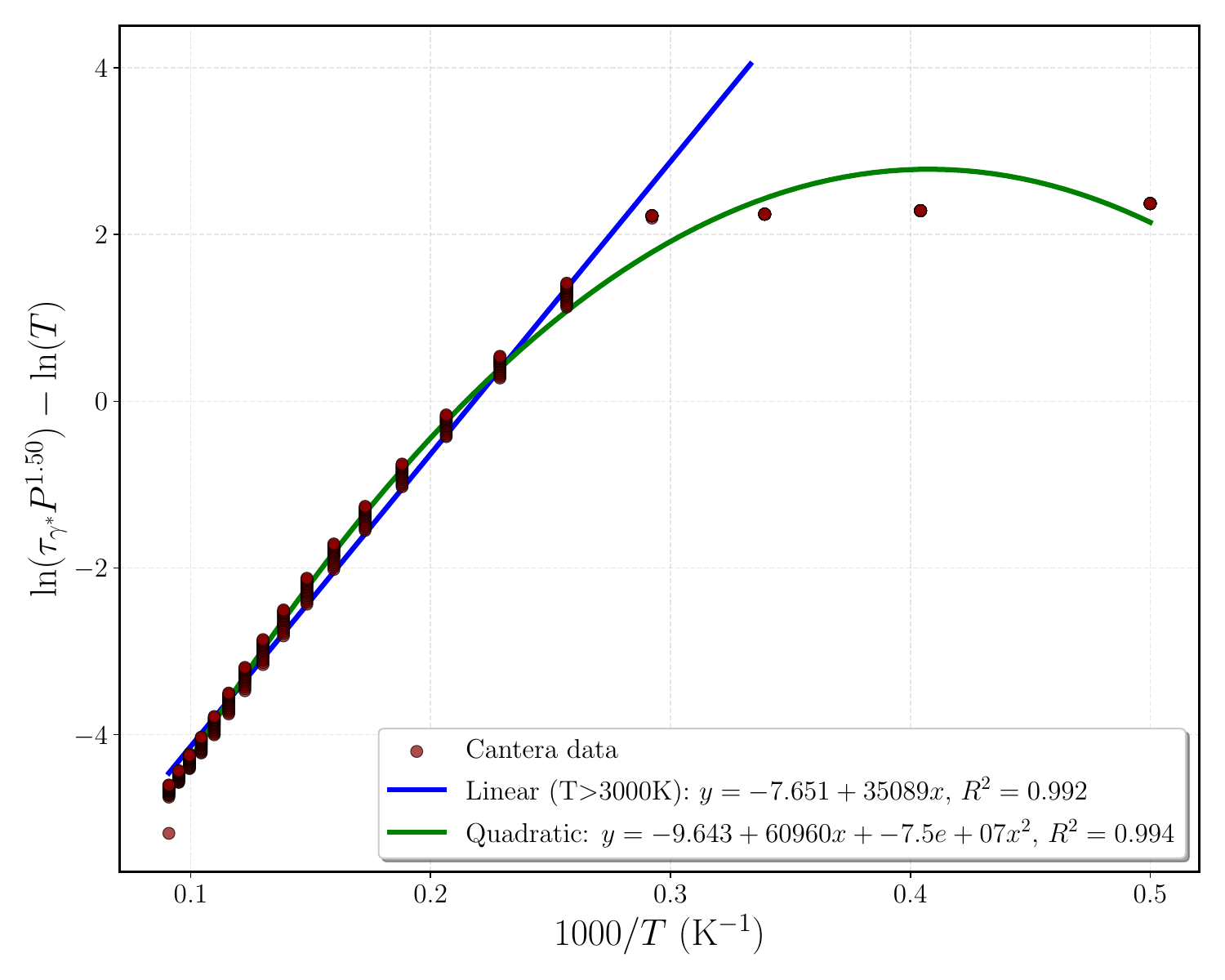}
\caption{}
\label{fig:arrhenius_fits_1}
\end{subfigure}
\begin{subfigure}[t]{0.9\columnwidth}
\includegraphics[width=\columnwidth]{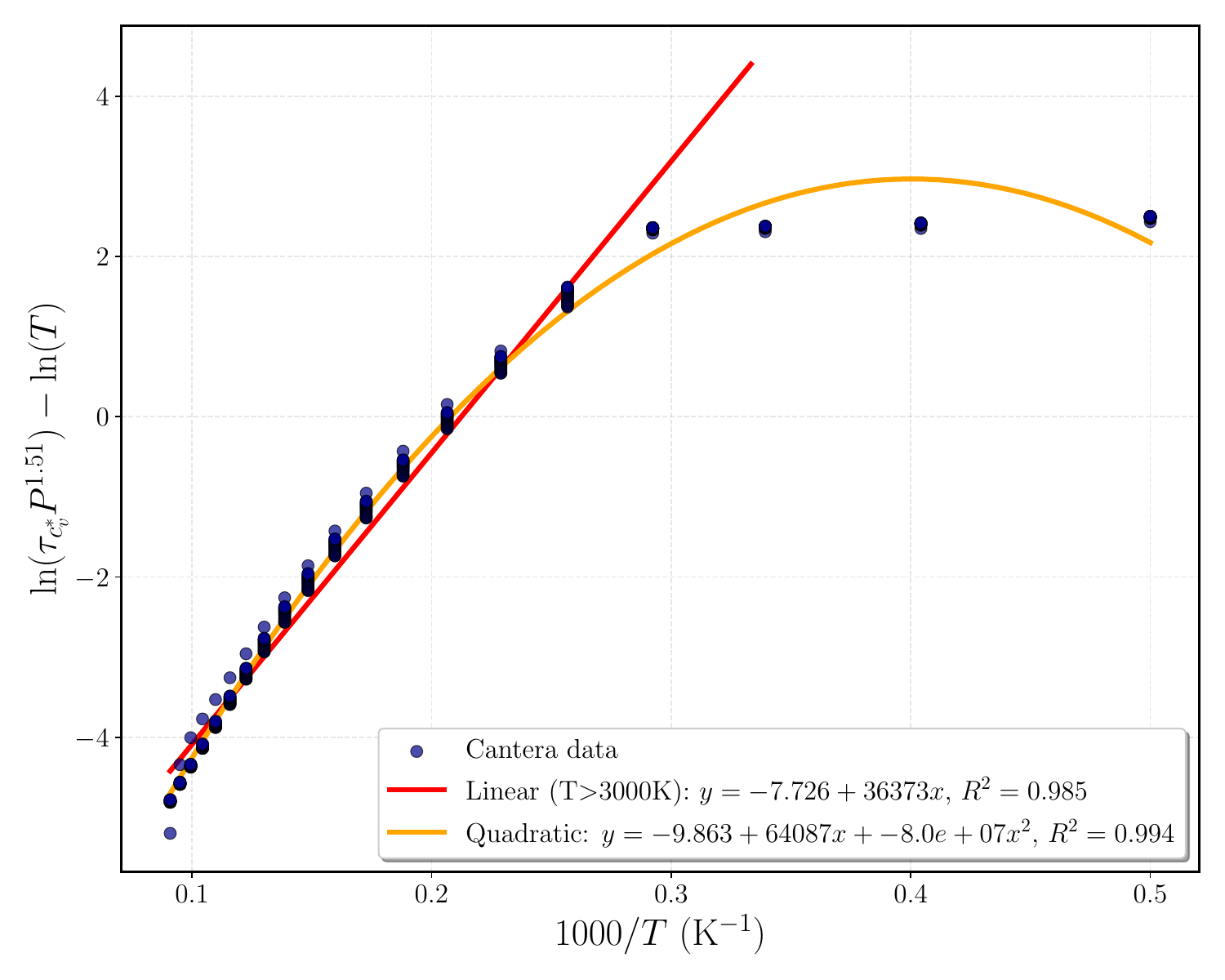}
\caption{}
\label{fig:arrhenius_fits_2}
\end{subfigure}
\caption{Arrhenius fits for the relaxation times of $\gamma^*$ and
  $c_\mathrm{v}^*$ for Martian atmospheric composition. The linear fit
  is performed only for $T > 3{,}000\,\mathrm{K}$, or equivalently
  $(1{,}000\,\mathrm{K})/T < 0.33$, corresponding to the temperature
  range in which chemical reactions become relevant on the time scales
  of interest. (\textit{a}) Relaxation time of $\gamma^*$,
  $\tau^{\gamma^*}$. (\textit{b}) Relaxation time of $c_\mathrm{v}^*$,
  $\tau^{c_\mathrm{v}^*}$.}
\label{fig:arrhenius_fits}
\end{figure}

We construct a reduced-order model for chemical non-equilibrium under
the assumption of
translational\hyp{}rotational\hyp{}vibrational\hyp{}electronic
equilibrium (NonEq-RTVE). This model is essential to make stability
calculations computationally efficient. Following an approach
analogous to the Landau--Teller non-equilibrium
model~\citep{landau1936schalldispersion}, the total derivatives of the
effective thermodynamic quantities used to represent composition
effects are assumed to relax toward equilibrium, denoted by the
subscript $\mathrm{eq}$, over characteristic times $\tau^{\gamma^*}$
and $\tau^{c_\mathrm{v}^*}$:
\begin{equation}
    \frac{D \gamma^*}{D t} = \dfrac{\partial \gamma^*}{\partial t} +
    u_i \dfrac{\partial \gamma^* }{\partial x_i} =
    \frac{\gamma^*_{\mathrm{eq}} - \gamma^*}{\tau^{\gamma^*}}
    \label{eq::non_eq_gamma}
\end{equation}
\begin{equation}
    \frac{D c_\mathrm{v}^*}{D t} =
    \dfrac{\partial c_\mathrm{v}^*}{\partial t}
    + u_i \dfrac{\partial c_\mathrm{v}^* }{\partial x_i} =
    \frac{{c_\mathrm{v}^*}_{\mathrm{eq}} - c_\mathrm{v}^*}
    {\tau^{c_\mathrm{v}^*}}.
    \label{eq::non_eq_cv}
\end{equation}
The characteristic relaxation time $\tau$ is motivated using
kinetic-theory scaling arguments.  The characteristic time associated
with a molecular collision is
\begin{equation*}
    \tau_c  \sim  \frac{\lambda}{v_r},
\end{equation*}
where $v_r$ is the mean relative intermolecular velocity and $\lambda$
is the mean free path. From kinetic theory
\citep{hill1986introduction}, and assuming an approximately constant
effective collision cross section,
\begin{equation*}
    \lambda \sim \frac{T}{p}, \quad \ v_r \sim \sqrt{T}.
\end{equation*}
In addition, only a fraction of molecular collisions can lead to
reaction. As a first approximation, reactive events are associated
with molecules whose energy exceeds the activation threshold $E_a$.
The fraction $P_r$ of collisions energetic enough to overcome the
activation threshold is assumed to follow an Arrhenius-type scaling
based on the Maxwell--Boltzmann distribution
\citep{hill1986introduction}, with dominant exponential dependence
\begin{equation*}
    P_r
    \sim
    \frac{1}{\sqrt{T}}
    \exp\left(-\frac{E_a}{R_g T}\right),
\end{equation*}
where $R_g$ is the gas constant. Combining the previous scaling
relations yields the following estimate for the characteristic
reaction time:
\begin{equation}
    \tau
    \sim
    \frac{\tau_c}{P_r}
    \sim
    \frac{T}{p}
    \exp\left(\frac{E_a}{R_g T}\right)
    \quad\Rightarrow\quad
    \tau
    =
    C_1\frac{T}{p}
    \exp\left(\frac{C_2}{T}\right),
\end{equation}
where $C_1$ and $C_2$ are constants to be determined.  In practice,
three-body reactions can introduce an additional pressure dependence
in the effective relaxation time~\citep{croiset1996influence}. To
account for this effect, we adopt the more general form
\begin{equation}
    \tau = C_1\frac{T}{p^m} \exp\left(\frac{C_2}{T}\right),
\end{equation}
where $m$ is also a model parameter.  Several initial compositions
$X_i$ are then relaxed to equilibrium by solving the full species
equations and associated reaction mechanisms with Cantera.  The
relaxation process is modeled as an adiabatic, constant-pressure,
constant-enthalpy reaction. This procedure is repeated for different
values of the initial non-equilibrium state, characterized by $T$ and
$p$. In particular, $\tau$ is estimated from the settling time
$t_{95}$ required to reach within $5\%$ of the equilibrium state, with
different decay times for $c_\mathrm{v}^*$ and $\gamma^*$. This
quantity is computed in Cantera for a set of initial conditions:
\begin{align}
0.05
&=
\frac{\gamma^*(t_{95}^{\gamma^*})-\gamma^*_{\mathrm{eq}}}
{\gamma^*(0)-\gamma^*_{\mathrm{eq}}}
=
\exp\left(
-\frac{t_{95}^{\gamma^*}}{\tau^{\gamma^*}}
\right),
\nonumber \\
0.05
&=
\frac{
c_\mathrm{v}^*(t_{95}^{c_\mathrm{v}^*})
-
{c_\mathrm{v}^*}_{\mathrm{eq}}
}
{
c_\mathrm{v}^*(0)
-
{c_\mathrm{v}^*}_{\mathrm{eq}}
}
=
\exp\left(
-\frac{t_{95}^{c_\mathrm{v}^*}}{\tau^{c_\mathrm{v}^*}}
\right),
\nonumber \\
\tau^{\gamma^*}(p,T,X_i)
&=
-\frac{t_{95}^{\gamma^*}}{\ln(0.05)},
\nonumber \\
\tau^{c_\mathrm{v}^*}(p,T,X_i)
&=
-\frac{t_{95}^{c_\mathrm{v}^*}}{\ln(0.05)} .
\end{align}
Once these relaxation times have been computed, $C_1$ and $C_2$ are
fitted separately for $c_\mathrm{v}^*$ and $\gamma^*$ using an Arrhenius plot:
\begin{equation}
\ln(\tau p^m) - \ln(T) = \ln(C_1) + \frac{C_2}{T}  \Rightarrow y = C_1^* + C_2 x,
\end{equation}
where $y = \ln(\tau p^m) - \ln(T)$, $x = 1/T$, and $C_1^* =
\ln(C_1)$. The pressure coefficient $m$ is adjusted to maximize the
regression coefficient $R^2$. The results for Martian composition are
shown in figure~\ref{fig:arrhenius_fits}. For high temperatures,
$T>3{,}000\,\mathrm{K}$, or equivalently $1000/T <
0.33\,\mathrm{K}^{-1}$ with $T$ expressed in kelvin, the data are well
approximated by a linear fit, consistent with the kinetic-theory
scaling. This is the regime in which chemical reactions become
important.  For $T<3{,}000\,\mathrm{K}$, or equivalently
$(1000\,\mathrm{K})/T>0.33$, reaction rates vary only weakly with
temperature and the reaction times become extremely long.  In this
lower-temperature regime, a quadratic fit is used to capture the
departure from the high-temperature kinetic scaling. The resulting
reduced-order model is implemented within the solver as a first-order
approximation of finite-rate chemistry effects through the effective
thermodynamic variables $\gamma^*$ and $c_\mathrm{v}^*$. Verification
of the model against a full thermochemical non-equilibrium solver is
presented in section~\ref{sec::verification_real_gas}.

Note that the current stability analyses are performed for closures in
which $\gamma^*$ and $c_\mathrm{v}^*$ are algebraic functions of the
conservative state. The reduced NonEq-RTVE model is
available for
nonlinear base-flow/time-marching studies but is not included in the
linear stability operator unless the state is explicitly augmented by
$\gamma^*$ and $c_\mathrm{v}^*$.

\section{Details of numerical implementation}
\label{sec::Details_of_numerical_implementation}

\subsection{Time-marching implementation}
\label{sec:time-marching}

The semi-discrete governing equations are written, for the vector of interior
conservative unknowns $\bm{q}=(\rho,\rho u,\rho v,\rho E)^{\mathsf T}$ on the
structured curvilinear grid, as
\begin{equation}
  \frac{\mathrm{d}\bm{q}}{\mathrm{d}t} = \bm{N}(\bm{q}),
  \label{eq:semi-discrete}
\end{equation}
where $\bm{N}$ is the nonlinear spatial operator. Time integration uses either the
explicit RK4 scheme (for time-accurate runs) or one of the two implicit solvers
described below (for steady base flows). The time step can be set to a fixed value
or determined by an adaptive CFL criterion.

\subsection{Implicit steady-state solvers}
\label{sec:implicit}

The solver provides two steady-state solvers: a Picard fixed-point solver and a Newton solver. Both implicit solvers converge to the steady state $\bm{N}(\bm{q})=\bm{0}$, which
is also the $\dt\!\to\!\infty$ fixed point of the backward-Euler discretization of
\eqref{eq:semi-discrete},
\begin{equation}
  \frac{\bm{q}^{n+1}-\bm{q}^{n}}{\dt}
  \;=\; \bm{N}\!\left(\bm{q}^{n+1}\right).
  \label{eq:backward-euler}
\end{equation}
The Picard solver advances the backward-Euler map \eqref{eq:backward-euler} by an
inner (pseudo-time) iteration at a \emph{fixed} step; the Newton solver instead
applies pseudo-transient continuation directly to the steady residual.

\paragraph{Picard fixed-point with adaptive under-relaxation}

The \texttt{"Relaxation"} solver treats \eqref{eq:backward-euler} as a
fixed-point map at a \emph{fixed} CFL and sweeps the inner loop iteration $k$ until the update
residual falls below \texttt{tolerance} or \texttt{max\_iter\_implicit} is
reached:
\begin{equation}
  \bm{q}^{k+1} = \bm{q}^{k}
    + \omega\,\big(\bm{\mathcal{G}}(\bm{q}^{k})-\bm{q}^{k}\big),
\end{equation}
with $\bm{\mathcal{G}}(\bm{q}^{k}) = \bm{q}^{n} +
\dt\,\bm{N}(\bm{q}^{k})$ the backward-Euler update, whose fixed point
$\bm{q}^{k}=\bm{\mathcal{G}}(\bm{q}^{k})$ solves
\eqref{eq:backward-euler}. The under-relaxation factor $\omega$ is
adapted from the residual history through a sigmoid map and clamped to
$\omega\in[0.5,\,0.99]$: $\omega$ is reduced when the residual stalls
or grows and is relaxed back toward its upper bound as the residual
decays.

\paragraph{Newton solver with pseudo-transient continuation}

The default \texttt{"Newton"} solver linearizes the steady residual
$\bm{N}(\bm{q})$ at every iteration and takes one damped Newton step using the
sparse Jacobian $\bm{L}=\partial\bm{N}/\partial\bm{q}$
(see appendix~\ref{sec::linearization_details}), solving for the increment
$\delta\bm{q}$
\begin{equation}
  \left(\frac{\bm{I}}{\dtau}-\bm{L}\right)\delta\bm{q}
  \;=\; \bm{N}(\bm{q}^{k}),
  \qquad
  \bm{q}^{k+1} = \bm{q}^{k} + \alpha\,\delta\bm{q},
  \label{eq:newton-step}
\end{equation}
by a direct sparse LU factorization (iterative methods can also be
used for larger grids). The $\bm{I}/\dtau$ term is a pseudo-transient
regularization \citep{kelley1998convergence}: for small $\dtau$ the
step is a stable under-relaxed march, and as $\dtau\!\to\!\infty$ it
recovers the full Newton step on $\bm{N}(\bm{q})=\bm{0}$.  Because the
Jacobian provides a consistent finite-difference linearization of the
discrete residual, the method can converge to steady solutions even
when those solutions are linearly unstable, provided that the
nonlinear iteration remains in the basin of attraction. This is
essential for subsequent stability analysis; an explicit time
integrator would not converge to such unstable steady states.

To amortize cost, $\bm{L}$ may be frozen for
\texttt{jacobian\_refresh} iterations (modified Newton).  For the
Newton solver, the pseudo-time step is grown automatically as the
solution converges, following the Switched Evolution Relaxation (SER)
rule of \citet{mulder1985experiments}.  Writing
$R_k=\lVert\bm{N}(\bm{q}^{k})\rVert$ for the current global residual
norm and $R_0$ for the reference norm recorded on the first step,
\begin{equation}
  \mathrm{CFL}^{k} \;=\;
  \mathrm{CFL}_0\,\left(\frac{R_0}{R_k}\right)^{p},
  \qquad
  \dtau^{k} = \dt\,\frac{\mathrm{CFL}^{k}}{\mathrm{CFL}_0},
  \label{eq:ser}
\end{equation}
where $\dt$ is the explicit-stable step (fixed time step or given by
CFL estimate) and $p$ is the ramp exponent
(\texttt{CFL\_ramp\_exponent}, default $1$). The CFL therefore starts
at the user value $\mathrm{CFL}_0$ (\texttt{time\_integration.CFL})
when the residual is large and increases without bound as
$R_k\!\to\!0$, so the Newton iteration takes increasingly large steps
near steady state. Note that the Picard solver cannot perform this CFL
ramp-up, as its inner iteration loop is still limited by the explicit
CFL stability criterion.

\subsection{Rankine--Hugoniot jump conditions}
\label{sec::shock_jump_both}

The Rankine--Hugoniot relations from Eq.~(\ref{eq:RankineHugoniot})
can be solved in different ways depending on which variables are
prescribed. In practice, two cases arise. In the first case, all
upstream flow conditions, including the shock propagation speed, are
known, and the downstream state must be determined.  This procedure is
used to initialize the flow field across the shock. In the second
case, the shock speed is unknown, but a characteristic quantity of the
post-shock flow, namely a Riemann invariant, is prescribed.  This
second approach is used at every time step to determine the updated
post-shock state and shock speed. Each case requires a different
numerical procedure, which is described in the following
subsections. Both procedures incorporate the thermochemical model
defined in section~\ref{sec::Thermochemical and transport models}, so
that real-gas effects are incorporated through the adopted
thermochemical closure across the shock.

\subsubsection{Shock initialization}
\label{sec::all_known}

In this case, the freestream conditions and the shock speed are
assumed to be known: $(U_1, e_1, \rho_1, \gamma_1^*(\rho_1,e_1),
u_s)$.  During the initialization step, the shock speed is usually set
to zero. Under these assumptions, Eq.~(\ref{eq:RankineHugoniot})
reduces to a system of two nonlinear equations:
\begin{align}
(\gamma_1^* - 1)\frac{e_1}{(U_1+u_s)^2}+1
&=
\left(\gamma_2^*(r_\rho,\theta)-1\right) r_\rho \theta
+\frac{1}{r_\rho},
\nonumber \\
\gamma_1^*\frac{e_1}{(U_1+u_s)^2}+\frac{1}{2}
&=
\gamma_2^*(r_\rho,\theta)\theta
+\frac{1}{2r_\rho^2},
\end{align}
where
\[
r_\rho=\frac{\rho_2}{\rho_1},
\qquad
\theta=\frac{e_2}{(U_1+u_s)^2},
\]
\[
\gamma_2^*(r_\rho,\theta)
=
\gamma^*\!\left((U_1+u_s)^2\theta,\rho_1 r_\rho\right).
\]
The system is solved iteratively by reducing it to a scalar nonlinear
equation and applying a bisection algorithm. In the present
implementation, the nonlinear solver only requires evaluating fitted
functions (see appendix~\ref{sec:eq_models}). This results in a
substantial speed-up, as shown in
table~\ref{tab:SD_Toolbox_speedup}. Since the Rankine--Hugoniot
relations must be solved many times during each iteration of the
shock-fitting algorithm, this acceleration reduces the overall
computational cost significantly.
\begin{table}
\caption{Average time required to compute one Rankine--Hugoniot
  solution with the Chemical-RTVE equilibrium model. Earth
  atmosphere: $X_{N_2}:0.7812$, $X_{O_2}:0.2095$, $X_{Ar}:0.0093$.}
\label{tab:SD_Toolbox_speedup}
{\setlength{\aboverulesep}{0pt}
\setlength{\belowrulesep}{0pt}
\begin{tabular*}{\tblwidth}{@{} L|LL@{} }
 & SD Toolbox & Present Solver \\
\midrule
Time per solution (ms) & 205 & 3 \\
\bottomrule
\end{tabular*}}
\end{table}

\subsubsection{Time marching of the shock}
\label{sec::shock_solution}

In the shock-fitting solver, the shock can move in response to the
jump conditions together with compatibility information from the
downstream flow field.  At a given time step, the upstream variables
entering Eq.~(\ref{eq:RankineHugoniot}) are assumed to be known:
$U_1$, $\rho_1$, $p_1$, and $e_1$. In addition, the post-shock
solution from the previous time step is available: $U_{2,0}$,
$\rho_{2,0}$, $p_{2,0}$, and $e_{2,0}$. The goal is then to determine
the updated post-shock state $(U_2, \rho_2, p_2, e_2)$ together with
the shock speed $u_s$.

Downstream of the shock, the Euler characteristic analysis implies
that the linearized Riemann invariant $J_2^-=\Delta
p_2-\rho_{2,0}a_{2,0}\Delta U_2$ should be preserved, where $\Delta
p_2=p_2-p_{2,0}$ and $\Delta U_2=U_2-U_{2,0}$. Here $a_{2,0}$ is the
characteristic sound speed used by the selected thermodynamic closure.
First, an initial guess is introduced for the pressure ratio,
\[
y = \frac{p_2}{p_1},
\]
obtained by interpolation from the flow field. By combining the
conservation equations, the density ratio can then be expressed as a
function of $y$:
\begin{equation}
\frac{\rho_2}{\rho_1} =
\frac{y\left(\frac{\gamma^*_2+1}{\gamma^*_2-1}\right) + 1}{y +
  \left(\frac{\gamma^*_1+1}{\gamma^*_1-1}\right)} \label{eq:density_ratio_final}
\end{equation}
The shock speed follows from the mass and momentum conservation
equations:
\begin{equation}
u_s = - U_1 \pm \sqrt{ \frac{p_1(y - 1)}{\rho_1} \left(
  \frac{y\left(\frac{\gamma^*_2+1}{\gamma^*_2-1}\right) +
    1}{\frac{2y}{\gamma^*_2-1} - \frac{2}{\gamma^*_1-1}} \right)
} \label{sol:us}
\end{equation}
The appropriate sign is selected according to the propagation
direction and the physically admissible compressive branch.

Once $u_s$ is known, the downstream velocity is obtained from mass
conservation:
\begin{equation}
U_2 = - u_s + \frac{\rho_1}{\rho_2} (U_1 + u_s)
\label{sol:U_2}
\end{equation}
This procedure defines a nonlinear relation between downstream
pressure and velocity, that is, $U_2 = f(p_2)$. In general, the value
of $U_2$ obtained from the initial guess for $p_2$ does not satisfy
the Riemann-invariant constraint
$J_2^- = \Delta p_2 - \rho_{2,0} a_{2,0} \Delta U_2$.
For this reason, a Newton--Raphson algorithm is used to converge to
the consistent solution:
\begin{algorithmic}[1]
\STATE Initialize $p_2^{(0)}$ from interpolation of the flow field
\WHILE{not converged}
    \STATE Compute $\rho_2$ from equation~(\ref{eq:density_ratio_final})
    \STATE Compute $u_s$ from equation~(\ref{sol:us})
    \STATE Compute $U_2 = f(p_2)$ from equation~(\ref{sol:U_2})
    \STATE Evaluate residual: $\mathcal{R} = \Delta p_2 - \rho_{2,0} a_{2,0} \Delta U_2$
    \STATE Compute derivative: $\dfrac{\partial \mathcal{R}}{\partial p_2} = \dfrac{\mathcal{R}(p_2 + \varepsilon) - \mathcal{R}(p_2)}{\varepsilon}$
    \STATE Update: $p_2^{(n+1)} = p_2^{(n)} - \dfrac{\mathcal{R}}{\partial \mathcal{R}/\partial p_2}$
    \STATE Check convergence: $|\mathcal{R}| < \text{tol}$
\ENDWHILE
\RETURN $p_2$, $\rho_2$, $U_2$, $u_s$
\end{algorithmic}
In practice, the initial guess is usually very close to the exact
solution because the flow field changes only slightly between
consecutive time steps. As a result, the algorithm typically
converges to machine precision within 3--5 iterations.

\subsection{Linearization of the non-linear system}
\label{sec::linearization_details}

This appendix details how the linearized operator $\bm{L} = \delta
\bm{N}/\delta\bm{q}|_{\bm{q}_0}$ introduced in
equation~\eqref{eq:linearized_system} is constructed and verified,
covering the finite-difference construction of the operator and the
role of the perturbation parameter.

\subsubsection{Finite-difference construction of \texorpdfstring{$\bm{L}$}{L}}
\label{sec::linearization_appendix_construction}

The operator $\bm{L}$ is built column by column using a first-order
forward finite difference of the discrete nonlinear residual
$\bm{N}$. Denoting by $\bm{e}_j$ the $j$-th canonical unit vector in
state space and by $\varepsilon$ the perturbation amplitude, each
column of the Jacobian is approximated as
\begin{equation}
  \bm{L}\,\bm{e}_j
  \;\approx\;
  \frac{\bm{N}(\bm{q}_{e,0} + \varepsilon\,\bm{e}_j) - \bm{N}(\bm{q}_{e,0})}
       {\varepsilon}.
  \label{eq:fd_column}
\end{equation}
Crucially, $\bm{N}$ in equation~\eqref{eq:fd_column} is the very same
discrete spatial operator that is used to march the base flow in time,
including the boundary-condition application and, in the shock-fitting
formulation, the shock-jump and metric updates.  The linearization is
therefore designed to be discretely consistent: up to
finite-difference, nonlinear-solver, and roundoff tolerances, the
reported eigenvalues, transient-growth gains, and receptivity
responses correspond to the linear dynamics of the discrete system
being integrated.

A naive evaluation of equation~\eqref{eq:fd_column} would require one
residual evaluation per column, i.e. $O(N^2)$ evaluations for an
$N\times N$ grid, which is prohibitive. Instead, the assembly exploits
the compact stencil of the finite-volume discretization: the residual
in any cell depends only on a $3\times 3$ block of neighbouring cells
(a nine-point domain of dependence). Two cells whose perturbations do
not share a common downstream residual cell can therefore be perturbed
simultaneously without polluting each other's columns. The grid is
partitioned into nine interleaved colours (every third cell in each
index direction), so that a single residual evaluation per colour
recovers an entire set of non-overlapping Jacobian columns. As a
result, the complete flow--flow block requires only nine residual
evaluations per conservative variable, i.e. $36$ evaluations in total,
\emph{independently of the grid size}.

In the shock-fitting formulation the state is augmented by the shock
displacements $\bm{\eta}'$, and $\bm{L}$ acquires the block structure
\begin{equation*}
  \bm{L} =
  \begin{bmatrix}
    \bm{L}_{qq} & \bm{L}_{q\eta} \\[2pt]
    \bm{L}_{\eta q} & \bm{L}_{\eta \eta}
  \end{bmatrix},
\end{equation*}
where $\bm{L}_{qq}$ is the flow--flow block described above,
$\bm{L}_{q\eta}$ and $\bm{L}_{\eta q}$ couple flow and shock-front motion, and
$\bm{L}_{\eta \eta}$ governs the self-interaction of the shock
displacement. The coupling blocks are obtained with the same
finite-difference procedure, perturbing either the post-shock cells
adjacent to the front or the shock-point coordinates in the local
shock-normal direction; this adds $O(N_\chi)$ residual evaluations,
which remains negligible compared with the cost of the eigenvalue
iteration. The verification of the linearized shock--disturbance
coupling is reported separately in
section~\ref{sec::Verification_Disturbances_Shock}.

\subsubsection{Perturbation parameter and verification}
\label{sec::linearization_appendix_epsilon}

The finite-difference construction in equation~\eqref{eq:fd_column}
introduces a single tunable parameter, the perturbation amplitude
$\varepsilon$, exposed to the user through the input field
\texttt{perturbation\_magnitude} and set to $\varepsilon = 10^{-8}$ by
default in all cases reported here.

The choice $\varepsilon = 10^{-8}$ follows the standard balance between
the two competing error sources of a forward difference. The
truncation error of equation~\eqref{eq:fd_column} grows linearly with
$\varepsilon$, whereas the round-off error grows as
$\varepsilon_{\mathrm{mach}}/\varepsilon$, with $\varepsilon_{\mathrm{mach}}
\approx 10^{-16}$ in double precision:
\begin{equation}
  \big\| \bm{L}^{\text{FD}} - \bm{L} \big\|
  \;\sim\;
  \underbrace{\tfrac{1}{2}\,\varepsilon\,\big\|\partial^2 \bm{N}\big\|}_{\text{truncation}}
  \;+\;
  \underbrace{\frac{2\,\varepsilon_{\mathrm{mach}}}{\varepsilon}\,\big\|\bm{N}\big\|}_{\text{round-off}} .
\end{equation}
For state and residual quantities of order unity this estimate is
minimised at $\varepsilon^\star \sim \sqrt{\varepsilon_{\mathrm{mach}}}
\approx 10^{-8}$, so that the default lies close to the
optimum and within the flat region of the error curve, which spans
roughly an order of magnitude in each direction.

Rather than relying on this estimate alone, the code verifies the
assembled operator a~posteriori every time it is constructed. A random
perturbation $\bm{q}'$ is applied to all conservative variables (and,
when active, to the shock displacements) and the linear operator action is compared against a fresh evaluation
of the nonlinear residual,
\begin{equation}
  \mathcal{E}
  \;=\;
  \frac{1}{\sqrt{n}}
  \left\|
  \frac{
  \bm{L}\,\bm{q}'_e
  -
  \left[
  \bm{N}(\bm{q}_{e,0}+\bm{q}'_e)
  -
  \bm{N}(\bm{q}_{e,0})
  \right]
  }
  { \big|\bm{L}\,\bm{q}'_e\big|+\delta }
  \right\|_2 .
\end{equation} 
where $n$ is the state dimension, the division is taken element-wise,
and $\delta = 10^{-14}$ guards against division by zero in
near-quiescent cells. A small value of $\mathcal{E}$ (in practice
$\mathcal{E} < 10^{-3}$) provides a consistency check that $\bm{L}$
reproduces the action of the nonlinear operator in an arbitrary
direction. Because this check uses a random direction and the full
nonlinear residual, it detects both an ill-chosen $\varepsilon$ and
any inconsistency in the assembly, and it does so for the exact flow,
grid, and thermochemical model at hand. In all cases reported in this
paper the verification error remained around $10^{-8}$--$10^{-6}$, and
did not affect numerical results.

\subsection{Global transient growth}
\label{sec::transient_growth_appendix}

This section provides a more detailed description of the transient
growth analysis for post-shock disturbances $q'$. To evaluate Chu's
energy norm, the perturbation variables are first transformed into the
variables in which the norm is naturally defined, namely $\tilde{q}' =
\left[p', u', v', \frac{s'}{R_{g,0}} \right]$:
\begin{equation*}
    \tilde{q}' = Q_V q',
\end{equation*}
\begin{equation*}
\setlength{\arraycolsep}{2.5pt}
    Q_V = 
\begin{bmatrix}
(\gamma^*_0 - 1)\frac{u_0^2 + v_0^2}{2} & -u_0(\gamma_0^* - 1) & -v_0(\gamma_0^* - 1) & \gamma_0^* - 1  \\
-\frac{u_0}{\rho_0} & \frac{1}{\rho_0} & 0 & 0  \\ 
-\frac{v_0}{\rho_0} & 0 & \frac{1}{\rho_0} & 0  \\
\frac{u_0^2 + v_0^2}{2p_0} - \frac{\gamma^*_0}{\gamma_0^* - 1} \frac{1}{\rho_0} & -\frac{u_0}{p_0} & -\frac{v_0}{p_0} & \frac{1}{p_0}
\end{bmatrix}.
\end{equation*}

Once the transformation matrix $Q_V$ has been defined, Chu's energy
norm can be written in matrix form as
\begin{equation*}
W_V = \begin{bmatrix}
\frac{\rho_0 a_0^2}{2 (\gamma_0^* p_0)^2} & 0 & 0 & 0  \\
0 & \frac{\rho_0}{2} & 0 & 0  \\
0 & 0 & \frac{\rho_0}{2} & 0  \\
0 & 0 & 0 & \frac{(\gamma_0^* - 1)p_0}{2 \gamma_0^*}
\end{bmatrix}
\end{equation*}
\begin{equation}
E = \int \left[ q'^T Q_V^T W_V Q_V \ q'\right] dV.
\end{equation}

This integral is then discretized by weighting the energy norm in each
cell, denoted by subscripts $_{ij}$, with the corresponding cell
volume $V_{ij}$:
\begin{equation*}
{\bm{M}}_{ij} = {W_V}_{ij} V_{ij}, \ \bm{Q}_{ij} = {Q_V}_{ij}
, \ E =  {\bm{q}'}^T \bm{Q}^T \bm{M} \bm{Q} \ \bm{q}'.
\end{equation*}

The linearization employs the same discretized operator used to
compute the base flow. Its action on an arbitrary perturbation
direction $\bm{v}$ is approximated numerically by finite differences
as
\begin{equation}
    \bm{L}\bm{v} \approx
    \frac{
    \bm{N}(\bm{q}_{e,0}+\varepsilon \bm{v})
    -
    \bm{N}(\bm{q}_{e,0})
    }{\varepsilon},
    \label{eq:linearization}
\end{equation}
This linearization allows the transient-growth problem to be written
as a generalized Rayleigh quotient:
\begin{equation*}
   \dfrac{\mathrm{d} \bm{q}_e'}{\mathrm{d} t} = \bm{L}\bm{q}_e', \ \bm{q}' = \bm{P} \bm{q}_e' \Rightarrow \bm{q}'(t) = \bm{P} \exp(\bm{L}t)\bm{q}_e'(0),
\end{equation*}
\begin{equation} \label{eq:GD}
\begin{split}
    G_D(t) &= \frac{\bm{q}'(t)^T \bm{Q}^T \bm{M} \bm{Q} \ \bm{q}'(t)}{\bm{q}'(0)^T \bm{Q}^T \bm{M} \bm{Q} \ \bm{q}'(0)} \\
    &= \frac{\bm{q}'_e(0)^T \exp(\bm{L}t)^T \bm{P}^T \bm{Q}^T \bm{M} \bm{Q} \bm{P} \ \exp(\bm{L}t) \ \bm{q}'_e(0)}{\bm{q}'_e(0)^T \bm{P}^T \bm{Q}^T \bm{M} \bm{Q} \bm{P} \ \bm{q}'_e(0)} \\
    &= \frac{\bm{q}'_e(0)^T \bm{C}(t) \ \bm{q}'_e(0)}{\bm{q}'_e(0)^T \bm{D} \ \bm{q}'_e(0)},
\end{split}
\end{equation}
where $\bm{C}(t) =  \exp(\bm{L}t)^T \bm{P}^T \bm{Q}^T \bm{M} \bm{Q}
\bm{P} \exp(\bm{L}t)$ and $\bm{D} = \bm{P}^T \bm{Q}^T \bm{M} \bm{Q} \bm{P}$. For a given
time $t$, $\bm{C}(t)$ can be evaluated, and the associated
generalized eigenvalue problem can then be solved to obtain the
optimal amplification factor $\lambda(t)$ and the corresponding
initial perturbation $\bm{q}_e'(0)$ that maximizes the growth at that
time.

By construction, both $\bm{C}(t)$ and $\bm{D}$ are symmetric. This
makes it possible to use the Lanczos algorithm to solve the Rayleigh
quotient problem for a given time $t$ \citep{li2015rayleigh}.  For
symmetric problems, the Lanczos algorithm is preferred over the
Arnoldi iteration because it exploits symmetry and is typically more
efficient in both memory usage and computational cost.  In particular,
the ARPACK \citep{lehoucq1998arpack} implementation of the Implicitly
Restarted Lanczos Method is employed.

The matrix exponential is approximated in a manner similar to that
used in section~\ref{sec::lin_stab_analysis_implementation}. In the
present case, an additional outer loop is introduced to advance the
time integration:
\begin{equation}
    \exp(\bm{L} t)
    \approx
    \prod_{m=0}^{N_t-1}
    \left[
    \sum_{k=0}^{M}
    \frac{(\bm{L}\Delta t_{\mathrm{CFL}})^k}{k!}
    \right],
    \qquad
    t=N_t\Delta t_{\mathrm{CFL}} .
    \label{eq::exp_integration}
\end{equation}
This procedure is the linear counterpart of an $M^{th}$-order accurate
time-integration scheme. As in
section~\ref{sec::lin_stab_analysis_implementation}, the parameters
$\text{CFL} = \frac{1}{2}$ and $M = 5$ are chosen to ensure an
accurate approximation of the matrix exponential. Matrix $\bm{D}$ is
computed explicitly because $\bm{P}$, $\bm{Q}$ and $\bm{M}$ are sparse, and
their product requires $\mathcal{O}(N^2)$ FLOPS and preserves the
same sparsity pattern, with a memory cost of $\mathcal{O}(N^2)$.
Matrix $\bm{C}(t)$, however, is never formed explicitly. As in
section~\ref{sec::lin_stab_analysis_implementation}, a matrix-free formulation is implemented, where the computations
are organized in terms of sparse matrix-vector products, thereby
retaining computational and memory scaling of $\mathcal{O}(N^2)$. All
sparse matrix-vector products required to approximate the exponential
are offloaded to the GPU, when available, within the inner loop of the
Lanczos iteration.

\subsection{Freestream receptivity}
\label{sec::freestream_receptivity_appendix}

Consider a set of freestream disturbances defined on the
upstream side of the fitted shock, and parametrized by a
user-defined set of temporal frequencies:
\begin{equation*}
    q'_\infty(s,t) = \Re\left\{\sum_{l=0}^{N_\omega} \hat{q}'_{\infty,l}(s) \exp\left(-i\omega_l t\right)\right\}.
\end{equation*}
This formulation explicitly separates the time-harmonic behavior
(captured by $\exp(-i\omega_l t)$) from the spatial structure along the
shock (contained in $\hat{q}'_{\infty,l}(s)$).

In frequency space, each mode satisfies:
\begin{equation*}
    \dfrac{\partial q'_{\infty,l}(s,t)}{\partial t} = -i\omega_l
    q'_{\infty,l}(s,t),
\end{equation*}
which has the desired exponential solution $q'_{\infty,l}(s,t) =
\hat{q}'_{\infty,l}(s)\exp(-i\omega_l t)$.  The discretized advection
equation can be rewritten in matrix form as:
\begin{equation}
(\bm{\widehat{q}}'_\infty)_{lk} = \begin{bmatrix}
        (\widehat{\rho})'_{lk} \\
        (\widehat{\rho u})'_{lk} \\
        (\widehat{\rho v})'_{lk} \\
        (\widehat{\rho E})'_{lk}
    \end{bmatrix} \Rightarrow
 \dfrac{\partial \bm{\widehat{q}}'_\infty}{\partial t} = -i \bm{\Omega} \bm{\widehat{q}}'_\infty.
\end{equation}
The frequency operator is block diagonal:
\[
\bm{\Omega}
=
\operatorname{diag}
\left(
\omega_0\bm{I}_{4N_k},
\omega_1\bm{I}_{4N_k},
\ldots,
\omega_{N_\omega}\bm{I}_{4N_k}
\right),
\]
where $N_k$ is the number of shock points and the factor $4$
corresponds to the conservative variables $(\rho,\rho u,\rho v,\rho
E)$.  To map the harmonics back to physical space, the real parts of
all frequency contributions are summed:
\begin{equation}
(\bm{q}'_\infty)_{k} = \begin{bmatrix}
        (\rho)'_{k} \\
        (\rho u)'_{k} \\
        (\rho v)'_{k} \\
        (\rho E)'_{k}
    \end{bmatrix}, 
 \bm{q}_\infty' = \bm{\Sigma} \bm{\widehat{q}}_\infty'.
\end{equation}
This makes it possible to define a new discrete operator that evolves
both downstream perturbations $\bm{q}_e'$ and the imposed freestream
disturbances $\bm{q}'_\infty$:
\begin{equation}
    \frac{\mathrm{d}}{\mathrm{d}t}
    \begin{bmatrix}
        \widehat{\bm q}'_\infty \\
        \bm q'_e
    \end{bmatrix}
    =
    \begin{bmatrix}
        -i\bm{\Omega} & \bm{0} \\
        \bm{B}\bm{\Sigma} & \bm{L}
    \end{bmatrix}
    \begin{bmatrix}
        \widehat{\bm q}'_\infty \\
        \bm q'_e
    \end{bmatrix}
    \equiv
    \frac{\mathrm{d}\overline{\bm q}'}{\mathrm{d}t}
    =
    \overline{\bm L}\,\overline{\bm q}' .
\end{equation}
where $\overline{\bm{L}}$ and $\overline{\bm{q}}'$ denote the
operator and state of the extended system.

Matrix $\bm{B}$ appears because freestream perturbations can perturb
the flow downstream from the shock. For a given upstream perturbation
direction $\bm{v}_\infty$, its action is computed as:
\begin{equation}
    \bm{B}\bm{v}_\infty =
    \frac{
    \bm{N}(\bm{q}_{e,0},\bm{q}_{\infty,0}
    +\varepsilon\bm{v}_\infty)
    -
    \bm{N}(\bm{q}_{e,0},\bm{q}_{\infty,0})
    }{\varepsilon}.
    \label{eq:linearization_B}
\end{equation}
The previous linearization is performed by applying infinitesimal
perturbations to the upstream quantities and computing the downstream
response with the Rankine--Hugoniot jump conditions; see
appendix~\ref{sec::shock_solution}. This is effectively an automated
procedure for linearizing the shock jump conditions using the
analytical solution; more details are provided in
appendix~\ref{sec::linearization_details}. Section~\ref{sec::Verification_Disturbances_Shock}
presents the verification of the linearized shock--disturbance
interaction model.

We then define the freestream receptivity gain $G_T$, which measures
how much disturbance energy the post-shock region develops relative to
the freestream energy supplied to it. It compares the Chu energy of
the post-shock field at time $t$, $E(t)$, with the freestream Chu
energy $E_\infty^\text{ref}$ that has entered the domain over a
reference time $t^\text{ref}$,
\begin{equation*}
  G_T(t) = \frac{E(t)}{E_\infty^\text{ref}}.
\end{equation*}


$E(t)$ is the Chu energy of the post-shock perturbation,
\begin{equation*}
  E(t) = \bm{q}'(t)^\dagger \bm{Q}^\dagger \bm{M} \ \bm{Q} \ \bm{q}'(t),
\end{equation*}
where $E(t)$ measures the disturbance energy that has developed downstream of the shock.
To carry the optimization directly over the
incoming wave, the freestream disturbance is
mapped into the extended system by the prolongation operator
\begin{equation*}
  \bm{P}_\infty =
  \begin{bmatrix}
  \bm{I} \\
  \bm{0}
  \end{bmatrix},
  \qquad
  \bm{P}_\infty\,\widehat{\bm{q}}'_\infty(0) =
  \begin{bmatrix}
  \widehat{\bm{q}}'_\infty(0) \\
  \bm{0}
  \end{bmatrix},
\end{equation*}
which places it in the upper block and sets the post-shock field
initially to rest. Additionally, the disturbances downstream of the shock are retrieved from the extended system by the projection operator
\begin{equation*}
  \bm{P}_D =
  \begin{bmatrix}
  \bm{0} \ \bm{I} 
  \end{bmatrix},
  \qquad
  \bm{P}_D\,\bm{\overline{q}'} =
  \bm{q}'_e.
\end{equation*}
The state evolution of the downstream disturbance can then be expressed in terms of the freestream disturbances
\begin{equation*}
  \bm{q}'_e(t) = \bm{R}(t)\,\widehat{\bm{q}}'_\infty(0),
  \qquad
  \bm{R}(t) = \bm{P}_D \exp\left(\bm{\overline{L}}t\right)\,\bm{P}_\infty,
\end{equation*}
where $\bm{\overline{L}}$ is the extended (autonomous
harmonic-forcing) operator. The action of $\bm{R}(t)$ is evaluated matrix-free
with the same Taylor-series approximation of the matrix exponential
employed in the transient-growth calculation (equation~\ref{eq::exp_integration}).
This allows us to define the energy downstream of the shock solely in terms of the
freestream disturbance
\begin{equation}
  E(t) = \widehat{\bm{q}}'_\infty(0)^\dagger \ \bm{C}_\infty(t) \ \widehat{\bm{q}}'_\infty(0),
  \label{eq:numerator_freestream_gain}
\end{equation}
where $ \bm{C}_\infty(t) = \bm{R}(t)^\dagger  \ \bm{P}^\dagger \bm{Q}^\dagger \bm{M} \ \bm{Q} \ \bm{P} \ \bm{R}(t)$ and $^\dagger$ denotes the conjugate transpose.

The incident input energy is defined from the Chu-energy flux of the
prescribed freestream disturbance through the fitted shock. Let
$\mathcal{E}_\infty$ denote the freestream version of the energy
density in \eqref{eq::Chu_norm}. The instantaneous incident energy flux is
\begin{equation*}
  \dot E_\infty(t)
  =
  \int_{S_s}
  \mathcal{E}_\infty(t)\,
  U_\infty
  \left|\bm{e}_x\cdot\bm{n}_s\right|
  \,\mathrm{d} S ,
\end{equation*}
where $S_s$ is the fitted-shock surface, $\bm{n}_s$ is the shock
normal and $\bm{e}_x$ is the unit vector in the freestream direction.
The reference input energy is
\begin{equation*}
  E_\infty^{\mathrm{ref}}
  =
  \overline{\dot E_\infty}\,t^{\mathrm{ref}},
  \qquad
  t^{\mathrm{ref}}
  =
  \frac{m_D}{\dot m_\infty},
\end{equation*}
where the overbar denotes averaging over the harmonic input, $m_D =
\int_{V_D}\rho_0\,\mathrm{d} V$ is the base-flow mass contained in the
post-shock domain, and
\begin{equation*}
  \dot m_\infty
  =
  \int_{S_s}
  \rho_\infty U_\infty
  \left|\bm{e}_x\cdot\bm{n}_s\right|
  \,\mathrm{d} S
\end{equation*}
is the freestream mass flux through the fitted shock. Thus
$t^{\mathrm{ref}}$ is the time required for the base-flow mass flux
entering through the shock to replenish the mass contained in
$V_D$.  The flux is
assembled over the discrete elements of the shock front,
\begin{equation*}
    \overline{\dot{E}_\infty}
    =
    \sum_k
    U_\infty A_k n_k
    \overline{\mathcal{E}_{\infty,k}},
    \qquad
    n_k=\left|\bm e_x\cdot\bm n_{s,k}\right|.
\end{equation*}
Here $A_k$ is the area of shock element $k$, $\bm n_{s,k}$ is the unit
normal to the fitted shock, and $n_k$ is the nonnegative projection of
that normal onto the freestream direction.  Using the previous
discrete definition, the reference energy can be expressed as
\begin{equation}
  E_\infty^{\mathrm{ref}}
  =
  \overline{\dot E_\infty}\,t^{\mathrm{ref}} =
  \widehat{\bm{q}}'_\infty(0)^{\dagger} \bm{D}_\infty \ \widehat{\bm{q}}'_\infty(0), \ \bm{D}_\infty = \bm{Q}^\dagger_\infty \bm{M}_\infty \bm{Q}_\infty,
  \label{eq:reference_energy_discrete}
\end{equation}
where $\bm{Q}_\infty$ is the coordinate transformation matrix for the freestream disturbances $(\bm{Q}_\infty)_{lk} = {Q_V}_k$. The associated weight matrix
$\bm{M}_\infty$ computes the weighted integral over all harmonics,
\begin{equation*}
(\bm{M}_\infty)_{lk}
=
\begin{cases}
{W_V}_k A_k n_k U_\infty t^{\mathrm{ref}},
& l=0, \\[4pt]
\dfrac{{W_V}_k A_k n_k U_\infty t^{\mathrm{ref}}}{2},
& l>0,
\end{cases}
\qquad
n_k=\left|\bm e_x\cdot\bm n_{s,k}\right|.
\end{equation*}
where the factor $\tfrac{1}{2}$ for $l > 0$ is the infinite-time
average of the harmonic flux $\left(\overline{\sin^2} = \tfrac{1}{2}\right)$.
Finally, combining equations~\ref{eq:numerator_freestream_gain} and \ref{eq:reference_energy_discrete}
\begin{equation*}
G_T(t) =  \frac{\widehat{\bm{q}}'_\infty(0)^\dagger \bm{C}_\infty(t) \ \widehat{\bm{q}}'_\infty(0)}{\widehat{\bm{q}}'_\infty(0)^\dagger \bm{D}_\infty \ \widehat{\bm{q}}'_\infty(0)}.
\end{equation*}
Since $\bm{C}_\infty(t)$ and $\bm{D}_\infty$ are Hermitian matrices, the
Rayleigh quotient problem at a given time $t$ is again solved with the
Lanczos algorithm, using the ARPACK \citep{lehoucq1998arpack}
implementation of the Implicitly Restarted Lanczos Method, as in the
previous section. The matrix $\bm{D}_\infty$ is formed explicitly: since
$\bm{Q}_\infty$ and $\bm{M}_\infty$ are sparse, their product costs
$\mathcal{O}(N^2)$ FLOPS, preserves the same sparsity pattern, and
requires $\mathcal{O}(N^2)$ memory. The matrix $\bm{C}_\infty(t)$, by
contrast, is never formed explicitly. Following
appendix~\ref{sec::transient_growth_appendix}, we adopt a matrix-free
formulation in which all computations are expressed as sparse
matrix--vector products, thereby retaining the $\mathcal{O}(N^2)$
computational and memory scaling. Within the inner loop of the Lanczos
iteration, the sparse matrix--vector products needed to approximate the
matrix exponential are offloaded to the GPU when one is available.

\bibliographystyle{cas-model2-names}
\bibliography{cas-refs}

\end{document}